\documentclass[a4paper,11pt]{article}
\pdfoutput=1 

\usepackage{jheppub} 
\usepackage[T1]{fontenc}
\usepackage{booktabs}
\usepackage{xspace}
\usepackage{dcolumn}
\usepackage{hyperref}
\usepackage{caption}
\usepackage
[subrefformat=parens,position=top,skip=-15pt,margin=15pt,justification=justified,singlelinecheck=false]
{subcaption}
\usepackage{slashed}
\usepackage{comment}
\usepackage[colorinlistoftodos, shadow]{todonotes}
\usepackage{amsmath}
\usepackage{graphicx}

\allowdisplaybreaks[1]

\def\refeq#1{\mbox{(\ref{#1})}}
\def\reffi#1{\mbox{Figure~\ref{#1}}}

\def\refta#1{\mbox{Table~\ref{#1}}}
\def\reftas#1{\mbox{Tables~\ref{#1}}}
\def\refse#1{\mbox{Section~\ref{#1}}}
\def\refses#1{\mbox{Sections~\ref{#1}}}
\def\refapp#1{\mbox{App.~\ref{#1}}}
\def\reffi#1{\mbox{Fig.~\ref{#1}}}

\def\refta#1{\mbox{Table~\ref{#1}}}
\def\reftas#1{\mbox{Tables~\ref{#1}}}
\def\refse#1{\mbox{Section~\ref{#1}}}
\def\refses#1{\mbox{Sections~\ref{#1}}}
\def\refapp#1{\mbox{App.~\ref{#1}}}

\def\citere#1{\mbox{Ref.~\cite{#1}}}
\def\citeres#1{\mbox{Refs.~\cite{#1}}}

\newcommand{\cmm}[1]{\ensuremath{#1}\ifmmode\else{}\fi}
\newcommand{\nmc}[2]{\newcommand{#1}{\cmm{#2}}}
\nmc{\al}{\alpha}
\nmc{\be}{\beta}
\nmc{\de}{\delta}
\nmc{\si}{\sigma}

\newcommand{\ri}{\mathrm i}
\newcommand{\ie}{{i.e.}\ }
\newcommand{\eg}{{e.g.}\ }
\newcommand{\cf}{{c.f.}\ }
\nmc{\rd}{\mathrm{d}}

\def\beq{\begin{equation}}
\def\eeq{\end{equation}}

\newcommand{\PH}{\ensuremath{\text{H}}\xspace}
\newcommand{\PHone}{\ensuremath{{\text{H}_1}}\xspace}
\newcommand{\PHtwo}{\ensuremath{{\text{H}_2}}\xspace}
\newcommand{\PHpm}{\ensuremath{\text{H}^\pm}\xspace}
\newcommand{\PHa}{\ensuremath{A_0}\xspace}

\newcommand{\Ph}{\ensuremath{\text{h}}\xspace}
\newcommand{\Pj}{\ensuremath{\text{j}}\xspace}
\newcommand{\Pp}{\ensuremath{\text{p}}\xspace}

\newcommand{\Pm}{\ensuremath{\mu}\xspace}
\newcommand{\Pnm}{\change{\ensuremath{\nu_\mu}}}

\newcommand{\Pt}{\ensuremath{\text{t}}\xspace}

\newcommand{\PW}{\ensuremath{\text{W}}\xspace}
\newcommand{\PZ}{\ensuremath{\text{Z}}\xspace}

\newcommand{\MH}{\ensuremath{M_\PH}\xspace}

\newcommand{\MW}{\ensuremath{M_\PW}\xspace}

\newcommand{\MZ}{\ensuremath{M_\PZ}\xspace}

\newcommand{\GeV}{\ensuremath{\,\text{GeV}}\xspace}
\newcommand{\TeV}{\ensuremath{\,\text{TeV}}\xspace}

\newcommand{\UV}{{\mathrm{UV}}}
\newcommand{\rw}{{\mathrm{w}}}
\nmc{\alem}{\alpha_{\mathrm{em}}}
\nmc{\sw}{s_{\rw}}
\nmc{\cw}{c_{\rw}}

\nmc{\g}{g_2}
\nmc{\gy}{g_1}
\nmc{\Gf}{G_\mathrm{F}}

\newcommand{\cL}{\ensuremath{\mathcal{L}}\xspace}
\newcommand{\cM}{\ensuremath{\mathcal{M}}\xspace}
\newcommand{\rT}{\ensuremath{\text{T}}\xspace}
\newcommand{\rL}{\ensuremath{\text{L}}\xspace}
\newcommand{\rR}{\ensuremath{\text{R}}\xspace}

\newcommand{\hc}{\ensuremath{\text{h.c.}}\xspace}
\nmc{\rB}{{\rm B}}
\nmc{\rs}{{\rm s}}

\newcommand{\fjts}{{\it FJ~Tadpole Scheme}\xspace}

\nmc{\Msb}{M_{\rm sb}}

\nmc{\ftone}{t_{s}}
\nmc{\tab}{t_{\alpha\beta}}

\nmc{\tHoneHone}{t_{H_1H_1}}
\nmc{\tHoneHtwo}{t_{H_1H_2}}
\nmc{\tHtwoHtwo}{t_{H_2H_2}}
\nmc{\tHaHa}{t_{\Ha\Ha}}
\nmc{\tHpmHpm}{t_{H^\pm H^\pm}}
\nmc{\tGzGz}{t_{G_0G_0}}
\nmc{\tGpmGpm}{t_{G^\pm G^\pm}}
\nmc{\tGzHa}{t_{G_0\Ha}}
\nmc{\tGpmHpm}{t_{G^\pm H^\pm}}

\newcommand{\nunuone}{{\nu_1}\bar\nu_1}
\newcommand{\nunutwo}{{\nu_2}\bar\nu_2}
\newcommand{\nunui}{{\nu_i}\bar\nu_i}
\newcommand{\nunuj}{{\nu_j}\bar\nu_j}

\newcommand{\Prophecy}{{\sc Prophecy4f}}

\let\lsim\lesssim

\def\draftdate{\relax}
\def\mda{\relax}
\def\mua{\relax}
\def\mla{\relax}
\def\draft{
\def\thtystars{******************************}
\def\sixtystars{\thtystars\thtystars}
\typeout{}
\typeout{\sixtystars**}
\typeout{* Draft mode!
         For final version remove \protect\draft\space in source file *}
\typeout{\sixtystars**}
\typeout{}
\def\draftdate{\today}
\def\mua{\marginpar[\boldmath\hfil$\uparrow$]%
                   {\boldmath$\uparrow$\hfil}%
                    \typeout{marginpar: $\uparrow$}\ignorespaces}
\def\mda{\marginpar[\boldmath\hfil$\downarrow$]%
                   {\boldmath$\downarrow$\hfil}%
                    \typeout{marginpar: $\downarrow$}\ignorespaces}
\def\mla{\marginpar[\boldmath\hfil$\rightarrow$]%
                   {\boldmath$\leftarrow $\hfil}%
                    \typeout{marginpar: $\leftrightarrow$}\ignorespaces}
\def\Mua{\marginpar[\boldmath\hfil$\Uparrow$]%
                   {\boldmath$\Uparrow$\hfil}%
                    \typeout{marginpar: $\uparrow$}\ignorespaces}
\def\Mda{\marginpar[\boldmath\hfil$\Downarrow$]%
                   {\boldmath$\Downarrow$\hfil}%
                    \typeout{marginpar: $\downarrow$}\ignorespaces}
\def\Mla{\marginpar[\boldmath\hfil{$\Rightarrow$}]%
                   {\boldmath{$\Leftarrow $}\hfil}%
                    \typeout{marginpar:$\leftrightarrow$}\ignorespaces}
\def\muanick{\marginpar[\boldmath\hfil$\uparrow$]%
                   {\boldmath$\textcolor{blue}\uparrow$\hfil}%
                    \typeout{marginpar:\textcolor{blue} $\uparrow$}\ignorespaces}
\def\mdanick{\marginpar[\boldmath\hfil$\downarrow$]%
                   {\boldmath$\textcolor{blue}\downarrow$\hfil}%
                    \typeout{marginpar: $\downarrow$}\ignorespaces}
\def\mlanick{\marginpar[\boldmath\hfil$\rightarrow$]%
                   {\boldmath$\textcolor{blue}\leftarrow $\hfil}%
                    \typeout{marginpar: $\leftrightarrow$}\ignorespaces}
\overfullrule 5pt
\oddsidemargin 15mm
\marginparwidth 29mm
}

\nmc{\Hhat}{\hat H}
\nmc{\hhat}{\hat h}
\nmc{\Phihat}{\hat \Phi}
\nmc{\phihat}{\hat \phi}
\nmc{\chihat}{\hat \chi}
\nmc{\etahat}{\hat \eta}
\nmc{\rhohat}{\hat \rho}
\nmc{\thetahat}{\hat \theta}
\nmc{\sihat}{\hat\sigma}
\nmc{\Phione}{\Phi_1}
\nmc{\Phitwo}{\Phi_2}
\nmc{\Zhat}{\hat Z}
\nmc{\Hahat}{\hat A_0}
\nmc{\Gzhat}{\hat G_0}
\nmc{\Xhat}{\hat X}
\nmc{\Yhat}{\hat Y}
\nmc{\Honehat}{\hat H_1}
\nmc{\Htwohat}{\hat H_2}

\nmc{\vone}{v_1}
\nmc{\vtwo}{v_2}
\nmc{\etaone}{\eta_1}
\nmc{\etatwo}{\eta_2}
\nmc{\etaplet}{\boldsymbol{\eta}}
\nmc{\chione}{\chi_1}
\nmc{\chitwo}{\chi_2}
\nmc{\chiplet}{\boldsymbol{\chi}}
\nmc{\Hone}{H_1}
\nmc{\Htwo}{H_2}
\nmc{\Hplet}{\mathbf{H}}
\nmc{\phionepm}{\phi_1^\pm}
\nmc{\phitwopm}{\phi_2^\pm}
\nmc{\Ha}{A_0}

\nmc{\tb}{t_\be}
\nmc{\ca}{c_\al}
\nmc{\catwo}{c^2_\al}
\nmc{\satwo}{s^2_\al}
\nmc{\sa}{s_\al}
\nmc{\stwoa}{s_\al}
\nmc{\cbe}{c_\be}
\nmc{\ctwobe}{c_{2\be}}
\nmc{\cbetwo}{c^2_\be}
\nmc{\sbe}{s_\be}
\nmc{\sbetwo}{s^2_\be}
\nmc{\cab}{c_{\al\be}}
\newcommand{\sab}{\ensuremath{s_{\alpha\beta}}}

\nmc{\dth}{\delta t_{H}}
\nmc{\dthhat}{\delta t_{\hat{H}}}
\nmc{\Th}{T^{H}}
\nmc{\Thhat}{T^{\Hhat}}
\nmc{\THone}{{T}^{H_1}}
\nmc{\THtwo}{{T}^{H_2}}
\nmc{\dtHone}{{\delta t}_{H_1}}
\nmc{\dtHtwo}{{\delta t}_{H_2}}
\nmc{\dtHhatone}{{\delta t}_{\hat H_1}}
\nmc{\dtHhattwo}{{\delta t}_{\hat H_2}}
\nmc{\MHone}{{M}_\PHone}
\nmc{\MHtwo}{{M}_\PHtwo}
\nmc{\MHa}{{M}_{\PHa}}
\nmc{\MHpm}{{M}_{\text{H}^\pm}}
\nmc{\MHsone}{{M}^2_\PHone}
\nmc{\MHstwo}{{M}^2_\PHtwo}
\nmc{\MHsa}{{M}^2_{\PHa}}
\nmc{\MHspm}{{M}^2_{\PHpm}}

\newcommand{\THDM}{THDM\xspace}
\newcommand{\HSESM}{HSESM\xspace}
\newcommand{\MSSM}{MSSM\xspace}
\newcommand{\MSbar}{\ensuremath{\overline{\text{MS}}}\xspace}
\newcommand{\PRTS}{PRTS\xspace}
\newcommand{\FJTS}{FJTS\xspace}
\newcommand{\MSbarPRTS}{{\MSbar}(\text{\PRTS})\xspace}
\newcommand{\MSbarFJTS}{{\MSbar}(\text{\FJTS})\xspace}
\newcommand{\OS}{OS\xspace}
\newcommand{\OSone}{OS1\xspace}
\newcommand{\OStwo}{OS2\xspace}
\newcommand{\OSonetwo}{OS12\xspace}
\newcommand{\RSBFM}{BFMS\xspace}

\nmc{\vshift}{\bar{v}}

\newcommand{\HAWKTwo}{{\sc Hawk~2.0}\xspace}
\newcommand{\Recola}{{\sc Recola}\xspace}
\newcommand{\RecolaTwo}{{\sc Recola2}\xspace}

\newcommand{\change}[1]{{#1}}

\title{Renormalization of mixing angles}
\subheader{\today}

\author{Ansgar Denner$^1$,}
\author{Stefan Dittmaier$^2$,}
\author{Jean-Nicolas Lang$^3$}

\affiliation{$^1$ %
        Universit\"at W\"urzburg, %
        Institut f\"ur Theoretische Physik und Astrophysik, \\  %
        97074 W\"urzburg, %
        Germany%
}
\affiliation{$^2$ %
        Albert-Ludwigs-Universit\"at Freiburg, %
        Physikalisches Institut, %
        79104 Freiburg, %
        Germany%
}
\affiliation{$^3$ %
        Universit\"at Z\"urich, 
        Physik-Institut, 
        CH-8057 Z\"urich,
        Switzerland%
}

\emailAdd{ansgar.denner@physik.uni-wuerzburg.de}
\emailAdd{stefan.dittmaier@physik.uni-freiburg.de}
\emailAdd{jlang@physik.uzh.ch}

\abstract{We discuss the renormalization of mixing angles for
  theories with extended scalar sectors. 
  Motivated by shortcomings of existing schemes for mixing angles,
  we review existing renormalization schemes and introduce new ones
  based on on-shell conditions or symmetry requirements such as rigid
  or background-field gauge invariance.  Considering in particular the
  renormalization of the mixing angles in the Two-Higgs-Doublet Model
  and the Higgs-Singlet Extension of the Standard Model, we compare
  electroweak corrections within these models for a selection of
  renormalization schemes.  As specific examples, we present
  next-to-leading-order results on the four-fermion decays of heavy
  and light CP-even Higgs bosons, $\PH_1/\PH_2\to\PW\PW/\PZ\PZ\to4f$,
  and on electroweak Higgs-boson production processes, i.e.\ 
  Higgs-strahlung and vector-boson fusion.  We find that our new
  proposals for on-shell and symmetry-based renormalization conditions
  are well-behaved for the considered benchmark scenarios in both
  models.  }

\begin{document} 

\mbox{}\hfill FR-PHENO-2018-10\\
\mbox{}\hfill ZU-TH 31/18

\maketitle
\flushbottom

\section{Introduction}
\label{se:introduction}

After the discovery of a Higgs boson at the Large Hadron Collider
(LHC)~\cite{Chatrchyan:2012xdj,Aad:2012tfa}, the investigation of the
Higgs sector is still of prime importance for particle
physics. Theories with extended Higgs sectors typically contain
additional scalar multiplets leading to physical scalar states that
mix. Simple examples of such extensions are the Two-Higgs-Doublet Model (\THDM) 
\cite{Gunion:2002zf,Branco:2011iw} and the Higgs-Singlet Extension of
the Standard Model (\HSESM)
\cite{Schabinger:2005ei,Patt:2006fw,Bowen:2007ia}. For a precise study
of such theories, next-to-leading-order (NLO) QCD and electroweak (EW)
corrections have to be taken into account. This requires a
renormalization of these models and thus the renormalization of mixing
angles or, more generally, of mixing matrices.

The need for renormalization of mixing matrices already appears in the
Standard Model (SM) where the quark-mixing matrix has to be
renormalized. While this is phenomenologically unimportant owing to
the smallness of the down-type quark masses, the problem has
nevertheless found quite some interest in the literature, and the
corresponding theoretical developments have also influenced the work
on the renormalization of mixing matrices in scalar sectors, which is
the subject of this paper.  A first renormalization condition for the
quark-mixing matrix based on on-shell field-renormalization constants
of the quark fields was proposed in
\citeres{Denner:1990yz,Denner:1991kt}.  This prescription is simple,
symmetric in the fields that mix, and smoothly connected to the limit
of degenerate quark masses. Later it was discovered
\cite{Gambino:1998ec} that the straightforward use of the
renormalization condition of \citeres{Denner:1990yz,Denner:1991kt}
gives rise to gauge-parameter-dependent counterterms for the
quark-mixing matrix and thus to a gauge-parameter-dependent
parametrization of $S$-matrix elements in terms of renormalized
parameters. In the sequel, various proposals were made for a
gauge-parameter-independent renormalization of the quark-mixing matrix
\cite{Gambino:1998ec,Balzereit:1998id,Pilaftsis:2002nc,Diener:2001qt,Kniehl:2006rc,Kniehl:2009kk}.
Typically, these are cumbersome to apply, their generalization beyond
one-loop order remains unclear, and/or they potentially lead to
singularities in the $S$-matrix elements for degenerate quark masses.
The last problem occurs, in particular, in the modified minimal
subtraction ($\MSbar$) scheme. A gauge-independent, symmetric,
physical renormalization condition was proposed in
\citere{Denner:2004bm}.  It was also suggested to define the
quark-mixing matrix counterterm from the quark-field renormalization
constants calculated in the 't~Hooft--Feynman gauge
\cite{Yamada:2001px}.  Generalizing this idea, it was argued in
\citere{Pilaftsis:2002nc} that any renormalization scheme for the
quark-mixing matrix may be viewed as a gauge-invariant scheme by
definition, in the sense that $S$-matrix elements remain invariant if
the gauge used in the calculation of the loop corrections and all
other renormalization constants is changed, while keeping the defining
gauge for the renormalization constants of the quark-mixing matrix
fixed.

The need for suitable renormalization schemes for mixing angles
becomes more important in extensions of the SM. Specific examples are
models with additional Higgs bosons, additional vector bosons, or
additional fermions. In particular, for the renormalization of mixing
angles in the scalar sector a variety of schemes 
were used in the literature.  
Specifically, the renormalization of the mixing angle $\beta$ in the
Minimal Supersymmetric Standard Model (MSSM) 
was discussed in
\citeres{Chankowski:1992er,Dabelstein:1994hb,Freitas:2002um,Baro:2008bg,Baro:2009gn}.

The renormalization of the mixing angles $\alpha$ and $\beta$ in the \THDM\ 
was considered in
\citeres{Kanemura:2004mg,LopezVal:2009qy,Kanemura:2014dja,Krause:2016oke,Denner:2016etu,Denner:2017vms,Altenkamp:2017ldc}.
Kanemura and collaborators
\cite{Kanemura:2004mg,Kanemura:2014dja,Kanemura:2015mxa} used the
vanishing of the renormalized non-diagonal ``on-shell'' scalar 2-point
functions to fix the mixing-angle renormalization.%
\footnote{We put ``on-shell'' in quotation marks here, since we want
to reserve this word to conditions that are based on $S$-matrix elements
rather than simply taking momenta on their mass shell in more general
quantities such as Green functions, self-energies, etc..}
Despite the choice
of ``on-shell'' momenta these conditions do not derive from $S$-matrix
elements and are gauge dependent. In \citere{LopezVal:2009qy}, the
mixing angle $\alpha$ was fixed 
by the condition that the mixing self-energy of the CP-even Higgs
bosons vanishes ``on-shell'', while $\beta$ was renormalized requiring
that the ratio of vacuum expectation values (vevs), $\vtwo/\vone$, is
expressed in terms of the ``true'' vacua following the treatment of
\citeres{Chankowski:1992er,Dabelstein:1994hb} in the \MSSM.  The
authors of \citere{Krause:2016oke} employ the renormalization
conditions of \citere{Kanemura:2004mg} within the \fjts by Fleischer
and Jegerlehner~\cite{Fleischer:1980ub} and define gauge-independent
counterterms based on the ``pinch-technique''
prescription~\cite{Binosi:2004qe}.%
\footnote{Following the arguments of
  \citeres{Denner:1994nn,Denner:1994xt} we consider the ``pinch
  technique'' just as one of many physically equivalent choices to fix
  the gauge arbitrariness in off-shell quantities (related to the
  't~Hooft--Feynman gauge of the quantum fields in the
  background-field method) rather than singling out ``its
  gauge-invariant part'' in any sense.}

In the series of papers~\cite{Denner:2016etu,Denner:2017vms} the gauge
dependence of the $\MSbar$ definition of mixing angles with respect to
different tadpole counterterm schemes was investigated,
and predictions were compared against before-mentioned schemes based
on mixing energies.  Finally, in \citere{Altenkamp:2017ldc} new
(gauge-independent) $\MSbar$ schemes were introduced, replacing a
mixing-angle definition by the $\MSbar$ renormalization of a coupling
parameter of the Higgs potential.

For the \HSESM the mixing-angle renormalization was discussed in
\citeres{Kanemura:2015fra,Bojarski:2015kra,Denner:2017vms,Altenkamp:2018bcs}.
In \citere{Kanemura:2015fra}, the renormalization scheme of
\citere{Kanemura:2004mg} was transferred to the \HSESM.  The authors
of \citere{Bojarski:2015kra} discuss different renormalization schemes
based on conditions on the scalar mixing energy or $\MSbar$
renormalization. While in \citere{Denner:2017vms} an $\MSbar$ scheme
was compared with schemes based on on-shell self-energies, in
\citeres{Altenkamp:2018bcs} different $\MSbar$ schemes were studied.

The purpose of this paper is a discussion of renormalization
prescriptions and schemes for mixing angles in general scalar sectors
of gauge theories and a comparison of different schemes in concrete
phenomenological applications in the \THDM and the \HSESM.  We
critically review existing renormalization prescriptions and introduce
new ones that exhibit several desirable properties
\cite{Freitas:2002um}.  In particular, we introduce genuine on-shell
renormalization conditions for mixing angles based on combinations of
suitable $S$-matrix elements and put renormalization schemes based on
symmetry requirements such as rigid invariance or background-field
gauge invariance on a general footing.

This paper is organized as follows. In \refse{se:prelim} we specify
some useful definitions and conventions. In \refse{se:schemes} we
review and introduce renormalization schemes for mixing angles in
scalar sectors.  We begin with a discussion of existing $\MSbar$
renormalization schemes, followed by sections where we construct new
on-shell renormalization schemes and renormalization schemes based on
symmetries.  In \refse{se:results} we provide a numerical discussion
of renormalization schemes in applications to Higgs decays into
4~fermions as well as Higgs production at the LHC via Higgs-strahlung
or weak vector-boson fusion.  After the conclusion in
\refse{se:Conclusions}, we give translation rules of our conventions
to other formulations in the literature in \refapp{app:conventions}.
Further appendices provide explicit analytical results for quantities
used in the various renormalization schemes, including scalar
self-energies in the background-field method, the tadpole
contributions to the scalar self-energies, vertex corrections for
on-shell schemes, and a discussion of background-field Ward identities
in different tadpole counterterm schemes.

\section{Preliminaries}
\label{se:prelim}

\subsection{Renormalization transformation for mixing fields}

As specific examples for the renormalization of mixing angles we
consider the mixing of scalar fields in the \THDM and the \HSESM.
Both theories involve two physical CP-even scalar bosons. Let the
corresponding fields in the symmetric basis be $\etaone$ and $\etatwo$
and the fields in the physical mass-eigenstate basis $H_1$ and $H_2$.
The fields are related via a rotation
\begin{align}
  \etaplet = \left(\begin{array}{c}
    \etaone\\
    \etatwo
  \end{array}\right)=
  R(\al)
  \left(\begin{array}{c}
    \Hone\\
    \Htwo
  \end{array}\right) =  R(\al) \Hplet, \qquad
  R(\alpha)=
  \left(\begin{array}{cc}
        \ca & -\sa\\
        \sa & ~\ca
  \end{array}\right),
  \label{eq:scalar_rotation}
\end{align}
where we use the shorthand notations $\ca=\cos\alpha$ and
$\sa=\sin\al$. In the general case of more than two mixing fields
$H_i$, the matrix $R(\boldsymbol{\al})$ depends on a set of mixing
angles $\boldsymbol{\al}=\{\al_i\}$.

Performing the renormalization in the physical basis in the complete
on-shell scheme \cite{Aoki:1982ed,Denner:1991kt}, the
renormalization transformations for the mixing angle and the field
renormalization constants of the scalar fields read
\begin{align}
\al_{\rB} ={}& \al+\de\al,\\
\Hplet_{\rB} ={}& (Z^H)^{1/2} \Hplet
\end{align}
with
\begin{align}
  (Z^H)^{1/2}={}&
  \left(\begin{array}{cc}
        (Z^H)_{11}^{1/2} &  (Z^H)_{12}^{1/2}\\
        (Z^H)_{21}^{1/2} &  (Z^H)_{22}^{1/2}
  \end{array}\right)
= {\bf 1} + \frac{1}{2}\de Z^H
= \left(\begin{array}{cc}
        1 + \frac{1}{2}\de Z^H_{11} &   \frac{1}{2}\de Z^H_{12}\\
         \frac{1}{2}\de Z^H_{21} &  1+\frac{1}{2}\de Z^H_{22}
  \end{array}\right),
\label{eq:matrix_renormalization_scalars}
\end{align}
where bare quantities carry an index $\rB$, and $\de\al$ and $\de
Z^H_{ij}$ represent the 
renormalization constants for the mixing angle $\alpha$ and
the scalar fields corresponding to mass eigenstates. In the complete on-shell scheme of
\citeres{Aoki:1982ed,Denner:1991kt}, the non-diagonal
field-renormalization constants are fixed as
\beq\label{eq:zij_onshell}
\delta Z^H_{ij} = \frac{2}{M_{\PH_i}^2-M_{\PH_j}^2}\Sigma_{ij}(M_{\PH_j}^2), \quad i\ne j,
\eeq
where $M_{\PH_i}$ and $M_{\PH_j}$ denote the masses of the
corresponding scalar bosons $\PH_i$ and $\PH_j$, and $\Sigma_{ij}$
their mixing energy.  In this paper we consistently use the convention
that the self and mixing energies include explicit and implicit
tadpole contributions, \ie they are defined as the higher-order
contributions to the inverse propagators which at one-loop order can
be depicted as
\begin{align}
  \Sigma_{ij} = 
\setlength{\unitlength}{1pt}
  \raisebox{-7pt}{\includegraphics{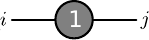}} 
  \;
  +
  \;
  \raisebox{-4pt}{\includegraphics{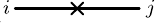}} 
  \;
  +
  \;
  \raisebox{-2.0pt}{\includegraphics{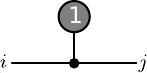}} 
  \;
  +
  \;
  \raisebox{-2.0pt}{\includegraphics{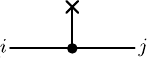}} 
  \label{eq:2ptadoles}.
\end{align}
The first contribution is the bare loop one-particle-irreducible (1PI)
energy and the second term the corresponding 2-point tadpole
counterterm.  The third and forth terms are the explicit tadpole loop
and tadpole counterterm contributions, respectively.  Note that no
counterterms other than the ones from the tadpoles (such as mass or
field renormalization constants) are included in $\Sigma_{ij}$.  In
the renormalization of the SM Higgs sector described in
\citeres{Denner:1994xt,Denner:1991kt}, the vev $v$ is renormalized in
such a way that the Higgs field has vanishing vev, and consequently,
all explicit tadpole contributions vanish, i.e.\ the third and forth
term in \eqref{eq:2ptadoles} add up to zero. However, implicit tadpole
counterterms may remain in the second term, but their explicit form
depends on the considered fields and on the tadpole counterterm scheme
in use. If we use self-energies {\it without any} tadpole
contributions, we will indicate this with an extra index 1PI for
one-particle irreducible.  
If not stated otherwise, the formulas are
given in the Fleischer and Jegerlehner tadpole counterterm scheme as
used in \citere{Denner:2016etu} for which the construction of the implicit tadpole
contributions to 2-point functions is given in
\refapp{app:tadpoles}.  This convention for the tadpole contributions
in $\Sigma_{ij}$ is in agreement with the one of
\citeres{Fleischer:1980ub,Actis:2006}.

In the presence of mixing, the well-known problem of degenerate states
in time-indepen\-dent perturbation theory of quantum mechanics also
appears in quantum field theory and becomes first apparent in the
one-loop renormalization.  The condition \eqref{eq:zij_onshell} shifts
corrections from the mixing of states that are induced by external
(non-diagonal) self-energy insertions to vertex counterterms.  For
degenerate masses, $M_{\PH_i}\to M_{\PH_j}$, the constants $\delta
Z^H_{ij}$ become singular, and thus also the $S$-matrix elements with
external $H_i,H_j$ fields unless this singularity is cancelled by some
other contribution.  As the loop diagrams are regular in this limit,
the cancellation should come from another counterterm. This is where
the mixing angles come into play, and one expects that an appropriate
renormalization of the mixing angles can make the $S$-matrix
well-behaved as we will sketch in the following.%
\footnote{In analogy to time-independent perturbation theory of
  quantum mechanics, the freedom of renormalization of the mixing
  angles corresponds to the freedom in choosing a basis for the
  (quasi) degenerate subsystem.}

Combining \refeq{eq:scalar_rotation} for bare fields with
\refeq{eq:matrix_renormalization_scalars} and using 
\beq
R(\boldsymbol{\alpha}_{\rB}) = R(\boldsymbol{\alpha}) +  \delta
R(\boldsymbol{\alpha},\de\boldsymbol{\alpha}),
\eeq
yields to one-loop accuracy
\beq
\etaplet_{\rB} = R(\boldsymbol{\alpha})\biggr(1+R^{\rT}(\boldsymbol{\alpha})\delta
R(\boldsymbol{\alpha},\de\boldsymbol{\alpha})+\frac{1}{2}\de Z^H\biggl)\Hplet.
\eeq
Thus, if the mixing matrix $R$ in the Lagrangian results only from the
rotation between symmetric and mass eigenstates as in
\refeq{eq:scalar_rotation}, the counterterm to the mixing matrix
appears only in the combination 
\beq\label{eq:dzmdR}
R^{\rT}(\boldsymbol{\alpha})\delta
R(\boldsymbol{\alpha},\de\boldsymbol{\alpha})+\frac{1}{2}\de Z^H
\eeq
 in $S$-matrix elements.
For the specific case of a single mixing angle this gives rise to the
two combinations
\beq\label{eq:dzmda}
{-\de\al} +\frac{1}{2}\de Z^H_{12}, \qquad\de\al +\frac{1}{2}\de Z^H_{21}.
\eeq
Choosing the mixing-angle counterterm appropriately allows one to
cancel the singularity for degenerate masses in $S$-matrix elements
arising from the denominators in \refeq{eq:zij_onshell}.

If the mixing angle $\al$ is promoted to a physical parameter upon
eliminating some bare parameter of the Lagrangian in its favour, the
counterterm $\de\al$ appears also independently of $\de Z^H_{12/21}$
in $S$-matrix elements.  This is the case in the \THDM\ and the
\HSESM\ if a quartic coupling parameter $\lambda_i$ of the Higgs
potential is eliminated in favour of the mixing angle $\alpha$ in the
sector of CP-even Higgs bosons.  However, it turns out that in this
case the mixing-angle counterterm appears in addition to
\refeq{eq:dzmda} only in the combination
$(M_{\PH_1}^2-M_{\PH_2}^2)\de\al$. This can easily be verified by
considering the coupling parameter $\lambda_i$ as a function of Higgs
masses and mixing angles (see, for instance, Eq.~(3.10) in
\citere{Denner:2016etu} or Eq.~(2.20) in \citere{Altenkamp:2017ldc}).
This statement can be generalized to mixing angles in more general
theories. Note, however, that the situation is different for mixing
angles defined via vevs, such as $\beta$ in the \THDM\ or the MSSM.

\subsection{Higgs-Singlet Extension of the Standard Model}

In the \HSESM we associate $\etatwo$ in Eq.~\eqref{eq:scalar_rotation} with the
scalar component of the Higgs doublet
\beq
\Phi=\left(\begin{array}{c} \phi^+ \\
\frac{1}{\sqrt{2}}(\vtwo+\etatwo+\ri\chi) 
\end{array}\right),
\eeq
where $\vtwo$ is the corresponding vev and $\phi^+$, $\chi$ the
would-be Goldstone-boson fields.
The field $\etaone$ is the field excitation of the (canonically
normalized) Higgs singlet field
\beq
\si = \vone+\etaone,
\eeq
which acquires the vev $\vone$.
The Higgs potential of the considered variant of the \HSESM\ is
given by \cite{Schabinger:2005ei,Patt:2006fw,Bowen:2007ia,Pruna:2013bma}
\begin{align}
  V_{\mathrm\HSESM} = -\mu_2^2 \Phi^\dagger \Phi  - \frac{1}{2}\mu_1^2 \si^2
  +\frac{\lambda_2}{4} \left(\Phi^\dagger \Phi\right)^2
  +\frac{\lambda_1}{16} \si^4
  +\frac{\lambda_{3}}{2} \Phi^\dagger \Phi\,\si^2,
  \label{eq:hsesmpot}
\end{align}
which possesses a $\mathbb{Z}_2$ symmetry under $\si\to-\si$.
Translation rules of these conventions to the ones used in
\citeres{Denner:2017vms,Altenkamp:2018bcs} are given in \refapp{app:conventions}.
The masses $\MW$ and $\MZ$ of the weak gauge bosons are
given by
\begin{align}
  \MW = \frac{1}{2} \g \vtwo, \qquad
  \MZ = \frac{1}{2} \sqrt{\g^2 + \gy^2}\; \vtwo,
\end{align}
where $\g$ and $\gy$ are the SU(2) and U(1) gauge couplings, respectively.

\subsection{Two-Higgs-Doublet Model}

In the \THDM, $\etaone$ and $\etatwo$ are the neutral CP-even
scalar components of the two Higgs-doublet fields
\begin{align}
\Phi_i = \left(\begin{array}{c}
\phi_i^+ \\  \frac{1}{\sqrt{2}}(v_i+\eta_i+\ri\chi_i)
\end{array}\right), \quad i=1,2,
\end{align}
and we associate the mass eigenstate $\PH_1=\PH_{\mathrm{h}}=\PH$ 
with the heavy scalar and $\PH_2=\PH_{\mathrm{l}}=\Ph$ with the light one.  
The self-interaction of the two Higgs doublets is induced by the Higgs
potential \cite{Gunion:2002zf,Branco:2011iw}%
\footnote{For more details on this
model, its parametrization and renormalization 
consider also \citeres{Denner:2017vms,Altenkamp:2017ldc}, which follow the
conventions of the original references.}
\begin{align}
  V_{\mathrm{\THDM}}={}&m_1^2\Phione^{\dagger}\Phione+m_2^2\Phitwo^{\dagger}\Phitwo
    -m_{12}^2\left(\Phione^{\dagger}\Phitwo+\Phitwo^{\dagger}\Phione\right)\notag\\
    &+\frac{\lambda_1}{2}\left(\Phione^{\dagger}\Phione\right)^2
    +\frac{\lambda_2}{2}\left(\Phitwo^{\dagger}\Phitwo\right)^2
    +\lambda_3\left(\Phione^{\dagger}\Phione\right)\left(\Phitwo^{\dagger}\Phitwo\right)
    +\lambda_4\left(\Phione^{\dagger}\Phitwo\right)\left(\Phitwo^{\dagger}\Phione\right)\notag\\
    &+\frac{\lambda_5}{2}\left[\left(\Phione^{\dagger}\Phitwo\right)^2
    +\left(\Phitwo^{\dagger}\Phione\right)^2\right],
\label{eq:thdmpot}
\end{align}
which has a $\mathbb{Z}_2$ symmetry w.r.t.\ $\Phi_1\to-\Phi_1$ and
$\Phi_2\to\Phi_2$ that is only softly broken by the $m_{12}^2$ term.

The two scalar doublets in the \THDM contain, in addition to the
scalar fields $\eta_i$, also pseudoscalar $\chi_i$ and charged scalar
$\phi^+_i$ fields. These are transformed to the physical basis as
follows,
\begin{align}
  \left(\begin{array}{c}
    \phionepm\\
    \phitwopm
  \end{array}\right)={}&
  R(\beta)
  \left(\begin{array}{c}
    G^{\pm}\\
    H^{\pm}
  \end{array}\right), \qquad
  \left(\begin{array}{c}
    \chione\\
    \chitwo
  \end{array}\right)=
  R(\be)
  \left(\begin{array}{c}
    G_0\\
    \Ha
  \end{array}\right) \quad
  \text{with}\quad
  R(\beta)=
  \left(\begin{array}{cc}
        \cbe & -\sbe\\
        \sbe & \cbe
  \end{array}\right).
  \label{eq:pseudochargedrotations}
\end{align}
Here $G^{\pm}$ and $G_0$ are the charged and neutral would-be
Goldstone-boson fields, and $H^{\pm}$ and $\Ha$ the physical charged
and pseudoscalar Higgs fields, respectively.  The mixing angle $\beta$
is related to the vevs $v_i$ of the two scalar doublets via $\tb
\equiv \tan\beta = \vtwo/\vone$, and we use the abbreviations
$\cbe=\cos\be$ and $\sbe=\sin\be$.  The masses $\MW$ and $\MZ$ of the
weak gauge bosons are given by
\begin{align}
  \MW = \frac{1}{2} \g v, \qquad
  \MZ = \frac{1}{2} \sqrt{\g^2 + \gy^2}\; v,\qquad
v=\sqrt{v_1^2+v_2^2}.  \label{SSB}
\end{align}

The renormalization transformations for the mixing angle $\beta$ and
the pseudoscalar fields in the complete on-shell scheme can be written
as
\begin{align}
\be_{\rB} ={}& \be+\de\be,\\
\left(\begin{array}{c} G_{0\rB} \\ \Ha{}_{\rB} \end{array}\right) 
={}& \left(\begin{array}{cc}
        1 + \frac{1}{2}\de Z_{G_0G_0} &   \frac{1}{2}\de Z_{G_0\Ha}\\
         \frac{1}{2}\de Z_{\Ha G_0} &  1+\frac{1}{2}\de Z_{\Ha\Ha}
  \end{array}\right)
\left(\begin{array}{c} G_{0} \\ \Ha \end{array}\right),
\label{eq:matrix_renormalization_ps}
\end{align}
where $\de\be$ and $\de
Z_{\ldots}$ represent the 
renormalization constants for the mixing angle $\be$ and
the physical pseudoscalar $\Ha$ and the would-be-Goldstone field $G_{0}$.
Similar equations can be written for the charged scalar fields.

\section{Renormalization schemes for mixing matrices}
\label{se:schemes}

In this section we review existing renormalization schemes for
mixing angles and propose and discuss new ones. In  \citere{Freitas:2002um}
desirable properties for the renormalization of mixing matrices
were formulated:
\begin{itemize}
\item 
The mixing-angle renormalization should be gauge independent, i.e.\
renormalized $S$-matrix elements should be gauge-independent functions
of the renormalized mixing angles.
\item The mixing-angle renormalization should be symmetric with
  respect to the mixing degrees of freedom. Moreover, the renormalized
  mixing angle should be independent of a specific physical process.
\item The mixing-angle renormalization should not spoil the numerical
  stability of the perturbative expansion; in particular, the running
  of parameters and radiative corrections to physical observables
  should be accessible via perturbation theory.
\end{itemize}
We add a further condition, which could be viewed as a
refinement of the third condition of \citere{Freitas:2002um}:
\begin{itemize}
\item In the limit of degenerate masses of the mixing particles or
  in the limit of extreme mixing angles, no singularities should be
  introduced in physical observables, \ie $S$-matrix elements should
  behave smoothly in these limits.  Furthermore, there should be no
  ``dead corners'' in the parameter space of the model where a
  renormalized input parameter nominally
goes to infinity.%
\footnote{See, e.g., the discussion of the
  $\overline{\mathrm{MS}}(\lambda_3)$ and FJ$(\lambda_3)$ schemes of
  the \THDM in \citeres{Altenkamp:2017ldc,Altenkamp:2017kxk}, where
  the parametrization of the mixing angle $\alpha$ by the coupling
  $\lambda_3$ develops a singularity for $\cos(2\alpha)\to0$.}
\end{itemize}
Focusing on these requirements, as far as possible, we propose new
renormalization schemes, specifically for, but not limited to the
\THDM\ and the \HSESM, and compare NLO predictions for some processes
obtained with some old and the new schemes.

We write the relations between renormalization constants and
self-energies without taking real parts leading to renormalization
constants with imaginary parts. This is appropriate for the
complex-mass scheme \cite{Denner:2005fg,Denner:2006ic}. In the usual
on-shell scheme, the real part of the self-energies should be taken in
all renormalization conditions.

\subsection[Renormalization of mixing angles in
   \texorpdfstring{$\overline{\mathrm{MS}}$}{MSbar} schemes]
  {Renormalization of mixing angles in
  \texorpdfstring{$\overline{\mathbf{MS}}$}{MSbar} schemes}
\label{se:msbar_renormalization}

A straightforward, universal renormalization scheme, which does not
distinguish a specific mass scale in the case of the renormalization
of a mixing angle, is provided by $\MSbar$ renormalization, where the
renormalization constants contain only ultraviolet(UV)-divergent parts
along with some universal finite constants, \ie the combination
\beq
\Delta=\frac{2}{4-D}-\gamma_{\mathrm{E}} + \ln(4\pi)
\eeq
in dimensional regularization, where $D$ is the space--time dimension
and $\gamma_{\mathrm{E}}$ the Euler--Mascheroni constant.  $\MSbar$
renormalization can be straightforwardly applied to arbitrary mixing
matrices.

The $\MSbar$ renormalization of mixing angles is, by construction,
symmetric in the fields that mix and does not depend on a specific
observable.  Since mixing angles are defined in the physical basis,
their $\MSbar$ renormalization depends on the precise treatment of
tadpoles~\cite{Fleischer:1980ub,Denner:2016etu}.  If tadpoles are
treated in a conventional way, \ie if {\it renormalized} tadpoles are
set to zero in the course of parameter renormalization and tadpole
counterterms partially absorbed into bare masses (see, e.g., the
renormalization of the SM in \citeres{Denner:1991kt,Actis:2006}), the
counterterms for the mixing angles become gauge dependent.  The
tadpole scheme based on \citere{Denner:1991kt} where the bare masses
are defined as the coefficients of the terms quadratic in the physical
fields in the Lagrangian is referred to as \PRTS
(Parameter-Renormalized Tadpole Scheme)
in the following.%
\footnote{This tadpole counterterm scheme differs from others by the
  fact that no implicit tadpole counterterms appear in 2-point
  functions with physical external fields.}  We denote the $\MSbar$
scheme applied to mixing angles based on the \PRTS as \MSbarPRTS in
the following.  If tadpoles are treated in the \fjts (\FJTS)
\cite{Fleischer:1980ub,Krause:2016oke,Denner:2016etu}, i.e.\ removed
by a suitable field redefinition, the resulting $\MSbar$
renormalization scheme applied to mixing angles, denoted in the
following as \MSbarFJTS, is by construction gauge independent.  The
\MSbarPRTS\ and \MSbarFJTS\ schemes have been worked out for the
\THDM\ and the \HSESM\ in different variants in
\citeres{Krause:2016oke,Denner:2016etu,Altenkamp:2017ldc,Denner:2017vms}
and \cite{Denner:2017vms,Altenkamp:2018bcs}, respectively.  The
``Tadpole scheme'' for the renormalization of $\tan\beta$ in the
Minimal Supersymmetric Standard Model studied in
\citere{Freitas:2002um} is equivalent to the \MSbarFJTS\ scheme for
the mixing angle~$\beta$. It was found that this scheme leads to an
unacceptably large scheme uncertainty in the renormalization of
$\tan\beta$. In \citere{Denner:2017vms} the \MSbarFJTS\ scheme has
been applied to the renormalization of mixing angles $\alpha$ and
$\beta$ in the \THDM\ and the \HSESM\ in an NLO study of heavy and
light Higgs production in Higgs-strahlung,
$\Pp\Pp\to\PH_{1,2}\mu^-\mu^+ +X$, and vector-boson fusion,
$\Pp\Pp\to\PH_{1,2}\Pj\Pj +X$. The results of the \MSbarFJTS\ scheme
turned out to be unstable and suffering from large scale uncertainties
in many scenarios, while results in schemes based on on-shell self-
and mixing energies remained well-behaved.  NLO results obtained with
the \MSbarPRTS\ and \MSbarFJTS\ schemes were compared in quite some
detail for the light Higgs-boson decays
$\PH_2\to\PW\PW/\PZ\PZ\to4\,$fermions in the
\THDM~\cite{Altenkamp:2017ldc,Altenkamp:2017kxk} and the
\HSESM~\cite{Altenkamp:2018bcs}.  While the results of the two schemes
are perturbatively stable in the \HSESM, with nice plateaus showing up
in the variation of the renormalization scale at NLO, the results on
$\PHtwo\to4f$ in the \MSbarFJTS\ scheme lead to serious perturbative
instabilities in \THDM\ scenarios that are away from the alignment
limit, where $|\cos(\beta-\alpha)|$ is not small.

All $\MSbar$ renormalization schemes for mixing angles give rise to
large corrections in the limit of degenerate masses.  This enhancement
results from terms of the form \refeq{eq:dzmda} in $S$-matrix
elements.  In $\MSbar$ schemes, $\delta\al$ cancels only the
UV-divergent parts, but the remaining UV-finite terms in
\refeq{eq:dzmda} resulting from the field (or wave-function)
renormalization become singular for $M_{\PH_i}\to M_{\PH_j}$.  The
size of these terms, which are also present in the \MSbarPRTS scheme,
is enhanced by additional tadpole contributions in the \MSbarFJTS\ 
scheme.  While for $\tan \beta$ similar enhancements due to additional
tadpole contributions take place, the limit $M_{\PH_i}\to M_{\PH_j}$
does not introduce singularities connected with the renormalization of
$\beta$, so that no corresponding singular contributions to $S$-matrix
elements can result.  In this sense, the situation for {\it true}
mixing angles (such as $\alpha$) is more involved.

Instead of imposing the $\MSbar$ condition on the mixing angles,
$\MSbar$ renormalization can be applied directly to parameters of the
Higgs potential.  This idea was, e.g., pursued in the ``$\lambda_3$''
schemes of \citeres{Altenkamp:2017ldc,Altenkamp:2017kxk} as an
alternative to the $\MSbar$ renormalization of the mixing angle
$\alpha$ in the \THDM; specifically, $\alpha$ was replaced as input
parameter by the coupling $\lambda_3$, which was $\MSbar$
renormalized.  While such a renormalization condition is gauge
independent and does not lead to singularities for degenerate masses,
problematic regions (``dead corners'') in the parameter space show up
where the relation between the distinguished coupling and the mixing
angle cannot be inverted.  In the ``$\lambda_3$'' schemes of
\citeres{Altenkamp:2017ldc,Altenkamp:2017kxk}, for instance, a
singularity in the parametrization of observables by $\lambda_3$
occurs in scenarios in which $\cos(2\alpha)\to0$.
To circumvent this problem in the \THDM, one would have to patch the
parameter space by switching from $\lambda_3$ to another scalar coupling
as renormalized parameter.
 
In summary, $\MSbar$ renormalization schemes for mixing angles have
some desirable properties (simplicity, symmetry, process
independence), but suffer, in general, from problems with perturbative
stability in certain parameter regions, such as for mass degeneracy of
the mixing fields.  Moreover, care has to be taken in view of gauge
dependence.  On the other hand, it should be mentioned that $\MSbar$
renormalization offers a simple way to estimate perturbative stability
by varying the renormalization scale in predictions and checking for a
stabilization of results in the transition from leading order (LO) to
NLO.

\subsection{Physical (on-shell) renormalization conditions for mixing angles}
\label{se:RSphys}

The renormalization of mixing angles can be directly fixed from
observables or $S$-matrix elements that depend on these mixing angles
at LO.  Such on-shell renormalization conditions
are evidently gauge independent.%
\footnote{Note that we do not consider renormalization conditions as
  ``on shell'' that are based on mixing energies, Green functions, or
  formfactors as well as ``matrix elements'' involving unphysical
  degrees of freedom at some ``on-shell'' configurations of momenta.
  This includes, in particular, mixing energies of scalar
  bosons, of  would-be Goldstone
  bosons with Higgs bosons or of  would-be Goldstone
  bosons with gauge bosons. In the literature, schemes of
  this kind are often called ``on shell'' as well.}
Fixing mixing angles by specific processes, however, has a number of
potential disadvantages:
\begin{itemize}
\item
By construction, on-shell renormalization conditions are process dependent and
often destroy the symmetries between the particles that mix. 
\item
On-shell conditions
are only directly applicable to $S$-matrix elements that do not
involve charged particles; otherwise 
the counterterms would become infrared (IR) singular.%
\footnote{Considering the analytic structure of one-loop 3-point
  functions, it can be seen that the corresponding IR singularities in
  decay $S$-matrix elements can only be cancelled by those of the
  field-renormalization constants of the external charged particles if
  one of the three external particles is massless and neutral, as for
  the photon in the electron--positron--photon vertex in QED. In other
  words, potential IR singularities in the $S$-matrix element used to
  fix a renormalization constant only cancel in this very specific
  case. In all other cases, IR singularities would enter the parameter
  renormalization constants.}  The problem of IR singularities can, in
principle, be avoided by imposing the renormalization condition on a
full physical observable, \eg by demanding that a partial decay width
does not receive any correction, but this procedure shifts
process-specific real-radiation effects into the renormalization
constant.
\item Typically, observables and $S$-matrix elements depend not only
  on a mixing angle, but on other parameters as well.  Upon defining
  the mixing-angle renormalization from such quantities, one thus
  absorbs corrections to the considered observable or $S$-matrix
  element into the mixing-angle counterterm that are related to other
  parameters of the model.  This can be a source for unnaturally large
  corrections.  In \citere{Krause:2016oke} it was, e.g., demonstrated
  that on-shell renormalization conditions based on specific
  observables lead to numerically unstable results in the \THDM.
\end{itemize}

The situation can be improved by considering combinations of physical
observables or $S$-matrix elements that depend exclusively on a
specific mixing angle and on no other parameters, so that
renormalization contributions of other parameters or normalization
effects systematically drop out.  For the quark-mixing matrix such a
renormalization scheme was proposed in \citere{Denner:2004bm}.

In order to fix the renormalization of the Higgs mixing angle $\al$
introduced in \refeq{eq:scalar_rotation}, we consider a set of
processes involving the fields $H_i$ that have a simple dependence on
the mixing angle $\alpha$ in LO.  If the dependence on $\alpha$ in the
considered observables only results from the transformation
\refeq{eq:scalar_rotation}, this is typically the case.

\subsubsection{Higgs-Singlet Extension of the Standard Model}

As a first specific example, we choose the decay of
scalar bosons $\Hone$ and $\Htwo$ into pairs of \PZ~bosons in the \HSESM.%
\footnote{In the \THDM the corresponding vertices involve $\al-\be$
  instead of $\al$ and thus can be used to renormalize this
  difference.}  The LO vertices read
\begin{equation}
\setlength{\unitlength}{1pt}
  \raisebox{-22pt}{\includegraphics{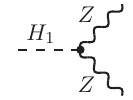}} 
= \frac{\ri e \sa}{\sw \cw^2}
  {\MW} g^{\mu\nu}, \qquad
  \raisebox{-22pt}{\includegraphics{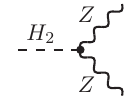}} 
= \frac{\ri e \ca}{\sw \cw^2}
  {\MW} g^{\mu\nu}.
\end{equation} 
The LO matrix elements for the decays of the two Higgs bosons into a
pair of Z~bosons read
\begin{align}
\cM_0^{H_1\to \PZ\PZ} = \frac{e \sa}{\sw \cw^2} {\MW} (\varepsilon_1^*\cdot\varepsilon_2^*), \qquad
\cM_0^{H_2\to \PZ\PZ} = \frac{e \ca}{\sw \cw^2} {\MW} (\varepsilon_1^*\cdot\varepsilon_2^*),
\end{align}
where $\varepsilon_{1,2}$ denote the polarization vectors of the two Z~bosons.

At LO, the ratio of the matrix elements $\cM^{H_i\to \PZ\PZ}$ for the
decays of the heavy and light scalar into a pair of Z~bosons is given
by $\sa/\ca$.  A possible renormalization condition is, thus, to
require that this ratio is equal to its LO value, \ie
\beq\label{eq:RC_phys_HZZ}
\frac{\cM_0^{H_1\to \PZ\PZ}}{\cM_0^{H_2\to \PZ\PZ}} = \frac{\sa}{\ca} \overset{!}{=} 
\frac{\cM^{H_1\to \PZ\PZ}}{\cM^{H_2\to \PZ\PZ}}.
\eeq
Using the complete on-shell scheme, the renormalized NLO matrix
elements can be written as
\begin{align}
\cM^{H_1\to \PZ\PZ}={}&
\cM_0^{H_1\to \PZ\PZ}\biggl(1+\delta_{H_1ZZ}
+\delta Z_e + \frac{1}{2}\frac{\delta \MW^2}{\MW^2} -  
      \frac{\de\cw ^2}{\cw^2} - \frac{\de \sw
        }{\sw} +\frac{\delta\sa}{\sa} 
\notag\\ &{}
+ \delta Z_{ZZ} 
       +\frac{1}{2} \delta Z^H_{11}  + \frac{1}{2} \delta Z^H_{21}\frac{\ca}{\sa} 
\biggr),
\notag\\  
\cM^{H_2\to \PZ\PZ}={}&
\cM_0^{H_2\to \PZ\PZ}\biggl(1+\delta_{H_2ZZ}
+\delta Z_e + \frac{1}{2}\frac{\delta \MW^2}{\MW^2} -  
      \frac{\de\cw ^2}{\cw^2} - \frac{\de \sw
        }{\sw} +\frac{\delta\ca}{\ca} 
\notag\\ &{}
+ \delta Z_{ZZ} 
       +\frac{1}{2} \delta Z^H_{22}  + \frac{1}{2} \delta Z^H_{12}\frac{\sa}{\ca} 
\biggr),
\label{eq:MHiZZ}
\end{align} 
where $\de_{H_iZZ}=\de_{H_iZZ}(M_{\PH_i}^2)$ are the unrenormalized
relative one-loop corrections to the respective decays, and the
counterterms have been written explicitly. In particular, the two
scalar fields are renormalized according to
\refeq{eq:matrix_renormalization_scalars}.  Inserting \refeq{eq:MHiZZ}
into \refeq{eq:RC_phys_HZZ} and expanding to NLO yields for the
counterterm of the mixing angle $\alpha$:
\begin{align}
\de\al={}& \ca\sa(\delta_{H_2ZZ}-\delta_{H_1ZZ}) +\frac{1}{2}\ca\sa(\delta Z^H_{22}-\delta Z^H_{11})
+\frac{1}{2}(\delta Z^H_{12}\satwo- \delta Z^H_{21}\catwo).
\label{eq:da_HZZ}
\end{align}
Using this counterterm, the renormalized NLO matrix elements become 
\begin{align}
\cM^{H_i\to \PZ\PZ}={}&
\cM_0^{H_i\to \PZ\PZ}\biggl(1+\delta_{H_1ZZ}\sa^2+\delta_{H_2ZZ}\ca^2
+\delta Z_e + \frac{1}{2}\frac{\delta \MW^2}{\MW^2} -  
      \frac{\de\cw ^2}{\cw^2} - \frac{\de \sw
        }{\sw} 
\notag\\ &
+ \delta Z_{ZZ} 
       +\frac{1}{2} \delta Z^H_{11}\sa^2  +\frac{1}{2} \delta Z^H_{22}\ca^2  + \frac{1}{2} (\delta Z^H_{21}+\delta Z^H_{12})\ca\sa 
\biggr), \quad i=1,2.
\label{eq:MHiZZren}
\end{align} 
Since all differences have been incorporated in the
renormalization of $\alpha$,
the relative corrections to the two different Higgs-boson decays
become equal. 

The renormalization condition \refeq{eq:RC_phys_HZZ} has several
desirable properties:
\begin{itemize}
\item 
It is gauge independent, because it is based on physical $S$-matrix elements; 
\item 
it is symmetric with respect to the scalar fields $H_1$ and $H_2$;
\item 
it is numerically stable for degenerate masses $M_{\PH_1}\sim M_{\PH_2}$;
\item 
it has smooth limits for extreme mixing angles, \ie for
$\ca\to0$ or $\sa\to0$.
\end{itemize}
The singularities for degenerate masses $M_{\PH_1}\sim M_{\PH_2}$ cancel
in all $S$-matrix elements for the following reason.  All appearances
of $\alpha$ that result from rewriting parameters of the scalar
potential involve a prefactor $M_{\PH_1}^2-M_{\PH_2}^2$ that cancels the
singularity.  For all appearances of $\alpha$ introduced in the
Lagrangian via the rotation \refeq{eq:scalar_rotation} the
counterterm appears always in the combinations \refeq{eq:dzmda} which
after inserting the on-shell counterterm $\de\al$ only depend on $\de
Z^H_{12}$ and $\de Z^H_{21}$ via the sum
\beq
\de Z^H_{12} +  \de Z^H_{21}.
\eeq
Upon using \refeq{eq:zij_onshell},  this becomes
\beq\label{eq:mixing_ren}
\de Z^H_{12}+\de Z^H_{21}=
2\frac{\Sigma^H_{12}(M_{\PH_2}^2)-\Sigma^H_{12}(M_{\PH_1}^2)}{M_{\PH_1}^2-M_{\PH_2}^2},
\eeq
which is finite for degenerate masses, $M_{\PH_1}\to M_{\PH_2}$, and
moreover all mo\-men\-tum-in\-de\-pen\-dent contributions to the
mixing energy, such as tadpole contributions, cancel therein. These
statements hold for all other on-shell schemes for $\al$ discussed in
this section.

Still, the condition \refeq{eq:RC_phys_HZZ} has some
disadvantages:
\begin{itemize}
\item It can be directly applied  to decay processes involving
  only electrically neutral external particles. For charged external
  particles, the renormalization constant $\delta\al$ becomes
  IR singular.
\item Depending on the masses of the external particles, the form
  factors have to be evaluated  at phase-space points in the
  unphysical region.%
  \footnote{At least for decays at the one-loop level this does not
    constitute an obstacle. In this case the relevant 3-point
    functions can be analytically continued to the unphysical region
    as discussed in \citere{tHooft:1978jhc}.}
This is, for instance, the case in \refeq{eq:da_HZZ} if $\PH_1$ or $\PH_2$ 
is identified with the observed Higgs state of mass $125\GeV$.
\end{itemize}

All these drawbacks can be lifted upon introducing extra neutral fields
with a simple coupling structure that allows us to fix the
renormalization of the mixing angles while recovering the original
theory upon sending the extra couplings to zero.

As an example for this procedure, 
we consider the \HSESM and add an additional
fermion singlet field $\psi$ with the Lagrangian
\beq
\cL_\psi = \ri\bar\psi\slashed{\partial}\psi - y_\psi \sigma\bar\psi\psi.
\eeq
In terms of scalar fields in the basis of mass eigenstates, this becomes
\beq
\cL_\psi = \ri\bar\psi\slashed{\partial}\psi - y_\psi(\vone+H_1\ca-H_2\sa) \bar\psi\psi.
\eeq
Considering the limit of a vanishing Yukawa coupling, $y_\psi\to0$, we
recover the original \HSESM\ with an additional massless fermion
$\psi$ ($m_\psi=y_\psi \vone\to0$), which completely decouples from
all other particles, \ie we effectively recover the original theory.
We require that the ratio of matrix elements for the decays of the two
scalar Higgs bosons into a $\psi\bar\psi$ pair of singlets is equal to
its leading-order value $-\ca/\sa$ in the limit of vanishing coupling
$y_\psi$:
\beq\label{eq:RC_phys_Hchichi}
\frac{\cM^{H_1\to \psi\psi}}{\cM^{H_2\to \psi\psi}}
\overset{!}{=} 
\frac{\cM_0^{H_1\to \psi\psi}}{\cM_0^{H_2\to \psi\psi}}
\propto -\frac{\ca}{\sa},
\eeq
where the proportionality factor, which is not spelled out,
only contains the ratio of spinor chains (with different kinematics),
but no other model parameters.
Since the ratio is based on matrix elements for the decay of
massive scalars into massless neutral fermions these are in the
physical region, and no IR singularities occur.  Moreover, since all
NLO vertex corrections in $\cM^{H_i\to \psi\psi}$
tend to zero at least quadratically in $y_\psi$
and thus drop out in the limit $y_\psi\to0$%
\footnote{No CP-odd effective coupling 
$\propto\bar\psi\gamma_5\psi$ is induced by loops.}, 
we obtain for the
counterterm
\begin{align}
\de\al={}& \frac{1}{2}(\delta Z^H_{11}-\delta Z^H_{22}) \ca\sa
+\frac{1}{2}(\delta Z^H_{12}\catwo-\delta Z^H_{21}\satwo).
\label{eq:da_hsesm_os}
\end{align}
Thus, owing to the simple structure of the model, all vertex
corrections drop out, and the mixing-angle counterterm is fixed by a
gauge-independent combination of field-re\-nor\-ma\-li\-za\-tion
constants only.  Note also that the spinor chains suppressed in
\refeq{eq:RC_phys_Hchichi} do not enter the final result for $\de\al$,
since they cancel in the ratios $\cM^{H_i\to \psi\psi}/\cM_0^{H_i\to
  \psi\psi}$.

\subsubsection{Two-Higgs-Doublet Model}
\label{se:os_thdm}

In order to formulate on-shell renormalization conditions for the
\THDM, we add two right-handed fermion singlets to the Lagrangian:
$\nu_{1\rR}$ transforming under the extra $\mathbb{Z}_2$ symmetry as 
$\nu_{1\rR}\to-\nu_{1\rR}$ and $\nu_{2\rR}$ transforming as $\nu_{2\rR}\to\nu_{2\rR}$,
so that $\nu_{i\rR}$ can only receive a Yukawa coupling to $\Phi_i$.
The additional Lagrangian reads
\begin{align}
\cL_{\nu_{\rR}} ={}& 
\ri\bar\nu_{1\rR}\slashed{\partial}\nu_{1\rR}
+\ri\bar\nu_{2\rR}\slashed{\partial}\nu_{2\rR}
-\left[ y_{\nu_1} \bar{L}_{1\rL} (\ri\sigma_2\Phi^*_1)\nu_{1\rR} 
+ y_{\nu_2} \bar{L}_{2\rL} (\ri\sigma_2\Phi^*_2)\nu_{2\rR} +\hc  \right], 
\end{align}
where $y_{\nu_i}$ are new Yukawa couplings that are considered in the
limit $y_{\nu_i}\to0$.  The fields $L_{i\rL}=(\nu_i,l_i)_{\rL}^{\rT}$
are left-handed lepton doublets of the SM, say the electron--neutrino
and muon--neutrino doublets, $\Phi_i$ are the two Higgs-doublet
fields, and $\sigma_2$ the second Pauli matrix.  Upon inserting the
representations of the doublet fields this leads to
\begin{align}
\cL_{\nu_{\rR}} ={}& - \frac{1}{\sqrt{2}}\left[y_{\nu_1}
\bar{\nu}_{1\rL}\nu_{1\rR}(v\cbe  +H_1 \ca  - H_2\sa + \ri \Ha \sbe
-\ri G_0\cbe) +\hc\right]
\notag\\ & {}
- \frac{1}{\sqrt{2}}\left[y_{\nu_2}
\bar{\nu}_{2\rL}\nu_{2\rR}(v\sbe  + H_1 \sa  + H_2 \ca - \ri
\Ha\cbe -\ri G_0\sbe)  +\hc\right]
\notag\\& {}
+\ldots,
\label{eq:LnuR}
\end{align}
where we suppressed terms involving charged scalar Higgs and would-be
Goldstone-boson fields.  For non-zero couplings $y_{\nu_i}$, the
neutrinos $\nu_1$ and $\nu_2$ correspond to massive Dirac fermions
without generation mixing owing to the conserved lepton number.  In
the limit $y_{\nu_i}\to0$, the neutrinos become massless, and the
right-handed parts decouple.
 
\paragraph{Renormalization of \boldmath{$\alpha$}}

The renormalization of the angle $\alpha$ can be fixed upon requiring
that the ratio of the matrix elements for the decays
$H_i\to\nunuone$ is the same at LO and NLO,
\beq
\frac{\cM^{H_1\to \nu_1\bar\nu_1}}{\cM^{H_2\to \nu_1\bar\nu_1}}
\overset{!}{=} 
\frac{\cM_0^{H_1\to \nu_1\bar\nu_1}}{\cM_0^{H_2\to \nu_1\bar\nu_1}}.
 \label{eq:rc_da_nu1_ons}
\eeq
This leads to the renormalization constant
\begin{align}
\de\al={}&(\delta_{H_1\nunuone}-\delta_{H_2\nunuone}) \ca\sa 
+\frac{1}{2}(\delta Z^H_{11}-\delta Z^H_{22})\ca\sa
+\frac{1}{2}(\delta Z^H_{12}\catwo- \delta Z^H_{21}\satwo),
 \label{eq:da_nu1_ons}
\end{align}
where $\delta_{H_i\nunuone}$ represent the unrenormalized relative one-loop corrections to the
decays $H_i\to\nunuone$.  
Alternatively, if the ratio of the matrix elements for the decays
$H_i\to\nunutwo$ is used to fix the renormalization of $\alpha$, 
this leads to the renormalization constant
\begin{align}
\de\al={}&(\delta_{H_2\nunutwo}-\delta_{H_1\nunutwo}) \ca\sa 
+\frac{1}{2}(\delta Z^H_{22}-\delta Z^H_{11})\ca\sa
+\frac{1}{2}(\delta Z^H_{12}\satwo- \delta Z^H_{21}\catwo).
 \label{eq:da_nu2_ons}
\end{align}
The explicit form of the vertex correction factors $\delta_{H_1\nunuone}$, etc.,
is given in \refapp{app:vcos}.
At NLO, these loop corrections respect the chiral structures of the respective
underlying LO couplings; beyond NLO this might not be the case anymore, so that
one would have to write the renormalization conditions in terms of the form factors
that correspond to the LO couplings.
   
\paragraph{Renormalization of \boldmath{$\beta$}}

The renormalization of the angle $\beta$ can be fixed by demanding
that the ratio of the matrix elements for the decays
$H_1\to\nunui$ and $\Ha\to\nunui$ is
the same at LO and NLO for one of the two neutrinos $\nu_i$.
Specifying $\nu_i$ to $\nu_1$, means
\beq\label{eq:RC_phys_beta_1}
\frac{\cM^{\Ha\to\nunuone}}{\cM^{H_1\to\nunuone}}
\overset{!}{=} 
\frac{\cM_0^{\Ha\to\nunuone}}{\cM_0^{H_1\to\nunuone}}
\propto \frac{\sbe}{\ca},
\eeq
where again the suppressed proportionality factor is
given by a ratio of spinor chains, but does not contain further model
parameters.
This results in
\begin{align}
\de\be={}& \frac{\sbe}{\cbe}\biggl(\de_{H_1\nunuone}-\de_{\Ha\nunuone}-\frac{1}{2}(\de Z_{\Ha\Ha}
-\de Z^H_{11})-\frac{1}{2}\frac{\sa}{\ca}\de Z^H_{21}\biggr)
+\frac{1}{2}\de Z_{G_0\Ha}
-\frac{\sbe}{\cbe}\frac{\sa}{\ca}\delta\alpha,
\end{align}
where $\de_{\Ha\nunuone}$ denotes the unrenormalized relative one-loop corrections to
the matrix element for the decay $\Ha\to\nunuone$.
Upon inserting the renormalization constant $\de\al$ from \refeq{eq:da_nu1_ons} this
becomes 
\begin{align}
\de\be={}&
\frac{\sbe}{\cbe}\biggl(\ca^2\de_{H_1\nunuone}+\sa^2\de_{H_2\nunuone}
-\de_{\Ha\nunuone}
\notag\\&{}
-\frac{1}{2}\left[\de Z_{\Ha\Ha}
-\catwo\de Z^H_{11}-\satwo\de Z^H_{22}+\ca\sa(\de Z^H_{12}+\de Z^H_{21})\right] 
\biggr)
+\frac{1}{2}\de Z_{G_0\Ha},
 \label{eq:db_nu1_ons} 
\end{align}
which involves only vertex corrections for neutrino $\nu_1$.
Note that we would get the same relation if we had fixed $\de\be$
from the ratio of $\Ha\to\nunuone$ and $H_2\to\nunuone$ matrix elements
by virtue of \refeq{eq:rc_da_nu1_ons}.

Alternatively, $\de\al$ and $\de\be$ can be fixed from analogous
matrix elements involving only neutrino $\nu_2$. 
In this case, for $\de\be$ we demand
\beq\label{eq:RC_phys_beta_2}
\frac{\cM^{\Ha\to\nunutwo}}{\cM^{H_1\to\nunutwo}}
\overset{!}{=} 
\frac{\cM_0^{\Ha\to\nunutwo}}{\cM_0^{H_1\to\nunutwo}}
\propto -\frac{\cbe}{\sa},
\eeq
resulting in
\begin{align}
\de\be={}& \frac{\cbe}{\sbe}\biggl(\de_{\Ha\nunutwo}-\de_{H_1\nunutwo}+\frac{1}{2}(\de Z_{\Ha\Ha}
-\de Z^H_{11})-\frac{1}{2}\frac{\ca}{\sa}\de Z^H_{21}\biggr)
\notag\\&{}
+\frac{1}{2}\de Z_{G_0\Ha}
-\frac{\cbe}{\sbe}\frac{\ca}{\sa}\delta\alpha,
\end{align}
which can be further processed with $\de\al$ from
\refeq{eq:da_nu2_ons} to yield 
\begin{align}
\de\be={}&
\frac{\cbe}{\sbe}\biggl(\de_{\Ha\nunutwo}-\sa^2\de_{H_1\nunutwo}-\ca^2\de_{H_2\nunutwo}
\notag\\&{}
+\frac{1}{2}\left[\de Z_{\Ha\Ha}
-\satwo\de Z^H_{11}-\catwo\de Z^H_{22}-\ca\sa(\de Z^H_{12}+\de Z^H_{21})\right] 
\biggr)
+\frac{1}{2}\de Z_{G_0\Ha}.
 \label{eq:db_nu2_ons}
\end{align}
The conditions \refeq{eq:db_nu1_ons} and \refeq{eq:db_nu2_ons} become
singular for $\cbe\to0$ or $\sbe\to0$, respectively. Since in the
phenomenological applications of the \THDM, $\cbe$ and $\sbe$ are
always non-vanishing, this does not lead to a singularity but can
cause artificial enhancements.

The renormalization scheme based on the conditions
\refeq{eq:da_nu1_ons} and \refeq{eq:db_nu1_ons} involving only $\nu_1$
is called \OSone in the following, the one based on 
\refeq{eq:da_nu2_ons} and \refeq{eq:db_nu2_ons} involving only $\nu_2$
is called \OStwo.

The conditions \refeq{eq:RC_phys_beta_1} and \refeq{eq:RC_phys_beta_2}
do not directly apply to $\beta$, but to a combination of $\alpha$ and
$\beta$.  A condition that fixes $\beta$ directly can be obtained by
using the decays into both neutrino singlets and requiring
\begin{align}
\frac{\cM^{H_1\to\nunuone}}{\cM^{H_1\to\nunutwo}}\,
\frac{\cM^{H_2\to\nunuone}}{\cM^{H_2\to\nunutwo}}
\left(
\frac{\cM^{\Ha\to\nunutwo}}{\cM^{\Ha\to\nunuone}}
\right)^2
\overset{!}{=} 
\frac{\cM_0^{H_1\to\nunuone}}{\cM_0^{H_1\to\nunutwo}}\,
\frac{\cM_0^{H_2\to\nunuone}}{\cM_0^{H_2\to\nunutwo}}
\left(
\frac{\cM_0^{\Ha\to\nunutwo}}{\cM_0^{\Ha\to\nunuone}}
\right)^2
= -\frac{\cbe^2}{\sbe^2}.
\label{eq:RC_phys_beta_12}
\end{align}
Note that the spinor chains all cancel within the multiple ratio.
This condition leads to the counterterm
\begin{align}
\de\be={}&
\frac{1}{2}{\cbe}{\sbe}\left(\de_{H_1\nunuone}+\de_{H_2\nunuone}-2\de_{\Ha\nunuone}-\de_{H_1\nunutwo}-\de_{H_2\nunutwo}+2\de_{\Ha\nunutwo}\right)
\notag\\&{}
-\frac{\cbe\sbe}{4\ca\sa}(\de Z^H_{12}+\de Z^H_{21})
+\frac{1}{2}\de Z_{G_0\Ha},
\end{align}
which is regular for $\sbe\to0$ of $\cbe\to0$, but potentially singular
for  $\sa\to0$ or $\ca\to0$.

A condition leading to a counterterm that is regular in all these
limits can be constructed upon using linear combinations of observable
quantities from different processes. To this end, we consider on-shell form
factors instead of complete matrix elements. For the scalar decays
into (nearly) massless fermions we write the matrix elements as
\begin{align}
\cM^{H_i\to\nunuj} = [\bar{u}_\nu v_\nu]_{H_i} F^{H_i\to\nunuj},\qquad
\cM^{\Ha\to\nunuj} = [\bar{u}_\nu \ri\gamma_5 v_\nu]_{\Ha} F^{\Ha\to\nunuj},
\end{align}
where $\bar{u}_{\nu}$ and $v_\nu$ are the spinors of the final-state
fermions and antifermions, and 
$[\dots]_{H_i/\Ha}$ indicates the decay kinematics of the spinor chain.
The functions $F^{H_i\to\nunuj}$ and $F^{\Ha\to\nunuj}$ denote
the formfactors for the decays into neutrinos $\nu_j$ of type $j=1,2$.
The LO formfactors follow directly from the Lagrangian \refeq{eq:LnuR}:
\begin{align}
 F_0^{H_1\to\nunuone} ={}& -\frac{1}{\sqrt{2}}y_{\nu_1}\ca,\quad
& F_0^{H_2\to\nunuone} ={}&  \frac{1}{\sqrt{2}}y_{\nu_1}\sa,\quad
& F_0^{\Ha\to\nunuone} ={}& -\frac{1}{\sqrt{2}}y_{\nu_1}\sbe,\notag\\
 F_0^{H_1\to\nunutwo} ={}& -\frac{1}{\sqrt{2}}y_{\nu_2}\sa,\quad
& F_0^{H_2\to\nunutwo} ={}& -\frac{1}{\sqrt{2}}y_{\nu_2}\ca,\quad
& F_0^{\Ha\to\nunutwo} ={}&  \frac{1}{\sqrt{2}}y_{\nu_2}\cbe.
\end{align}

As renormalization condition we require that the following LO relation
holds also at higher orders:
\begin{align}
0 &{}=  \frac{F_0^{\Ha\to\nunuone}}{\ca F_0^{H_1\to\nunuone} - \sa
       F_0^{H_2\to\nunuone}}\cbe
 +\frac{F_0^{\Ha\to\nunutwo}}{\sa F_0^{H_1\to\nunutwo} + \ca
     F_0^{H_2\to\nunutwo}}\sbe
\notag\\
&\overset{!}{=}  \frac{F^{\Ha\to\nunuone}}{\ca F^{H_1\to\nunuone} - \sa
       F^{H_2\to\nunuone}}\cbe
 +\frac{F^{\Ha\to\nunutwo}}{\sa F^{H_1\to\nunutwo} + \ca
       F^{H_2\to\nunutwo}}\sbe.
\end{align}
This fixes the counterterm for the mixing angle $\beta$ to 
\begin{align}
 \de\beta={}&\frac{1}{2}\cbe\sbe\left[(\ca^2-\sa^2)(\de Z^H_{11}-\de
   Z^H_{22}) -2\ca\sa(\de Z^H_{12}+\de Z^H_{21})\right]
+ \frac{1}{2}\de Z_{G_0\Ha}
\notag\\&{}+
\cbe\sbe\left(
\delta_{\Ha\nunutwo}+\ca^2\de_{H_1\nunuone}+\sa^2\de_{H_2\nunuone}
-\delta_{\Ha\nunuone}-\sa^2\de_{H_1\nunutwo}-\ca^2\de_{H_2\nunutwo}
\right).
\label{eq:db_os12_thdm}
\end{align}
This result is non-singular in all limits $\sa\to0$, $\ca\to0$,
$\sbe\to0$, or $\cbe\to0$.

The above on-shell renormalization conditions for mixing angles in the
\HSESM and \THDM depend on the introduction of specific auxiliary
fields. While this method can in principle be generalized to more
general theories, we are not able to provide a simple specific recipe
for the on-shell renormalization of mixing angles or mixing matrices
in general. As far as we can see, this has to be investigated anew for
each theory, but the shown examples can certainly serve as guidelines.

\subsection{Renormalization of mixing angles based on symmetries}
\label{se:RSsymm}

\subsubsection{Rigid symmetry and wave-function renormalization for physical states}
\label{se:rigid_symmetry1}

The renormalization of mixing matrices can be related to the
wave-function renormalization of external fields upon using rigid
symmetry, \ie the symmetry under global gauge transformations, of the
Lagrangian.  Again, we specifically consider the \THDM and the \HSESM,
where the CP-even scalar fields in the symmetric and mass-eigenstate
bases are related via \refeq{eq:scalar_rotation}.

A theory with a spontaneously broken gauge symmetry can be
renormalized using the renormalization constants for fields and
dimensionless parameters from the symmetric phase
\cite{tHooft:1971qjg,tHooft:1971akt,Lee:1972fj,Lee:1974zg,Lee:1972yfa,Lee:1973fn,Lee:1973rb}.
In particular, the counterterms for the dimensionless parameters and
fields can be directly taken from the symmetric formulation.  Once the
counterterms for the mass parameters are adjusted appropriately, all UV
divergences cancel while the external lines in $S$-matrix elements 
still require additional
finite wave-function renormalization. For the SM such a
renormalization scheme 
was used in \citere{Bohm:1986rj}.  In the
case considered here, the relevant renormalization transformations
read
\begin{align}
\al_{\rB} ={}& \al+\de\al,\\
\etaplet_{\rB} ={}& (Z^{\eta})^{1/2} \etaplet, \\
  (Z^\eta)^{1/2}={}&
  \left(\begin{array}{cc}
        (Z^{\eta}_{1})^{1/2} &  0\\
        0 &  (Z^{\eta}_2)^{1/2}
  \end{array}\right)
=1+\frac{1}{2}\delta Z^{\eta}
= \left(\begin{array}{cc}
        1 + \frac{1}{2}\de Z^{\eta}_1 &  0\\
        0 &  1+\frac{1}{2}\de Z^{\eta}_2
  \end{array}\right),
\label{eq:symmetric_renormalization}
\end{align}
where the components $\eta_1$ and $\eta_2$ of $\etaplet$ 
belong to different multiplets of the gauge group.

Consistency of
Eqs.~\refeq{eq:scalar_rotation} and \refeq{eq:symmetric_renormalization} requires
that at least the divergent parts of the renormalization constants in
both schemes are related via
\begin{align}
 (Z^H)^{1/2}\big|_\UV = R^{\rT}(\al+\de\al)  (Z^{\eta})^{1/2}
R(\al)\big|_\UV =  [R^{\rT}(\al)+\de R^{\rT}(\al,\de\al)]  (Z^{\eta})^{1/2}
R(\al)\big|_\UV.
\end{align}
Using $0=\de(R^{\rT}R)=\de R^{\rT}R+R^{\rT}\de R$, we get
\begin{align}
 \de Z^H\big|_\UV = -2 R^{\rT}(\al)\de R(\al,\de\al)\big|_\UV 
+ R^{\rT}(\al)\de  Z^{\eta} R(\al)\big|_\UV,
\label{eq:Z-constraint_symmetry}
\end{align}
where
\begin{align}
\de  R(\al,\de\al) =
  \left(\begin{array}{cc}
        -\sa & -\ca\\
        \ca & -\sa
  \end{array}\right)\de\al, 
\qquad
R^{\rT}(\al) \de  R(\al,\de\al)=
  \left(\begin{array}{cc}
        0 & -\de\al\\
        \de\al & 0
  \end{array}\right). 
\end{align}
This implies
\begin{align}
\label{eq:Zsymrel1}
\de Z^H_{11}\big|_\UV ={}& \catwo\de Z^{\eta}_1\big|_\UV + \satwo\de Z^{\eta}_2\big|_\UV, \\
\label{eq:Zsymrel2}
\de Z^H_{22}\big|_\UV ={}& \satwo\de Z^{\eta}_1\big|_\UV + \catwo\de Z^{\eta}_2\big|_\UV, \\
\label{eq:Zsymrel3}
\de Z^H_{12}\big|_\UV+\de Z^H_{21}\big|_\UV ={}& 2\ca\sa(\de Z^{\eta}_2-\de Z^{\eta}_1)\big|_\UV, \\
\label{eq:Zsymrel4}
\de Z^H_{12}\big|_\UV-\de Z^H_{21}\big|_\UV ={}& 4\de\al\big|_\UV.
\end{align}
Thus, we find, in particular, a relation between the renormalization
constant of the mixing angle $\de\al$ and the non-diagonal field
renormalization constants of the scalar Higgs-field pair.  
While the 
relations \refeq{eq:Zsymrel1}--\refeq{eq:Zsymrel4} hold for the
UV-divergent parts, 
as for instance discussed in
\citere{Altenkamp:2017ldc}, not all of them can be required
simultaneously for the
finite parts if the field renormalization is fixed in the complete
on-shell scheme.

On the other hand, \refeq{eq:Zsymrel4} can be used to fix the
renormalization of the mixing angle $\alpha$ in terms of on-shell
field renormalization constants of the scalar Higgs fields, \ie we can
define
\begin{align}
\de\al = \frac{1}{4}\bigl(\de Z^H_{12}-\de Z^H_{21}\bigr) 
       = \frac{\Sigma^H_{12}(M_{\PH_2}^2)+\Sigma^H_{12}(M_{\PH_1}^2)}{2(M_{\PH_1}^2-M_{\PH_2}^2)},
\label{eq:da_rigid_symmetry}
\end{align}
where we expressed the non-diagonal field renormalization constants by
the non-diagonal mixing energy upon using the on-shell renormalization
conditions \refeq{eq:zij_onshell}.  This renormalization condition has
been introduced by Kanemura et al. in \citere{Kanemura:2004mg} and
used in \citere{Krause:2016oke} both in the tadpole scheme of
\citere{Kanemura:2004mg} and the \FJTS.

As discussed in \refse{se:prelim}, the counterterm to the mixing angle
appears, besides in the regular expression $(M_{\PH_1}^2-M_{\PH_2}^2)\de\al$, only
in the combinations \refeq{eq:dzmda} in $S$-matrix elements.  Thus,
when using the renormalization condition \refeq{eq:da_rigid_symmetry},
only the combination
\beq\label{eq:rel_mixing_ren_a}
\frac{1}{2}\de Z^H_{12}-\de\al = \de\al +\frac{1}{2}\de Z^H_{21} =
\frac{1}{4}\bigl(\de Z^H_{12}+\de Z^H_{21}\bigr)
\eeq
remains. According to \refeq{eq:mixing_ren}, this is finite for
degenerate masses, 
$M_{\PH_1}\to M_{\PH_2}$, and moreover all momentum-independent
contributions to the mixing energy, such as tadpole contributions,
cancel therein.  

Renormalizing mixing angles through appropriately chosen field
renormalization constants has several advantages:
\begin{itemize}
\item 
The symmetry between the different states is respected;
\item 
if the limit of vanishing mixing is protected by a symmetry
implying $\Sigma^H_{12}\to0$ for $\al\to0$,
this is not violated by the renormalization condition, \ie
$\de\al\to0$  for $\al\to0$;
\item 
there is a smooth limit for degenerate masses, \ie the
  renormalization is numerically stable and does not lead to enhanced corrections;
\item 
there is no problem with IR singularities since the mixing
  energies are free of such contributions.
\end{itemize}
An apparent drawback is the gauge dependence of the field
renormalization constants. However, we can choose a specific gauge to
calculate the counterterms for the mixing angles and fix 
these counterterms in this gauge.%
\footnote{The suggestions of \citeres{Pilaftsis:2002nc,Yamada:2001px}
  and the proposal of \citere{Kanemura:2015mxa} for the
  renormalization of $\beta$ follow a similar reasoning.}  Then, we
can vary the gauge as usual (but keeping $\de\al$ fixed in the
original gauge), and $S$-matrix elements are gauge independent for
fixed $\de\al$. It remains to pick a suitable gauge to fix the
mixing-angle counterterm.  In order not to introduce artificially
large parameters, the 't~Hooft--Feynman gauge ($\xi=1$) can be chosen.
This can be done in the conventional formalism or, preferably, in the
background-field method (BFM) where rigid symmetry holds for the
effective action by construction (see \refse{se:BFM}).  Once the
mixing-angle renormalization is fixed in this way, relations between
observables can be calculated as usual.  The gauge independence of
$S$-matrix elements ensures that no singularities appear if the
calculation is done in a different gauge.  This procedure relies on
the fact that unrenormalized $S$-matrix elements with appropriate LSZ
factors are gauge independent as functions of the bare parameters.
This requires that all relations between bare parameters are gauge
independent which is the case in the \FJTS\ scheme, but not in the
tadpole schemes of \citeres{Denner:1991kt,Actis:2006}.

For the sake of clarity, let us assume that the mixing angle
counterterm $\de\al(\xi_0)$ has been fixed from $\de Z^H_{ij}(\xi_0)$
in a specific gauge, with fixed gauge parameter $\xi_0$.  If the
$S$-matrix elements are calculated in a different gauge (with gauge
parameter $\xi\ne\xi_0$), but with the fixed counterterm
$\de\al(\xi_0)$, the cancellations of potential singularities for
degenerate masses occurring in \refeq{eq:rel_mixing_ren_a} and
\refeq{eq:mixing_ren} are not obvious anymore.  The contributions of
such terms can be studied by considering the gauge dependence of
\refeq{eq:da_rigid_symmetry}. The gauge-dependent parts of the
one-loop mixing energy obey the Nielsen identity~\cite{Gambino:1999ai}
\begin{align}
\partial_\xi \Sigma_{ij}(s) 
={}& \Lambda_{ij}(s)(s-M_{\PH_j}^2)
+ (s-M_{\PH_i}^2) \Lambda'_{ij}(s),
\end{align}
where  $\Lambda^{(\prime)}_{ij}$ are one-loop Green functions involving the
operators for the BRST transformations of the fields $H_i$.
This implies, 
\begin{align}
\partial_\xi \frac{\Sigma_{ij}(M_{\PH_i}^2)+\Sigma_{ij}(M_{\PH_j}^2)}{M_{\PH_i}^2-M_{\PH_j}^2}
=\Lambda_{ij}(M_{\PH_i}^2)-\Lambda'_{ij}(M_{\PH_j}^2)
,
\end{align}
\ie the gauge-dependent part is regular for $M_{\PH_i}\to M_{\PH_j}$ 
and free of momentum-independent contributions.

Using \refeq{eq:zij_onshell} and \refeq{eq:Z-constraint_symmetry} for
higher-dimensional matrices, the renormalization condition
\refeq{eq:da_rigid_symmetry} can be straightforwardly generalized to
models with more physical scalar fields as
\begin{align}
R^{\rT}(\boldsymbol{\al})\de R({\boldsymbol{\al},\de\boldsymbol{\al}}) 
= \frac{1}{4}\left[(\de Z^H)^{\rT } - \de Z^H\right],
\label{eq:da_rigid_symmetry_general}
\end{align}
which fixes the finite parts in $\de\boldsymbol{\al}$ by definition.
Moreover, \refeq{eq:Z-constraint_symmetry} implies
\begin{align}
 \frac{1}{2}\left[(\de Z^H)^{\rT } + \de Z^H\right]\big|_\UV =  
R^{\rT}(\boldsymbol{\al})\de  Z^{\eta} R(\boldsymbol{\al})\big|_\UV.
\end{align}

In the light of the discussion of this section, the original proposal
for the renormalization of the quark-mixing matrix
\cite{Denner:1990yz} in the SM turns out to be viable.  Strictly
speaking, this requires the use of the gauge-independent \FJTS\ 
scheme.  However, since there are no tadpole contributions to the
quark mixing energies and thus to the non-diagonal fermion field
renormalization constants in the SM, the renormalization condition of
\citere{Denner:1990yz} for the quark-mixing matrix is not affected by
the tadpole scheme.  Thus, after fixing the counterterms for the
quark-mixing matrix in the 't~Hooft--Feynman gauge,
(gauge-independent) observables can be calculated as usual.  Moreover,
for the relevant one-loop fermion self-energies the results of the BFM
are equivalent to those of the conventional formalism.

\subsubsection{Rigid symmetry and wave-function renormalization for unphysical states}
\label{se:rigid_symmetry2}

The renormalization condition \refeq{eq:da_rigid_symmetry}
can also be used for the mixing with the would-be Goldstone bosons as
for instance in the pseudoscalar and charged-scalar sector of the
\THDM. Using the fact that would-be Goldstone bosons are massless
upon disregarding the gauge-fixing term that is not renormalized in
linear gauges, we obtain for instance for mixing of the pseudoscalar
with the would-be Goldstone boson,
\begin{align}
\de Z_{G_0\Ha}={}&-\frac{2}{M_{\Ha}^2}\Sigma^{G_0\Ha}(M_{\Ha}^2), \qquad
\notag\\
\de
Z_{\Ha G_0}={}&\frac{2}{M_{\Ha}^2(M^2-M_{\Ha}^2)}\Bigl(M^2\Sigma^{G_0\Ha}(M_{\Ha}^2)-M_{\Ha}^2\Sigma^{G_0\Ha}(M^2)\Bigr),
\label{eq:zgj_onshell}
\end{align}
where $M^2$ is an arbitrary scale at which the $G_0\Ha$ mixing energy
$\Sigma^{G_0\Ha}$ is required to vanish.  Choosing, for instance,
$M=0$ leads to
\begin{align}
\de
Z_{\Ha G_0}={}&\frac{2}{M_{\Ha}^2}\Sigma^{G_0\Ha}(0),
\label{eq:zgj_onshell2}
\end{align}
and [in analogy to \refeq{eq:da_rigid_symmetry}]
\begin{align}
\de\be = \frac{1}{4}\bigl(\de Z_{G_0\Ha}-\de Z_{\Ha G_0}\bigr) 
       = -\frac{1}{2M_{\Ha}^2} \Bigl(\Sigma^{G_0\Ha}(M_{\Ha}^2)
+\Sigma^{G_0\Ha}(0)\Bigr).
\label{eq:db_rigid_symmetry_goldstone}
\end{align}
 
In the BFM (see \refse{se:BFM}) this simplifies owing to the 
Ward identity%
\footnote{For the validity of \refeq{eq:wi_chiHa_mdts} it is
  crucial that the self-energy is the quantity appearing in the
  inverse of the propagator, \ie that it includes tadpole
  contributions and corresponding counterterms.}
\refeq{eq:wi_chiHa_mdts}, $\Sigma^{\Gzhat\Hahat}(0)=0$,
to
\begin{align}
\de\be = \frac{1}{4}\bigl(\de Z_{\Gzhat\Hahat}-\de Z_{\Hahat \Gzhat}\bigr) 
       = -\frac{1}{2M_{\Ha}^2} \Sigma^{\Gzhat\Hahat}(M_{\Ha}^2).
\label{eq:db_rigid_symmetry}
\end{align}
This expression formally coincides with the renormalization of $\beta$
proposed in \citere{Kanemura:2004mg} within the Landau gauge of the
conventional formalism.  We stress that the simple form
\refeq{eq:db_rigid_symmetry} for the counterterm $\delta\beta$ holds
only in special gauges, while in general
\refeq{eq:db_rigid_symmetry_goldstone} has to be used. Moreover,
\refeq{eq:db_rigid_symmetry} requires the \PRTS\ scheme of
\citere{Denner:1991kt}, while in the \FJTS\ scheme extra tadpole
contributions appear [see Eq.~\refeq{eq:Z12fjts} below].

\subsubsection{Background-field gauge invariance}
\label{se:BFM}

The method of the previous sections is not applicable to the
renormalization of parameters that are not directly related to the
mixing of fields.  An example is the renormalization of the singlet
sector of the \HSESM.  Besides the mass of the second Higgs scalar, a
second parameter has to be selected. One can choose one of the quartic
couplings $\lambda_1$ or $\lambda_3$ of the Higgs-singlet field, its
vev $\vone$, or the quantity $\tan\beta = \vtwo/\vone$, in analogy to
the \THDM.  In such cases, the background-field gauge invariance that
appears when quantizing in the BFM can be exploited to fix
renormalization constants.
The BFM (see, e.g., \citeres{Abbott:1980hw,Weinberg:1996kr,Bohm:2001yx})
for the EW SM was introduced in \citere{Denner:1994xt}, and
its application to the \THDM and the \HSESM was described in
\citere{Denner:2017vms}.  Following these references we denote
background fields with carets.  All relations discussed in
\refses{se:rigid_symmetry1} and \ref{se:rigid_symmetry2} hold in the BFM 
for the corresponding
quantities of background fields as well.

In the conventional formalism, the gauge-fixing term breaks rigid
invariance.%
\footnote{This does not affect the discussion in
  \refses{se:rigid_symmetry1} and \ref{se:rigid_symmetry2}, since it
  relies only on the fact that a symmetric field renormalization is
  possible in the spontaneously broken phase.}  In the BFM, rigid
gauge invariance is maintained for the background fields upon choosing
an appropriate gauge-fixing term for the quantum
fields~\cite{Denner:1994xt,Denner:2017vms}. Rigid invariance for the
background field gives rise to Ward identities (\cf \refapp{app:wi})
and restrictions on the renormalization constants for the background
fields.  In particular, in the BFM the relations
\refeq{eq:Zsymrel1}--\refeq{eq:Zsymrel4} can all be maintained
including the finite parts.  The relations resulting from rigid gauge
invariance in the BFM were presented for the SM in the renormalization
scheme of \citere{Denner:1991kt} in \citere{Denner:1994xt}. In the
\FJTS\ scheme, as
defined in \citeres{Fleischer:1980ub,Denner:2016etu}, additional
tadpole contributions $\propto\Delta v$ appear in the Ward identities
and thus also in these relations. For instance, using
the conventions of \citere{Denner:2017vms} the last line of Eq.~(46)
of \citere{Denner:1994xt} for the SM in the 
\FJTS\ scheme becomes
\begin{align}
\label{eq:rcrelsm}
\delta Z_{\Phihat} = \delta Z_{\etahat} = \delta Z_{\chihat} = \delta Z_{\phihat} 
      = - 2 \delta Z_e - 
        \frac{\cw ^2}{\sw^2} \frac{\delta \cw ^2}{\cw ^2} + 
        \frac{\delta \MW^2}{\MW^2} +    2\frac{\Delta v}{v}, 
\end{align}
where the extra term, the shift 
$\Delta v$ in the vev, is related to the tadpole by
\beq
\Delta v = \frac{-\dthhat}{\MH^2} = \frac{\Thhat}{\MH^2},
\eeq
$\dthhat$ denotes the tadpole counterterm and $\Thhat$ the
unrenormalized tadpole, \ie the renormalized Higgs one-point vertex
function is given by $\Gamma^{\hat H}=(\Thhat+\dthhat)=0$.  The
counterterms $\delta Z_e$ and $\delta \cw ^2$ are fixed according to
(48) and (46) of \citere{Denner:1994xt} and can be shown to be
independent of the tadpole counterterm scheme. In the \FJTS\ scheme
the W-boson mass counterterm gets additional implicit tadpole
contributions $\propto\Delta v$ as described in
\citere{Denner:2016etu} which is equivalent to including explicit
tadpoles in the self-energy that determines the counterterm, \ie
\beq\label{eq:dmw2}
\delta \MW^2 = \Sigma_{\mathrm{T}}^{\hat W\hat W}(\MW^2) 
= \Sigma_{\mathrm{1PI,T}}^{\hat W\hat W}(\MW^2)  -\frac{e\MW}{\sw} \Delta v
= \Sigma_{\mathrm{1PI,T}}^{\hat W\hat W}(\MW^2)  -\frac{e\MW}{\sw\MH^2} \Thhat.
\eeq
As a consequence, in the \FJTS\ scheme the W-boson mass counterterm is
gauge independent, and the gauge dependence of $\delta Z_{\etahat}$
matches the one of $\Delta v$.  In the renormalization scheme of
\citere{Denner:1991kt} the tadpole contribution in \refeq{eq:dmw2} is
absent, and the mass counterterm is gauge dependent (since the
definition of the bare mass involves the shifted vev, see the
discussion in \citere{Denner:2017vms}).

Using 
\beq
\MW=\frac{e}{2\sw}v, \qquad {\MW}_{,\rB}=\frac{e_{\rB}}{2{\sw}_{,\rB}}v_{\rB},
\eeq
the definition 
\beq
v_{\rB}=Z_v v=v\left(1+\delta Z_v\right),
\eeq
as well as the renormalization transformations of the SM parameters,
the relation \refeq{eq:rcrelsm} implies
\beq
 \label{eq:rcrelsmb}
\delta Z_{\Phihat} =\delta Z_{\etahat} = \delta Z_{\chihat} = \delta Z_{\phihat} = 2\delta
Z_v + 2\frac{\Delta v}{v}.
\eeq 
Thus, the vev is renormalized as the corresponding scalar
doublet field apart from the explicit shift $\Delta v$ introduced to
ensure a vanishing tadpole at one-loop order.

In the rest of this section we use the \FJTS\ scheme.

\paragraph{Higgs Singlet Extension of the Standard Model}
\label{se:HSESM}

In the \HSESM within the BFM, \refeq{eq:rcrelsm} and
\refeq{eq:rcrelsmb} still hold with the components of the SM doublet
replaced by those of the doublet of this model,
\beq
\label{eq:rcrelhs1}
\delta Z_{\Phihat} =\de Z^{\etahat}_2 = \delta Z_{\etahat_2} = \delta Z_{\chihat} = \delta Z_{\phihat} = 2\delta
Z_{\vtwo} + 2\frac{\Delta \vtwo}{\vtwo},
\eeq 
where
\beq
\de Z_{\vtwo}=  - \delta Z_e - 
\frac{1}{2}\frac{\cw ^2}{\sw^2} \frac{\delta \cw ^2}{\cw ^2} + 
     \frac{1}{2}   \frac{\delta \MW^2}{\MW^2}.
\label{eq:dZv}
\eeq
In addition, we obtain a corresponding relation between the
renormalization of the component $\etaone$ of the singlet field $\si$ and
the corresponding vev $\vone$,
\beq
 \label{eq:rcrelhs2}
\delta Z_{\sihat} =\de Z^{\etahat}_1 = \delta Z_{\etahat_1} = 2\delta
Z_{\vone} + 2\frac{\Delta \vone}{\vone}.
\eeq 
The shifts of the vevs can be expressed in terms of the tadpole counterterms as
\begin{align}
\Delta \vone = -\frac{\dtHhatone}{\MHsone}\ca + \frac{\dtHhattwo}{\MHstwo}\sa,\qquad
\Delta \vtwo = -\frac{\dtHhatone}{\MHsone}\sa - \frac{\dtHhattwo}{\MHstwo}\ca.
\end{align}

Fixing $\de Z^{\etahat}_2$ from \refeq{eq:rcrelhs1}, using the
on-shell field renormalization constants \refeq{eq:zij_onshell} to
determine $\de Z^{\etahat}_1$ from \refeq{eq:Zsymrel3} including the
finite parts, 
\begin{align}
\de Z^H_{12}+\de Z^H_{21} ={}& 2\ca\sa(\de Z^{\eta}_2-\de Z^{\eta}_1), 
\end{align}
where $\delta Z^{\Hhat}_{ij}$ are the elements of the field
renormalization matrix for the scalars as defined in
\refeq{eq:matrix_renormalization_scalars},
we can
use \refeq{eq:rcrelhs2} to fix the renormalization of the singlet vev.
This results in
\begin{align}
\de Z_{\vone} ={}&
\frac{1}{2}(\de Z^{\etahat}_{1}-\de Z^{\etahat}_{2}) + \de Z_{\vtwo} 
+\frac{\Delta \vtwo}{\vtwo}
-\frac{\Delta \vone}{\vone}
\notag\\
={}&
-\frac{1}{4\ca\sa}(\de Z^{\Hhat}_{12}+\de Z^{\Hhat}_{21}) + \de Z_{\vtwo}
+\frac{\Delta \vtwo}{\vtwo}
-\frac{\Delta \vone}{\vone}.
\label{eq:de_vs_HSESM_BFM}
\end{align}

Defining
\beq
\tan\be_{\rB} = \frac{v_{2,\rB}}{v_{1,\rB}}, \qquad
\tan\be = \frac{\vtwo}{\vone},
\label{eq:tb_HSESM}
\eeq
this translates to
\begin{align}
\delta\tan\beta={}&\frac{1}{2} \tan\beta\left(\de Z^{\etahat}_2-\de
  Z^{\etahat}_1
+2\frac{\Delta \vone}{\vone}
-2\frac{\Delta \vtwo}{\vtwo}
\right)
\notag\\ 
\label{eq:de_beta_HSESM_BFM}
={}&\frac{1}{4} \frac{\tan\beta}{\ca\sa}
\left(\de Z^{\Hhat}_{12}+\de Z^{\Hhat}_{21}\right)
+ \tan\beta\left(\frac{\Delta \vone}{\vone}-\frac{\Delta \vtwo}{\vtwo}\right)
,
\end{align}
where
\beq
\frac{\Delta \vone}{\vone}-\frac{\Delta \vtwo}{\vtwo}
=\frac{e}{2\sw\MW}\left[\frac{\dtHhatone}{\MHsone}(\sa-\ca\tan\beta)
+\frac{\dtHhattwo}{\MHstwo}(\ca+\sa\tan\beta)\right].
\label{eq:Delta_v_HSESM}
\eeq

Alternative definitions of the counterterms can be obtained upon using
\refeq{eq:Zsymrel1} and \refeq{eq:Zsymrel2} instead of
\refeq{eq:Zsymrel3} to fix $(\de Z^{\etahat}_1-\de Z^{\etahat}_2)$.
\change{In particular, upon using an appropriate linear combination of
\refeq{eq:Zsymrel1}, \refeq{eq:Zsymrel2} and \refeq{eq:Zsymrel3}, one
finds
\beq\label{eq:Zsymrelreg}
\de Z^{\etahat}_1-\de Z^{\etahat}_2 =(\ca^2-\sa^2)\left(\de Z^{\Hhat}_{11}-\de Z^{\Hhat}_{22}\right)-2\sa\ca\left(\de Z^{\Hhat}_{12}+\de Z^{\Hhat}_{21}\right).
\eeq
Using this expression avoids the potential singularity for $\ca\to0$ or $\sa\to0$ in
\refeq{eq:de_beta_HSESM_BFM} and leads to  
\begin{align}
\de Z_{\vone} ={}&
\frac{1}{2}\left[(\ca^2-\sa^2)\left(\de Z^{\Hhat}_{11}-\de Z^{\Hhat}_{22}\right)-2\sa\ca\left(\de Z^{\Hhat}_{12}+\de Z^{\Hhat}_{21}\right)\right]
+ \de Z_{\vtwo}
+\frac{\Delta \vtwo}{\vtwo}
-\frac{\Delta \vone}{\vone}
\label{eq:de_vs_HSESM_BFM_new}
\end{align}
or
\begin{align}
\delta\tan\beta={}&
\label{eq:de_beta_HSESM_BFM_new}
\frac{1}{2} {\tan\beta}
\left[(\sa^2-\ca^2)\left(\de Z^{\Hhat}_{11}-\de Z^{\Hhat}_{22}\right)+2\sa\ca\left(\de Z^{\Hhat}_{12}+\de Z^{\Hhat}_{21}\right)\right]
\notag\\&{}
+ \tan\beta\left(\frac{\Delta \vone}{\vone}-\frac{\Delta \vtwo}{\vtwo}\right)
\end{align}}
instead of \refeq{eq:de_vs_HSESM_BFM} and \refeq{eq:de_beta_HSESM_BFM}.

\change{Equations
  \refeq{eq:de_vs_HSESM_BFM}--\refeq{eq:de_beta_HSESM_BFM_new}} can be
used to fix the renormalization of $v_{1}$ or $\beta$ upon using the
field renormalization constants $\de Z^H_{ij}$ for the scalar fields
\refeq{eq:zij_onshell} in the complete on-shell scheme.  Since these
are gauge dependent, a gauge needs to be fixed. A convenient choice is
the 't~Hooft-Feynman gauge of the BFM.  \change{The renormalization
  scheme based on \refeq{eq:de_beta_HSESM_BFM_new} and
  \refeq{eq:Delta_v_HSESM} for $\beta$ and on
  \refeq{eq:da_rigid_symmetry} for $\al$ in the \HSESM is denoted as
    \RSBFM in the following.%
    \footnote{Note that in the first preprint version of this paper the
      scheme \RSBFM in the \HSESM was based on
      \refeq{eq:de_beta_HSESM_BFM} and \refeq{eq:Delta_v_HSESM}. For
      the scenarios considered in \refse{se:results} the differences
      between the two choices in the numerical results are marginal.}}

\paragraph{Two-Higgs-Doublet Model}
\label{se:THDM}

Considering now the \THDM and using again the BFM and the
corresponding background-field gauge invariance, we obtain instead of
\refeq{eq:rcrelsm}:
\begin{align}
\label{eq:rcrel2hdm1}
\delta Z^{\etahat}_1 
      = - 2 \delta Z_e - 
        \frac{\cw ^2}{\sw^2} \frac{\delta \cw ^2}{\cw ^2} + 
        \frac{\delta \MW^2}{\MW^2} +    2\frac{\Delta \vone}{\vone}
        + 2\frac{\delta\cbe}{\cbe},\\ 
\label{eq:rcrel2hdm2}
\delta Z^{\etahat}_2 
      = - 2 \delta Z_e - 
        \frac{\cw ^2}{\sw^2} \frac{\delta \cw ^2}{\cw ^2} + 
        \frac{\delta \MW^2}{\MW^2} +    2\frac{\Delta \vtwo}{\vtwo}
        + 2\frac{\delta\sbe}{\sbe},
\end{align}
where the shifts of the vevs can be expressed via the tadpole
counterterms as
\begin{align}
\Delta \vone = -\frac{\dtHhatone}{\MHsone}\ca + \frac{\dtHhattwo}{\MHstwo}\sa,\qquad
\Delta \vtwo = -\frac{\dtHhatone}{\MHsone}\sa - \frac{\dtHhattwo}{\MHstwo}\ca.
\end{align}

Equations \refeq{eq:rcrel2hdm1} and \refeq{eq:rcrel2hdm2} imply
\begin{align}\
\label{eq:de_BFM_2HDMa}
\delta\be={}&\frac{1}{2} \cbe\sbe\left(\de Z^{\etahat}_2-\de
  Z^{\etahat}_1\right)
+\frac{e}{2\sw\MW}\left(\sbe\Delta \vone -\cbe\Delta \vtwo\right)
,\\
\cbetwo\de Z^{\etahat}_1 + \sbetwo\de Z^{\etahat}_2 
 ={}& - 2 \delta Z_e - 
        \frac{\cw ^2}{\sw^2} \frac{\delta \cw ^2}{\cw ^2} + 
        \frac{\delta \MW^2}{\MW^2} 
+\frac{e}{\sw\MW}\left(\cbe\Delta \vone +\sbe\Delta \vtwo\right).
\label{eq:de_BFM_2HDMb}
\end{align}
Using \refeq{eq:Zsymrel3} including finite parts, this yields
\begin{align}
\label{eq:debeta_BFM_2HDM}
\delta\be={}&\frac{1}{4}\frac{\cbe\sbe}{\ca\sa}\left(\de Z^{\Hhat}_{12}+\de Z^{\Hhat}_{21}\right)
+\frac{e}{2\sw\MW}\left(\sbe\Delta \vone -\cbe\Delta \vtwo\right).
\end{align}
The terms involving shifts in the vevs in \refeq{eq:debeta_BFM_2HDM}
can be expressed by the tadpoles as
\beq
\label{eq:deltav_2HDM}
\left.\sbe\Delta \vone -\cbe\Delta \vtwo\right.
=  \frac{\dtHhatone}{\MHsone}\sin(\al-\be) + \frac{\dtHhattwo}{\MHstwo}\cos(\al-\be).
\eeq
As for the \HSESM, alternative counterterms can be obtained using
\refeq{eq:Zsymrel1} and \refeq{eq:Zsymrel2} instead of
\refeq{eq:Zsymrel3} to fix $(\de Z^{\etahat}_1-\de Z^{\etahat}_2)$.
\change{If $\sa$ or $\ca$ become small, the counterterm defined by
  \refeq{eq:debeta_BFM_2HDM} becomes artificially large, which is
  actually the case in the THDM scenarios B1 and B2 considered in
  \refse{se:results} below. This can be avoided by using
  \refeq{eq:Zsymrelreg} in Eq.~\refeq{eq:de_BFM_2HDMa}, resulting in
\begin{align}
\label{eq:debeta_BFM_2HDM_new}
\delta\be={}&\frac{1}{2}\cbe\sbe\left[(\sa^2-\ca^2)\left(\de Z^{\Hhat}_{11}-\de Z^{\Hhat}_{22}\right)+2\ca\sa\left(\de Z^{\Hhat}_{12}+\de Z^{\Hhat}_{21}\right)\right]
\notag\\&{}
+\frac{e}{2\sw\MW}\left(\sbe\Delta \vone -\cbe\Delta \vtwo\right).
\end{align}
The renormalization scheme based on \change{\refeq{eq:debeta_BFM_2HDM_new}} and
\refeq{eq:deltav_2HDM} for $\beta$ and \refeq{eq:da_rigid_symmetry}
within the BFM for $\al$ in the \THDM is denoted as \RSBFM in the following.%
\footnote{Note that in the first preprint version of this paper the scheme
  \RSBFM in the \THDM was based on  Eq.~\refeq{eq:debeta_BFM_2HDM} and
  \refeq{eq:deltav_2HDM}.}}
We note that the renormalization of $\alpha$ in the ``on-shell tadpole-pinched
scheme'' of \citere{Krause:2016oke} is equivalent to the one based on
\refeq{eq:da_rigid_symmetry} within the BFM.

There are a number of further possibilities to fix the counterterm
$\de\beta$ using rigid invariance in the BFM.  The discussion of
\refses{se:rigid_symmetry1} and \ref{se:rigid_symmetry2} holds also for the mixing between the
pseudoscalar or charged scalar fields. This gives rise to the
relations
\begin{align}
\label{eq:Zsymrelg1}
\de Z_{11} ={}& \cbetwo\de Z^{\etahat}_1 + \sbetwo\de Z^{\etahat}_2, \\
\label{eq:Zsymrelg2}
\de Z_{22} ={}& \sbetwo\de Z^{\etahat}_1 + \cbetwo\de Z^{\etahat}_2, \\
\label{eq:Zsymrelg3}
\de Z_{12}+\de Z_{21} ={}& 2\cbe\sbe(\de Z^{\etahat}_2-\de Z^{\etahat}_1), \\
\label{eq:Zsymrelg4}
\de Z_{12}-\de Z_{21} ={}& 4\de\be,
\end{align}
where $\de Z_{ij}$ refer to the field renormalization constants of the
pseudoscalar $(G_0,\Ha)^{\rT}$ or of the charged scalar
$(G^\pm,H^\pm)^\rT$ fields.

From Eqs. \refeq{eq:de_BFM_2HDMa}, \refeq{eq:de_BFM_2HDMb}, \refeq{eq:Zsymrelg1}--\refeq{eq:Zsymrelg4} we can derive
\begin{align}
\label{eq:Z11fjts}
\delta Z_{11} ={}&  - 2 \delta Z_e - 
        \frac{\cw ^2}{\sw^2} \frac{\delta \cw ^2}{\cw ^2} + 
        \frac{\delta \MW^2}{\MW^2} 
+\frac{e}{\sw\MW}\left(\cbe\Delta \vone +\sbe\Delta \vtwo\right),\\
\label{eq:Z12fjts}
\delta Z_{21} ={}& 
\frac{e}{\sw\MW}\left(\cbe\Delta \vtwo -\sbe\Delta \vone\right),
\end{align}
as well as the relations for the counterterm to the mixing
angle $\be$
\begin{align}
\label{eq:delta_be_2hdm_1}
\delta\be={}&\frac{1}{4}\left(\de Z_{12}+\de Z_{21}\right)
+\frac{e}{2\sw\MW}\left(\sbe\Delta \vone -\cbe\Delta \vtwo\right),\\
\label{eq:delta_be_2hdm_2}
\delta\be={}&\frac{1}{4}\de Z_{12}
+\frac{e}{4\sw\MW}\left(\sbe\Delta \vone -\cbe\Delta \vtwo\right),
\end{align}
where we can use the field renormalization constants of either
$(G_0,\Ha)^{\rT}$ or $(G^\pm,H^\pm)^\rT$ in the complete on-shell scheme.
Note that \refeq{eq:Z12fjts} results from \refeq{eq:zij_onshell} and 
the Ward identity \refeq{eq:wi_chiHa_fjts} valid in the \FJTS\ scheme.

Equations \refeq{eq:debeta_BFM_2HDM}, \refeq{eq:debeta_BFM_2HDM_new},
\refeq{eq:delta_be_2hdm_1}, or
\refeq{eq:delta_be_2hdm_2} provide all the same divergent parts for
$\de\be$, but differ in the finite parts. Moreover, all conditions are
gauge dependent, since the on-shell field renormalization constants are
gauge dependent.  Specific renormalization schemes can be fixed upon
choosing a specific gauge, such as the 't~Hooft--Feynman gauge (within
the BFM) and a specific equation.

For the gauge dependence of the renormalization schemes based on the
BFM the same remarks as in \refse{se:rigid_symmetry1} apply.

\subsection{Summary of renormalization schemes}

The various renormalization schemes used for the \HSESM\ in this paper are summarized
in \refta{tab:HSESMschemes}. 
\begin{table}
\centerline{\renewcommand{\arraystretch}{1.2} 
\begin{tabular}{|c||c|c|c|}
\hline
Scheme & $\alpha$ & $\lambda_1$, $\vone$, or $\tan\beta$ & comments
\\
\hline
\hline
\MSbarPRTS & $\MSbar$ & $\MSbar$ for $\lambda_1$ & no tadpoles in mass terms
\\
\hline
\MSbarFJTS & $\MSbar$ & $\MSbar$ for $\lambda_1$ & FJ Tadpole Scheme
\\
\hline
\OS & \refeq{eq:da_hsesm_os} & $\MSbar$ for $\lambda_1$ &
on-shell renormalization of $\alpha$
\\
\hline
\RSBFM & \refeq{eq:da_rigid_symmetry} & 
$\vone$ via \refeq{eq:de_vs_HSESM_BFM}, &
$\Sigma_{12}^{\Hhat}$, $T^{\hat H_i}$  for $\delta\alpha$ from BFM 
\\
&  & 
or equivalently $\tan\beta$ via \refeq{eq:de_beta_HSESM_BFM} &
\\
\hline
\end{tabular}
}
\caption{Summary of renormalization schemes used in the \HSESM.}
\label{tab:HSESMschemes}
\end{table}
The $\MSbar$ and FJ schemes discussed in \citere{Altenkamp:2018bcs}
are identical to the \MSbarPRTS and \MSbarFJTS schemes of this paper,
up to the point that
$\lambda_{12}^{\mbox{\scriptsize\cite{Altenkamp:2018bcs}}}=\lambda_3/2$
is used as $\MSbar$-renormalized parameter instead of $\lambda_1$.
The $\MSbar$ scheme of \citere{Denner:2017vms} is identical to the
\MSbarFJTS scheme of this paper.

The renormalization schemes used for the \THDM\ in this paper are summarized
in \refta{tab:THDMschemes}.
\begin{table}
\centerline{\renewcommand{\arraystretch}{1.2} 
\begin{tabular}{|c||c|c|c|c|}
\hline
Scheme & $\alpha$ & $\beta$ & $\lambda_5$ & comments
\\
\hline
\hline
\MSbarPRTS & $\MSbar$ & $\MSbar$ & $\MSbar$ &
no tadpoles in mass terms; $\MSbar(\alpha)$ in \citere{Altenkamp:2017ldc}
\\
\hline
\MSbarFJTS & $\MSbar$ & $\MSbar$ & $\MSbar$ &
FJ Tadpole Scheme; FJ$(\alpha)$ in \citere{Altenkamp:2017ldc}
\\
\hline
\OSone & \refeq{eq:da_nu1_ons} & \refeq{eq:db_nu1_ons} &$\MSbar$ &
on-shell renormalization of $\alpha$ and $\beta$\\
\hline
\OStwo & \refeq{eq:da_nu2_ons} & \refeq{eq:db_nu2_ons} &$\MSbar$ &
on-shell renormalization of $\alpha$ and $\beta$\\
\hline
\OSonetwo & \refeq{eq:da_nu2_ons} & \refeq{eq:db_os12_thdm} &$\MSbar$ &
on-shell renormalization of $\alpha$ and $\beta$\\
\hline
\RSBFM & \refeq{eq:da_rigid_symmetry} & \refeq{eq:debeta_BFM_2HDM_new}
&$\MSbar$ & $\Sigma_{12}^{\Hhat}$, $T^{\hat H_i}$ for $\delta\alpha$
and $\delta\beta$ from BFM 
\\
\hline
\end{tabular}
}
\caption{Summary of renormalization schemes used in the \THDM.}
\label{tab:THDMschemes}
\end{table}
The $\MSbar(\lambda_3)$ and FJ$(\lambda_3)$ schemes of
\citere{Altenkamp:2017ldc}, where the scalar coupling $\lambda_3$
replaces the angle $\alpha$ as $\MSbar$-renormalized parameter of the
$\MSbar(\alpha)$ and FJ$(\alpha)$ schemes, are not considered in this
paper.
The $\MSbar$ scheme of \citere{Denner:2016etu} coincides with the
\MSbarFJTS scheme of this paper up to the fact that $\lambda_5$ is used 
 as $\MSbar$-renormalized parameter instead of $M_{\mathrm{sb}}^2 =
 M^2_{\PHa} +4\MW^2\sw^2\lambda_5/e^2$.

\subsection{Parameter conversion between renormalization schemes}
\label{se:conversion}

\subsubsection{Matching procedure and running couplings}

For a comparison of predictions based on different renormalization
schemes a conversion of renormalized parameters is necessary.  The
matching between different schemes is based on the fact that the bare
parameters defining the model are renormalization-scheme independent.
Following \citere{Altenkamp:2017ldc}, we describe two variants that
can be used to convert renormalized parameters at NLO.

Denoting a set of input parameters generically as $\{p_i\}$ defined in
two different renormalization schemes ``$(1)$'' and ``$(2)$'', the two
different renormalization schemes are connected via
\begin{align}
  p_{\mathrm{B},i} = p^{(1)}_i + \delta p^{(1)}_i (\{p^{(1)}_j\})
      = p^{(2)}_i + \delta p^{(2)}_i (\{p^{(2)}_j\}),
  \label{eq:exact_conversion}
\end{align}
where $\{p_{\mathrm{B},i}\}$ is the set of bare parameters which are
by definition renormalization-scheme independent.  There are basically
two possibilities to translate the renormalized parameters from scheme
$(1)$ into $(2)$: performing a ``full conversion'' upon solving
\refeq{eq:exact_conversion} numerically for $\{p^{(2)}_j\}$ with a
given set of parameters $\{p^{(1)}_j\}$, or linearizing
\refeq{eq:exact_conversion} in $p^{(2)}_i$ by replacing
$\{p^{(2)}_j\}$ by $\{p^{(1)}_j\}$ in the last term, so that
\begin{align}
p^{(2)}_i =
p^{(1)}_i + \delta p^{(1)}_i (\{p^{(1)}_j\})- \delta p^{(2)}_i (\{p^{(1)}_j\}) + \dots,
  \label{eq:linearized_conversion}
\end{align}
which is valid up to terms beyond NLO.  The results of the two
versions agree in NLO accuracy.  The advantage of the full conversion
is mainly the exact invertibility, i.e.\ converting from scheme $(1)$
into $(2)$ and back into $(1)$, reproduces the parameters
$\{p^{(1)}_j\}$ exactly, while the linearized version reproduces those
parameters only in NLO accuracy.

In the subsequent section, we make use of the full conversion to
translate benchmark scenarios between the various renormalization
schemes.  For the \THDM, this means that up to three parameters are
converted at a time; for the \HSESM, only up to two parameters are
concerned.  For each benchmark scenario we perform the conversion at a
chosen central scale $\mu=\mu_0$, where $\mu_0$ is chosen according to
the process and model.  In the extension of
\Prophecy~\cite{Bredenstein:2006rh,Bredenstein:2006ha} to the
\HSESM~\cite{Altenkamp:2018bcs} and
\THDM~\cite{Altenkamp:2017ldc,Altenkamp:2017kxk}, which is used to
calculate Higgs decay widths into four fermions at NLO,
\refeq{eq:exact_conversion} is solved numerically in double precision
by minimizing the $\chi^2$ of the error when solving
\begin{align}
p^{(2)}_i =
p^{(1)}_i + \delta p^{(1)}_i (\{p^{(1)}_j\})- \delta p^{(2)}_i (\{p^{(2)}_j\})
\end{align}
for $\{p^{(2)}_j\}$ by iteration.%
\footnote{For the mixing angles $\alpha$ and $\beta$, the matching equation
is written in the (NLO-correct) form 
$\alpha^{\MSbarPRTS}=\alpha^{\MSbarFJTS}-\Delta\alpha^t(\THone,\THtwo)|_{\mathrm{finite}}$
with finite tadpole terms quantified by the function
$\Delta\alpha^t$, and similarly for $\beta$.}
In \RecolaTwo{}, quasi-Newton methods (based on the L-BFGS algorithm) are used
to find the full solution compatible in double precision.  This procedure is
used to prepare run cards for the Monte Carlo program
\HAWKTwo~\cite{Denner:2014cla}, with the parameters already converted from a
specific input scheme and 
evolved to the scale, which are then
used to evaluate Higgs-production cross sections at NLO.

In order discuss scale uncertainties, we perform the scale variation by
including effects of running \MSbar-renormalized parameters.
The running is given by the following coupled system of differential equations
\begin{align}
  \frac{\partial}{\partial \mu^2} p_i (\mu^2) = \beta_{p_i} (\{ p_j (\mu^2) \}),
\end{align}
where $\beta_{p_i}$ denotes the $\beta$-function of $p_i$ which is non-zero if
$p_i$ is defined in an $\MSbar$ scheme. For the integration we use
standard Runge--Kutta techniques. 

\subsubsection{Higher-order ambiguities in the conversion}
\label{sec:exactconvambiguities}

The ``full'' parameter conversion described in the previous section suffers from
ambiguities that are connected to higher-order contributions beyond NLO accuracy.
Firstly, the whole renormalization programs addressed in this paper are worked out
to NLO, i.e.\ many relations between renormalizations constants are worked out
to linear order only. Beyond NLO, many missing terms should be completed.
Secondly, taking the matching equation~\refeq{eq:exact_conversion} at NLO
(or any fixed order) in the ``full'' conversion,
it should be noted that not all UV divergences cancel exactly, because the
parameters in the coefficients in front of the UV-divergent contributions in 
$\de p^{(i)}$ are not the same (but defined in different schemes).
That means that some UV scale has to be fixed in the evaluation of $\de p^{(i)}$
(which is typically set to the renormalization scale), the effect of which, however,
is beyond NLO.

We, thus, cannot claim that the full conversion is more precise than the linearized
version; both are equivalent at NLO. As already mentioned, the full conversion has
the advantage of being exactly invertible.
Another benefit of supporting both versions is the fact that the comparison of the
respective results gives the typical size of effects beyond NLO, which often helps to
identify scenarios which are perturbatively unstable.

Looking beyond NLO, it should be mentioned that again specific care
has to be taken with respect to the inclusion of tadpole contributions
in the matching equation~\refeq{eq:exact_conversion}.  For instance,
in the \PRTS\ scheme tadpole contributions enter the relations between
bare parameters, while in the \FJTS\ scheme they do not.  Thus, if the
matching of the schemes is done in the original parametrization of the
theory (i.e.\ at the level of $\mu^2$ and $\lambda$ parameters in the
scalar potential), in general there will be tadpole contributions in
$\de p_i^{(\mathrm{\PRTS})}$, but not in $\de p_i^{(\mathrm{\FJTS})}$,
so that
\begin{align}
  \de p_i^{(\mathrm{\FJTS})} 
= \de p_i^{(\mathrm{\PRTS})} + \Delta T_i\left(\de t_j^{(\mathrm{\PRTS})}\right),
\end{align}
where $\Delta T_i\left(\de t_j^{(\mathrm{\PRTS})}\right)$ is some
function of \PRTS\ tadpole counterterms $\de t_j^{\mathrm{\PRTS}}$.

Besides the treatment of tadpoles other sources of ambiguities exist.
Since the conversion is performed at fixed loop order
the precise choice of the independent parameters matters in higher orders.
For instance,
the matching of bare parameters for the mixing angles can be performed directly
on the mixing angles 
\begin{align}
  \alpha_\mathrm{B} = \alpha + \de \alpha, \quad
  \beta_\mathrm{B} = \beta + \de \beta,
\end{align}
or it can be performed on derived parameters such as
\begin{align}
  \tan \alpha_\mathrm{B} = \tan \alpha + \de \tan \alpha, \quad
  \tan \beta_\mathrm{B} = \tan \beta + \de \tan \beta,
\end{align}
which will result in a conversion equivalent only at NLO.  In
\refapp{se:conversiontables} the conversion tables for the input
parameters of \refse{se:results} are given.  The conversion is
performed in two variants which exemplify the before-mentioned
ambiguities and their impact on final results.

\section{Phenomenological results}
\label{se:results}

\subsection[Higgs-boson decays
\texorpdfstring{$\PH_{1,2}\to\PW\PW/\PZ\PZ\to4\,$fermions}{H1/2->WW/ZZ->4
  fermions}]%
{Higgs-boson decays
\boldmath{$\PH_{1,2}\to\PW\PW/\PZ\PZ\to4\,$fermions}}
\label{se:h4f}

The Monte Carlo program
\Prophecy~\cite{Bredenstein:2006rh,Bredenstein:2006ha}
provides a ``\textbf{PROP}er description of the \textbf{H}iggs d\textbf{EC}a\textbf{Y} 
into \textbf{4 F}ermions'' and calculates
observables for the decay process Higgs${} \to \PW\PW/\PZ\PZ \to 4\,$fermions at NLO EW+QCD.
Its first version~\cite{Bredenstein:2006rh,Bredenstein:2006ha} was designed for the SM
and slightly generalized to include a possible fourth fermion generation in
\citere{Denner:2011vt}.
In \citeres{Altenkamp:2017ldc,Altenkamp:2017kxk} and \cite{Altenkamp:2018bcs},
\Prophecy\ was extended to the corresponding decays of the light CP-even
Higgs bosons of the \THDM\ and the \HSESM, respectively,
keeping the functionality and applicability of the program basically the same.
The \THDM\ and the \HSESM\ were renormalized using $\MSbar$ schemes for the
mixing angles $\alpha$ and $\beta$.
In the following, we present first results from a further extension of
\Prophecy%
\footnote{The corresponding version of \Prophecy\ can be obtained from the
authors on request and will be available via {\tt www.hepforge.org} soon.}
which covers the decays of the heavy CP-even scalar bosons as well and
which additionally supports the on-shell and symmetry-inspired renormalization schemes
described in this paper.
All Higgs-boson decays via intermediate on- or off-shell EW gauge bosons W/Z into all
light fermions (all other than top quarks) are supported,
but potential decays of a heavy Higgs boson into a pair of light Higgs bosons
such as $\PH_1\to\PH_2\PH_2$, or to $\Pt\bar\Pt$~pairs 
are considered as separate processes and not
included in the calculation.

\subsubsection{Higgs-Singlet Extension of the  Standard Model}
\label{se:h4f_HSESM}

In \citere{Altenkamp:2018bcs}, the \HSESM\ was renormalized in two
schemes, called $\MSbar$ and FJ there, which are identical with the
\MSbarPRTS\ and \MSbarFJTS\ schemes of this paper up to the point that different
$\MSbar$-renormalized quartic scalar couplings 
$\lambda_i$ are used
to parametrize the Higgs sector.  In \citere{Altenkamp:2018bcs}, the
coupling $\lambda_{12}=\lambda_3/2$, 
which mixes the doublet and singlet scalars
was chosen, while we choose the
quartic coupling $\lambda_1$ of the singlet sector in this paper instead. For
the application to the decays $\PH_{1,2}\to4f$, this choice makes only
a marginal difference, which is even beyond NLO, since the
renormalization of those quartic Higgs couplings does not enter in
this case. Only a minor effect from the different running of the
couplings in some one-loop corrections remains.

In \citere{Altenkamp:2018bcs}, five different \HSESM\ scenarios were
considered, which are still compatible with current LHC results.
These scenarios, which are called BHM200$^\pm$, BHM400, BHM600, and
BHM800, identify the light Higgs boson $\PH_2$ with the discovered
state with a mass of $\sim125\GeV$ and contain a heavy Higgs boson
$\PH_1$ of mass $M_{\PH_1}=200\GeV,\dots,800\GeV$, as suggested by the
names of the scenarios.  The \HSESM\ parameters of scenarios
BHM200$^\pm$, BHM400, and BHM600 are summarized in \refta{tab:SESMinput}.
In the following, we take over these scenarios and refer to
\citere{Altenkamp:2018bcs} for the precise values of the SM-like input
parameters.
\begin{table}
\centerline{
\begin{tabular}{|c||c|c|c|}
\hline
Scenario & $M_{\PH_1}$[GeV] & $s_\alpha$ & $\lambda_3/2=\lambda_{12}^{\mbox{\scriptsize\cite{Altenkamp:2018bcs}}}$ 
\\
\hline
\hline
BHM200$^\pm$ & 200 & $\pm0.29$ & $\pm0.07$
\\
\hline
BHM400 & 400 & 0.26 & $0.17$
\\
\hline
BHM600 & 600 & 0.22 & $0.23$
\\
\hline
\end{tabular}
}
\caption{\HSESM\ input parameters in the considered \HSESM\ scenarios.
The light Higgs boson has mass $M_{\PH_2}=125.1\GeV$.}
\label{tab:SESMinput}
\end{table}

\begin{table}
\centerline{\renewcommand{\arraystretch}{1.25}
\begin{tabular}{|c||l|l|l|l|}
\hline
& \multicolumn{2}{c|}{BHM200$^+$} & \multicolumn{2}{c|}{BHM200$^-$} 
\\
Scheme & \multicolumn{1}{c|}{LO} & \multicolumn{1}{c|}{NLO}  & \multicolumn{1}{c|}{LO} & \multicolumn{1}{c|}{NLO}  
\\
\hline
\hline
\MSbarPRTS & $0.83361(3)_{-4.4\%}^{+3.2\%}$ & $0.90539(6)_{+0.5\%}^{+0.5\%}$ & 
        $0.83261(3)_{-4.5\%}^{+3.3\%}$ & $0.90546(7)_{+0.6\%}^{+0.5\%}$  
\\
\hline
\MSbarFJTS & $0.82292(3)_{+2.7\%}^{-3.5\%}$ & $0.90550(7)_{+0.0\%}^{+0.7\%}$ & 
        $0.82614(3)_{+0.7\%}^{-0.6\%}$ & $0.90558(7)_{-0.1\%}^{+0.0\%}$ 
\\
\hline
\OS & $0.84034(3)$ & $0.90553(6)_{-0.0\%}^{+0.0\%}$ & 
      $0.84034(3)$ & $0.90552(6)_{-0.0\%}^{+0.0\%}$
\\
\hline
\RSBFM & \change{$0.84036(3)$} & $0.90553(6)$  & 
         $0.84035(3)$ & $0.90552(6)$
\\
\hline
\multicolumn{5}{c}{}\\
\hline
& \multicolumn{2}{c|}{BHM400} & \multicolumn{2}{c|}{BHM600} 
\\
Scheme & \multicolumn{1}{c|}{LO} & \multicolumn{1}{c|}{NLO}  & \multicolumn{1}{c|}{LO} & \multicolumn{1}{c|}{NLO}  
\\
\hline
\hline
\MSbarPRTS &
$0.85209(3)_{-0.5\%}^{+0.5\%}$ & $0.92159(7)_{-0.0\%}^{+0.0\%}$ & 
$0.87067(3)_{-0.1\%}^{+0.1\%}$ & $0.94060(7)_{-0.0\%}^{+0.0\%}$ 
\\
\hline
\MSbarFJTS & 
$0.85349(3)_{+1.6\%}^{-2.1\%}$ & $0.92166(7)_{+0.3\%}^{+0.1\%}$ & 
$0.87608(3)_{+1.2\%}^{-1.5\%}$ & $0.94106(7)_{+0.3\%}^{-0.0\%}$
\\
\hline
\OS &
$0.85548(3)$ & $0.92178(6)_{-0.0\%}^{+0.0\%}$ & 
$0.87309(3)$ & $0.94078(7)_{-0.0\%}^{+0.0\%}$ 
\\
\hline
\RSBFM & \change{$0.85663(3)$} & \change{$0.92206(6)$} & 
\change{$0.87381(3)$} & \change{$0.94118(7)$}
\\
\hline
\end{tabular}
}
\caption{LO and NLO decay widths $\Gamma^{\PH_2\to4f}$[MeV]
of the light \HSESM\ Higgs boson $\PH_2$ for various 
\HSESM\ scenarios in different renormalization schemes,
with the \OS\ scheme as input scheme (and full conversion
of the input parameters into the other schemes).
The scale variation (given in percent) corresponds to
the scales $\mu=\mu_0/2$ and $\mu=2\mu_0$ with central
scale $\mu_0=M_{\PH_2}$.}
\label{tab:SESM-H24f}
\end{table}
\begin{figure}
\centerline{
\includegraphics[width=0.49\textwidth]{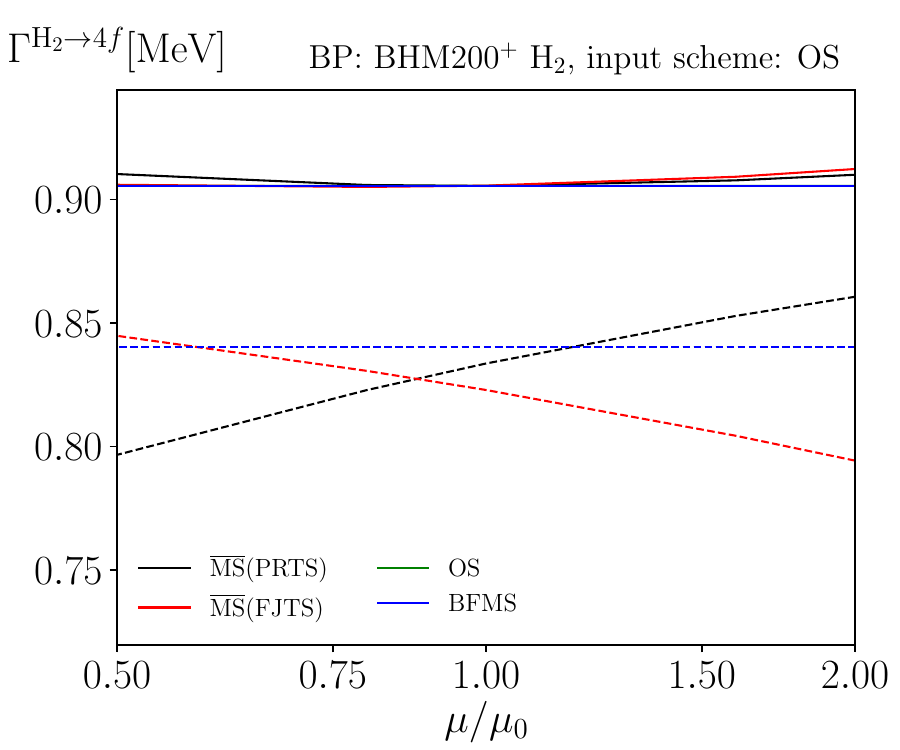}
\includegraphics[width=0.49\textwidth]{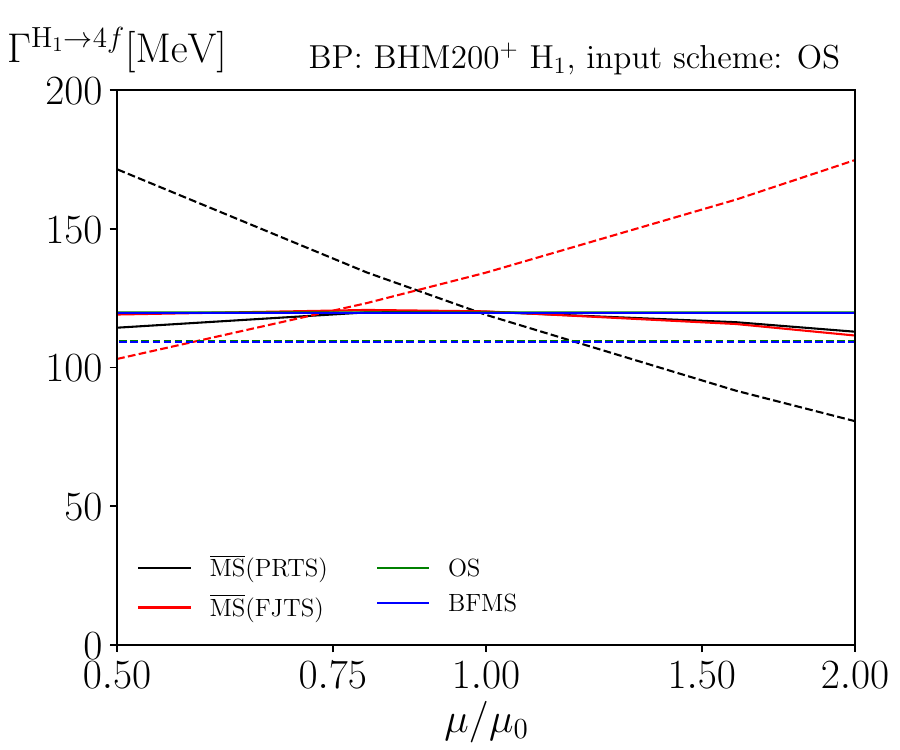}
}
\caption{Scale dependence of the decay widths 
  $\Gamma^{\PH_2\to4f}$ (left) and $\Gamma^{\PH_1\to4f}$ (right) of
  the light and heavy \HSESM\ Higgs bosons $\PH_2$ and $\PH_1$ for the
  \HSESM\ scenario BHM200$^+$ in different renormalization schemes,
  with the \OS\ scheme as input scheme (and full conversion of the
  input parameters into the other schemes).  LO results are shown as
  dashed, NLO as full lines; the central scale is set to
  $\mu_0=M_{\PH_2}$.}
\label{fig:SESM-H4f}
\end{figure}
\begin{table}
\centerline{\renewcommand{\arraystretch}{1.25}
\begin{tabular}{|c||l|l|l|l|}
\hline
& \multicolumn{2}{c|}{BHM200$^+$} & \multicolumn{2}{c|}{BHM200$^-$} 
\\
Scheme & \multicolumn{1}{c|}{LO} & \multicolumn{1}{c|}{NLO}  & \multicolumn{1}{c|}{LO} & \multicolumn{1}{c|}{NLO}  
\\
\hline
\hline
\MSbarPRTS & 
$118.976(4)_{+43.6\%}^{-31.9\%}$ & $120.240(7)_{-4.7\%}^{-5.9\%}$ & 
$120.397(4)_{+44.6\%}^{-32.4\%}$ & $120.114(7)_{-5.6\%}^{-5.7\%}$ 
\\
\hline
\MSbarFJTS & 
$134.126(4)_{-23.2\%}^{+30.1\%}$ & $120.403(9)_{-1.0\%}^{-7.1\%}$ & 
$129.570(4)_{-6.0\%}^{+5.6\%}$   & $120.132(8)_{+0.6\%}^{-0.1\%}$ 
\\
\hline
\OS & 
$109.430(4)$ & $119.847(8)_{+0.0\%}^{-0.0\%}$ & 
$109.430(4)$ & $119.812(8)_{+0.0\%}^{-0.0\%}$ 
\\
\hline
\RSBFM & 
\change{$109.393(4)$} & \change{$119.846(8)$} & 
$109.419(4)$ & $119.811(8)$
\\
\hline
\multicolumn{5}{c}{}\\
\hline
& \multicolumn{2}{c|}{BHM400} & \multicolumn{2}{c|}{BHM600} 
\\
Scheme & \multicolumn{1}{c|}{LO} & \multicolumn{1}{c|}{NLO}  & \multicolumn{1}{c|}{LO} & \multicolumn{1}{c|}{NLO}  
\\
\hline
\hline
\MSbarPRTS & 
$1617.26(4)_{+6.3\%}^{-6.2\%}$ & $1648.62(8)_{+0.6\%}^{-0.6\%}$ &
$4530.1(1)_{+2.1\%}^{-2.3\%}$ & $4546.0(2)_{+0.6\%}^{-0.4\%}$ 
\\
\hline
\MSbarFJTS & 
$1582.44(4)_{-21.7\%}^{+27.6\%}$ & $1646.83(8)_{-3.6\%}^{-1.5\%}$ & 
$4007.1(1)_{-24.8\%}^{+32.5\%}$ & $4509.4(3)_{-6.0\%}^{-0.3\%}$ 
\\
\hline
\OS & 
$1533.42(4)$ & $1643.86(8)_{+0.0\%}^{-0.0\%}$ &
$4295.9(1)$ & $4532.4(2)_{+0.0\%}^{-0.0\%}$
\\
\hline
\RSBFM & \change{$1505.02(4)$} & \change{$1636.86(9)$} &
\change{$4226.6(1)$} & \change{$4493.8(2)$}
\\
\hline
\end{tabular}
}
\caption{As in \refta{tab:SESM-H24f}, but for the decay width
$\Gamma^{\PH_1\to4f}$[MeV] of the heavy \HSESM\ Higgs boson $\PH_1$.}
\label{tab:SESM-H14f}
\end{table}
Table~\ref{tab:SESM-H24f} shows a comparison of the LO and NLO results
for the  decay width of the light Higgs boson, $\PH_2\to4f$,
obtained in the different renormalization schemes for each scenario.
Figure~\ref{fig:SESM-H4f} illustrates the scale dependence of
$\Gamma^{\PH_2\to4f}$ on the l.h.s.\ for scenario BHM200$^+$; the
results for the other scenarios look qualitatively similar.  The input
parameters of the scenarios are defined in the \OS\ scheme and
consistently converted into the other schemes, as described in
\refse{se:conversion} 
(full conversion).
The residual dependence of the LO and NLO decay widths on the
renormalization scale $\mu$, as obtained from a rescaling of the
central scale $\mu_0=M_{\PH_2}$ by factors of $1/2$ and $2$, is shown
in percent as lower and upper suffixes, respectively.  One crucial
observation in \refta{tab:SESM-H24f} is that all renormalization
schemes deliver central NLO results differing only below the permille
level, while the LO results deviate significantly by up to 2\%, \ie
the renormalization-scheme dependence reduces drastically in the
transition from LO to NLO.  Another important observation concerns the
dependence on the renormalization scale.  For the \MSbarPRTS\ and \MSbarFJTS\ 
schemes, in which the $\MSbar$ definitions of the angle $\alpha$ lead
to a running $\alpha(\mu)$, the residual scale dependence of the decay
widths is reduced from ${\lesssim5\%}$ at LO to $\lesssim0.7\%$ at NLO. This
observation was already made in \citere{Altenkamp:2018bcs}, since the
$\MSbar$ and FJ schemes from there almost coincide with the \MSbarPRTS\ and
\MSbarFJTS\ schemes used here.  The residual NLO scale uncertainty of
$\lesssim0.7\%$, as estimated from rescaling $\mu_0$ by factors of $1/2$
and $2$, is, thus, of the order of magnitude typically expected from a
well-behaved EW NLO calculation. 
Clearly, the higher-order scale
dependence of the \OS and \RSBFM schemes cannot be used as 
an estimate for
the theoretical uncertainty, while the difference of these schemes
reflects part of the renormalization-scheme dependence.

Table~\ref{tab:SESM-H14f} shows a comparison of the LO and NLO results
for the decay width of the heavy Higgs boson, $\PH_1\to4f$,
 obtained in the different renormalization schemes
for each scenario. 
Figure~\ref{fig:SESM-H4f} illustrates the scale dependence of 
$\Gamma^{\PH_1\to4f}$ on the r.h.s.\ for scenario BHM200$^+$, while again
the results for the other scenarios look qualitatively similar.
Note that the phenomenology of the decays $\PH_2\to4f$ and $\PH_1\to4f$
is rather different. Firstly, the chosen values for the mass $M_{\PH_1}$
of the heavy Higgs boson are larger than $2\MW$ and $2\MZ$, so that
the decays can proceed via two resonant W or Z~bosons, while
the decays $\PH_2\to4f$ involve at least one off-shell W/Z~boson.
This effect and the larger phase space for $\PH_1$ leads to a large
enhancement of the decay width for $\PH_1\to4f$.
This enhancement is somewhat damped by the second difference of the
two decay types. While the LO decay width of $\PH_2\to4f$ 
involves an explicit factor of $c_\alpha^2$ w.r.t.\ the SM case,
the LO decay width of $\PH_1\to4f$ contains the complementary 
factor $s_\alpha^2$. Since $\sa\sim 0.2{-}0.3$ in the
considered scenarios, $\Gamma^{\PH_1\to4f}$ is reduced by a factor of
$\sim0.04{-}0.1$ w.r.t.\ to the SM decay width with the same 
(hypothetical) Higgs-boson mass $M_{\PH_1}$.
The renormalization-scale dependence in the \MSbarPRTS and \MSbarFJTS
schemes is reduced by a factor $5{-}10$ at NLO as compared to LO.
A similar reduction is observed for the differences between the
renormalization schemes.

\subsubsection{Two-Higgs-Doublet Model}
\label{se:h4f_THDM}

In \citeres{Altenkamp:2017ldc,Altenkamp:2017kxk},
the \THDM\ was renormalized in four schemes,
called $\MSbar(\alpha)$, FJ$(\alpha)$, $\MSbar(\lambda_3)$, 
and FJ$(\lambda_3)$ there, 
where the first two are identical with the \MSbarPRTS\ and \MSbarFJTS\
schemes of this paper, respectively. 
In the other two schemes, $\MSbar(\lambda_3)$ and FJ$(\lambda_3)$,
the angle $\alpha$ is replaced by the scalar self-coupling $\lambda_3$
as $\MSbar$-renormalized input parameter; these two schemes are not
considered in the following.

In \citeres{Altenkamp:2017ldc,Altenkamp:2017kxk}, following suggestions
in the literature, various different \THDM\ scenarios
were considered, which are still compatible with current LHC results
and identify the light CP-even Higgs boson $\PH_2$ with the discovered
state with a mass of $125\GeV$.
In the following, we pick four of those scenarios, covering typical cases
with light or heavy Higgs bosons in addition to the known of mass $125\GeV$.
Table~\ref{tab:THDMinput} summarizes the corresponding \THDM\ input
parameters, while the remaining SM-like parameters are taken over from
\citeres{Altenkamp:2017ldc,Altenkamp:2017kxk}.
\begin{table}
\centerline{
\begin{tabular}{|c||c|c|c|c|c|c|}
\hline
Scenario & $M_{\PH_1}$ & $M_{\PH^+},M_{\Ha}$ & 
$\lambda_5$ & $t_\beta$ & $\cab$ & comment
\\
 & [GeV] & [GeV] & & & &
\\
\hline
\hline
A1 & 300 & 460 & $-1.9$ & 2 & 0.1 & Aa in \citeres{Altenkamp:2017ldc,Altenkamp:2017kxk}
\\
\hline
A2 & 300 & 460 & $-1.9$ & 2 & 0.2 & A$(\cab=0.2)$ \\
&&&&&& in \citeres{Altenkamp:2017ldc,Altenkamp:2017kxk}
\\
\hline
B1 & 600 & 690 & $-1.9$ & 4.5 & 0.15 & B1$(\cab=0.15)$ \\
&&&&&& in \citeres{Altenkamp:2017ldc,Altenkamp:2017kxk}
\\
\hline
B2 & 200 & 420 & $-2.5746$ & 3 & 0.3 & BP3$_{B1}$ in \citeres{Altenkamp:2017ldc,Altenkamp:2017kxk},
\\
&&&&&& BP3B1 in \citere{Denner:2017vms}
\\
\hline
\end{tabular}
}
\caption{\THDM\ input parameters in the considered scenarios of a \THDM\ of Type~I.
The light CP-even Higgs boson has mass $M_{\PH_2}=125\GeV$, and $\cab=\cos(\al-\be)$.}
\label{tab:THDMinput}
\end{table}

\begin{figure}
\centerline{
\includegraphics[width=0.49\textwidth]{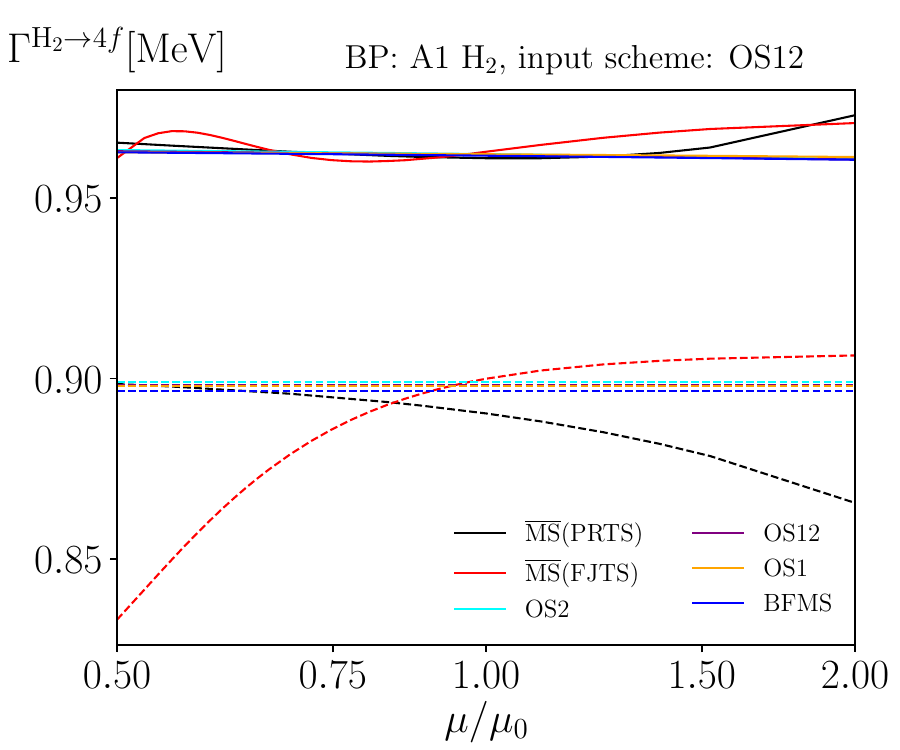}
\includegraphics[width=0.49\textwidth]{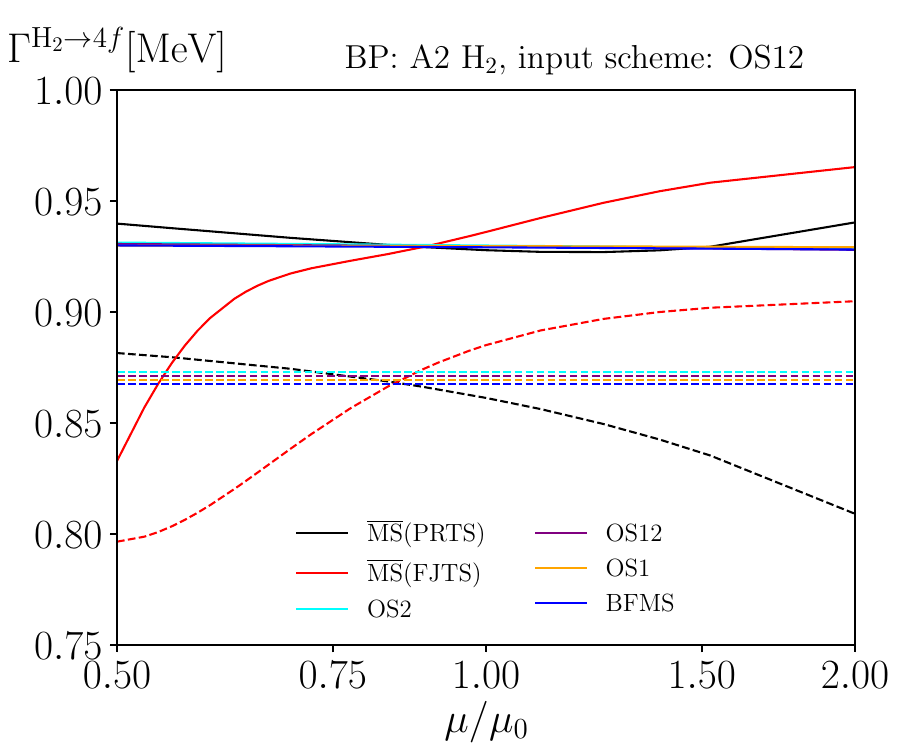}
}
\caption{Scale dependence of the decay widths 
  $\Gamma^{\PH_2\to4f}$ of the light \THDM\ Higgs boson $\PH_2$ for
  the \THDM\ scenarios A1 (left) and A2 (right) in different
  renormalization schemes, with the \OSonetwo\ scheme as input scheme
  (and full conversion of the input parameters into the other
  schemes).  LO results are shown as dashed, NLO as full lines; the
  central scale is set to
  $\mu_0=(M_{\PH_2}+M_{\PH_1}+M_{\Ha}+2M_{\PH^+})/5$.}
\label{fig:THDM-H24f}
\end{figure}

Table~\ref{tab:THDM-H24f} shows a comparison of the LO and NLO results
for the $\PH_2\to4f$ decay width obtained in the different
renormalization schemes for each scenario.  
Figure~\ref{fig:THDM-H24f}
illustrates the scale dependence of $\Gamma^{\PH_2\to4f}$ in greater
detail.
\begin{table}
\centerline{\renewcommand{\arraystretch}{1.25}
\begin{tabular}{|c||l|l|l|l|}
\hline
& \multicolumn{2}{c|}{A1} & \multicolumn{2}{c|}{A2} 
\\
Scheme & \multicolumn{1}{c|}{LO} & \multicolumn{1}{c|}{NLO}  & \multicolumn{1}{c|}{LO} & \multicolumn{1}{c|}{NLO}  
\\
\hline
\hline
\MSbarPRTS & 
$0.89035(3)_{+0.9\%}^{-2.8\%}$ & $0.96107(7)_{+0.4\%}^{+1.2\%}$ &
$0.86130(3)_{+2.3\%}^{-6.1\%}$ & $0.92784(7)_{+1.3\%}^{+1.3\%}$ 
\\
\hline
\MSbarFJTS & 
$0.89996(3)_{-7.4\%}^{+0.7\%}$ & $0.96286(7)_{-0.2\%}^{+0.8\%}$ &
$0.88508(3)_{-10.0\%}^{+2.2\%}$ & $0.93605(7)_{-11.0\%}^{+3.1\%}$ 
\\
\hline
\OSone & 
$0.89801(3)$ & $0.96218(7)_{+0.1\%}^{-0.1\%}$ &
$0.86917(3)$ & $0.92968(7)_{+0.0\%}^{-0.1\%}$
\\
\hline
\OStwo & 
$0.89911(3)$ & $0.96221(7)_{+0.1\%}^{-0.1\%}$ &
$0.87295(3)$ & $0.92995(7)_{+0.1\%}^{-0.2\%}$
\\
\hline
\OSonetwo & 
$0.89832(3)$ & $0.96197(7)_{+0.1\%}^{-0.1\%}$ &
$0.87110(3)$ & $0.92947(7)_{+0.1\%}^{-0.2\%}$
\\
\hline
\RSBFM &
\change{$0.89647(3)$} & \change{$0.96177(7)_{+0.1\%}^{-0.1\%}$} &
\change{$0.86764(3)$} & \change{$0.92914(7)_{+0.1\%}^{-0.1\%}$} 
\\
\hline
\multicolumn{5}{c}{}\\
\hline
& \multicolumn{2}{c|}{B1} & \multicolumn{2}{c|}{B2} 
\\
Scheme & \multicolumn{1}{c|}{LO} & \multicolumn{1}{c|}{NLO}  & \multicolumn{1}{c|}{LO} & \multicolumn{1}{c|}{NLO}  
\\
\hline
\hline
\MSbarPRTS &
$0.88609(3)_{+0.4\%}^{-4.9\%}$ & $0.94053(7)_{+1.0\%}^{+1.1\%}$ & $0.87941(3)_{+2.6\%}^{-42.7\%}$ & $0.924184(7)_{+4.2\%}^{-39.2\%}$
\\
\hline
\MSbarFJTS & 
$0.90406(3)_{+0.4\%}^{+0.4\%}$ & $0.96073(7)_{-1.1\%}^{+1.5\%}$ & $0.90654(3)_{-2.5\%}^{-87.0\%}$ &  $0.96980(7)_{-0.0\%}^{>+100\%}$
\\
\hline
\OSone &
$0.86829(3)$ & $0.9428(1)_{-0.2\%}^{-0.1\%}$ &  0.82069(3) & $0.87179(6)_{+0.3\%}^{-1.3\%}$
\\
\hline
\OStwo &
$0.88838(3)$ & $0.94136(7)_{+0.2\%}^{-0.4\%}$ & 0.82684(3) & $0.87242(6)_{+0.3\%}^{-0.4\%}$
\\
\hline
\OSonetwo &
$0.88698(3)$ & $0.94074(7)_{+0.2\%}^{-0.5\%}$ & $0.82573(3)$ & $0.87189(6)_{+0.3\%}^{-0.5\%}$ 
\\
\hline
\RSBFM & 
\change{$0.88721(3)$} & \change{$0.94113(8)_{+0.1\%}^{-0.2\%}$} & 
\change{$0.81262(3)$} & \change{$0.86741(6)_{+0.0\%}^{+0.1\%}$} 
\\
\hline
\end{tabular}
}
\caption{LO and NLO decay widths $\Gamma^{\PH_2\to4f}$[MeV]
of the light CP-even Higgs boson $\PH_2$ of the \THDM\ for various 
scenarios in different renormalization schemes,
with the \OSonetwo\ scheme as input scheme (and full conversion
of the input parameters into the other schemes).
The scale variation (given in percent) corresponds to the 
scales $\mu=\mu_0/2$ and $\mu=2\mu_0$ with central
scale $\mu_0=(M_{\PH_2}+M_{\PH_1}+M_{\Ha}+2M_{\PH^+})/5$.}
\label{tab:THDM-H24f}
\end{table}

Among the two schemes with $\MSbar$-renormalized mixing angles
$\alpha$ and $\beta$, the \MSbarPRTS\ scheme delivers NLO results with
a sufficiently well reduced renormalization-scale dependence of
$1{-}2\%$ in scenarios A1, A2, and B1, while this is the case for the
\MSbarFJTS\ scheme only in scenarios A1 and B1.  For scenario B2,
neither the \MSbarPRTS, nor the \MSbarFJTS\ scheme produces results
with a visible stability region in $\mu$.  A possible reason is the
relatively small difference $\MHone-\MHtwo$ in the scalar masses in
scenario B2 which can lead to enhanced corrections owing to the mixing
renormalization constants Eq.~\refeq{eq:zij_onshell} in the $\MSbar$
schemes. Moreover, the problems of those schemes in B2 can already be
expected from the parameter conversion (\cf
\refta{tab:conversion-THDM-OS12} in \refapp{se:conversiontables}) into
the schemes, which involves very large corrections and shows
significant differences between linearized and full conversions. It
should be mentioned, however, that the versions of these $\MSbar$-like
renormalization schemes in which $\alpha$ is replaced as input
parameter by the coupling parameter $\lambda_3$ (the schemes
$\MSbar(\lambda_3)$ and FJ$(\lambda_3)$ of
\citeres{Altenkamp:2017ldc,Altenkamp:2017kxk}) behave somewhat better
and at least produce some narrow plateau regions in the vicinity of
the scale $\mu_0$ (not shown in this paper).

The \OS\ schemes and the \RSBFM\ variant, on the other hand, 
deliver NLO results which agree within very few per mille, which is
also the size of the residual scale dependence resulting from the
running of $\lambda_5$ in the loop corrections.
An exception is scenario B2, which shows a residual scale dependence
of up to $1\%$.

\begin{figure}
\centerline{
\includegraphics[width=0.49\textwidth]{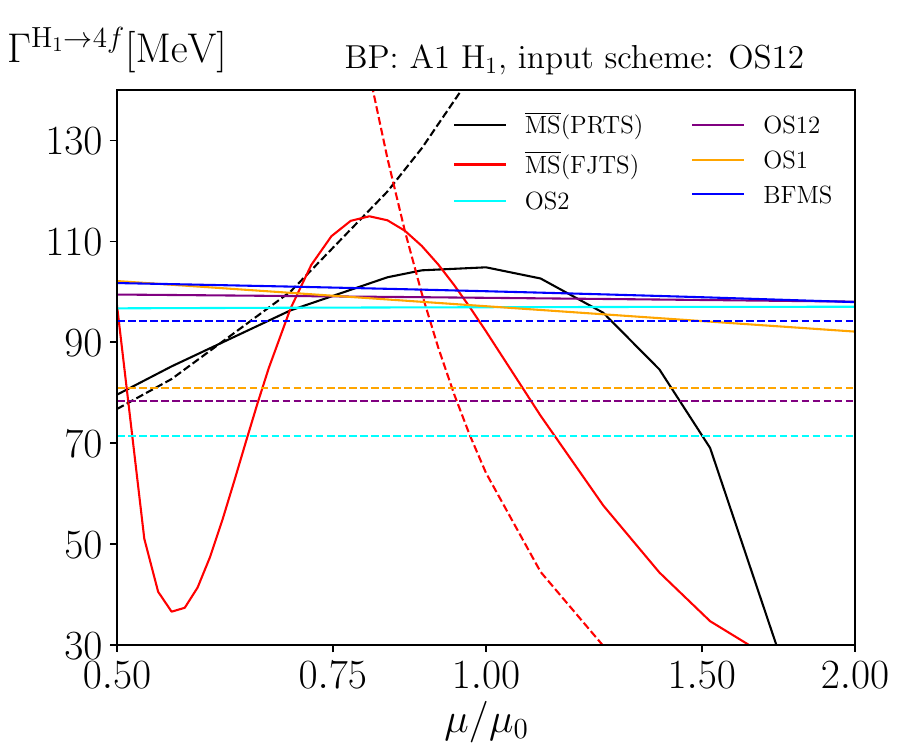}
\includegraphics[width=0.49\textwidth]{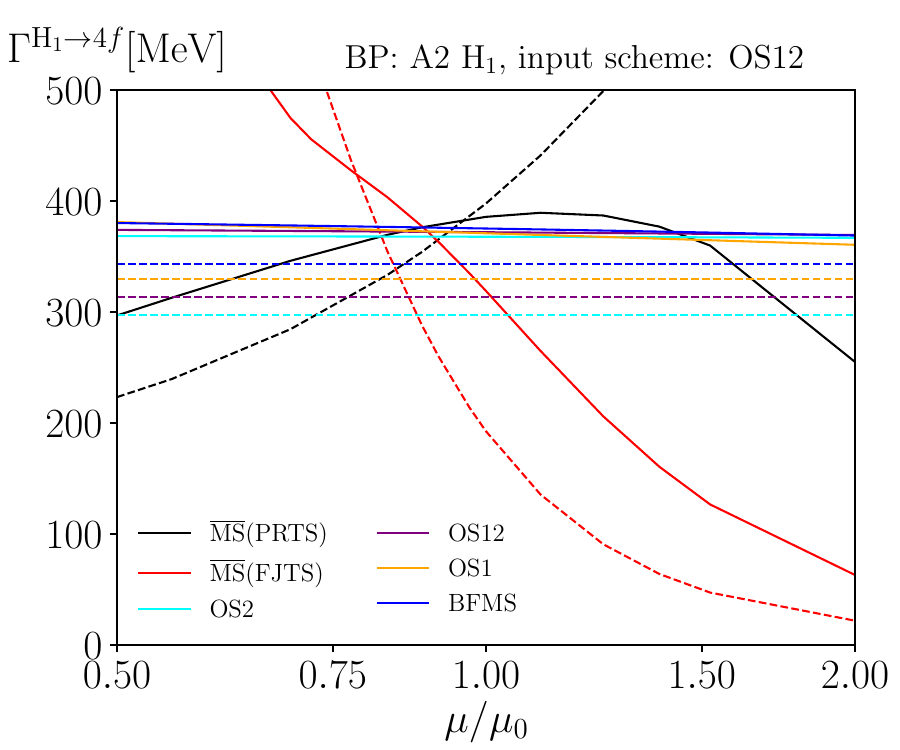}
}
\caption{As in \reffi{fig:THDM-H24f}, but for the decay width
$\Gamma^{\PH_1\to4f}$ of the heavy CP-even Higgs boson of the \THDM.}
\label{fig:THDM-H14f}
\end{figure}

Now we turn to the discussion of the decays of the heavy CP-even Higgs
boson $\PH_1$.  Table~\ref{tab:THDM-H14f} and \reffi{fig:THDM-H14f}
show the LO and NLO results and their scale dependence for
$\PH_1\to4f$ decay width obtained in the different renormalization
schemes for each scenario.
\begin{table}
\centerline{\renewcommand{\arraystretch}{1.25}
\begin{tabular}{|c||l|l|l|l|}
\hline
& \multicolumn{2}{c|}{A1} & \multicolumn{2}{c|}{A2} 
\\
Scheme & \multicolumn{1}{c|}{LO} & \multicolumn{1}{c|}{NLO}  & \multicolumn{1}{c|}{LO} & \multicolumn{1}{c|}{NLO}  
\\
\hline
\hline
\MSbarPRTS & 
$147.102(4)_{-47.8\%}^{>+100\%}$ & $104.86(2)_{-24.1\%}^{<-100\%}$ &
$397.75(1)_{-43.8\%}^{>+100\%}$ & $385.72(2)_{-23.0\%}^{-33.9\%}$
\\
\hline
\MSbarFJTS & 
$64.096(2)_{>+100\%}^{-86.9\%}$  & $92.17(1)_{+5.6\%}^{-81.4\%}$ &
$192.524(5)_{>+100\%}^{-88.6\%}$ & $318.95(5)_{>+100\%}^{-80.2\%}$
\\
\hline
\OSone & 
$80.992(2)$ & $97.145(7)_{+5.1\%}^{-5.2\%}$ & 
$329.800(9)$ & $370.93(2)_{+2.7\%}^{-2.8\%}$
\\
\hline
\OStwo & 
$71.429(2)$ & $96.95(1)_{-0.2\%}^{+0.1\%}$ &
$297.253(8)$ & $367.80(3)_{+0.1\%}^{-0.3\%}$
\\
\hline
\OSonetwo & 
$78.304(2)$ & $98.812(8)_{+0.7\%}^{-0.8\%}$ &
$313.217(8)$ & $371.86(3)_{+0.6\%}^{-0.7\%}$
\\
\hline
\RSBFM &
\change{$94.265(2)$} & \change{$100.117(5)_{+1.6\%}^{-2.2\%}$} &
\change{$343.049(9)$} & \change{$375.22(2)_{+1.3\%}^{-1.7\%}$}
\\
\hline
\multicolumn{5}{c}{}\\
\hline
& \multicolumn{2}{c|}{B1} & \multicolumn{2}{c|}{B2} 
\\
Scheme & \multicolumn{1}{c|}{LO} & \multicolumn{1}{c|}{NLO}  & \multicolumn{1}{c|}{LO} & \multicolumn{1}{c|}{NLO}  
\\
\hline
\hline
\MSbarPRTS &
$2083.68(5)_{-17.9\%}^{>+100\%}$ & $2162.3(1)_{-27.0\%}^{<-100\%}$ & $40.122(1)_{-82.7\%}^{>+100\%}$ & $57.73(1)_{-82.7\%}^{>+100\%}$
\\
\hline
\MSbarFJTS & 
$325.923(7)_{-100\%}^{-99.6\%}$ & $1179.8(3)_{-99.8\%}^{-99.5\%}$ &  $1.21460(4)_{>+100\%}^{>+100\%}$ & 
 $-5.226(3)_{>+100\%}^{<-100\%}$
\\
\hline
\OSone &
$3824.52(9)$ 
& 
$1542.9(9)_{+28.2\%}^{-21.5\%}$
&  $124.325(4)$ & $132.307(8)_{+1.9\%}^{+4.6\%}$
\\
\hline
\OStwo & 
$1860.03(4)$ & $2130.3(1)_{+0.1\%}^{-1.6\%}$ & $115.515(4)$ & $131.09(1)_{+2.4\%}^{-4.2\%}$
\\
\hline
\OSonetwo &
$1997.07(5)$ & $2155.2(1)_{+0.7\%}^{-1.4\%}$ & $117.108(4)$ & $131.88(1)_{+2.2\%}^{-2.9\%}$
\\
\hline
\RSBFM & 
\change{$1973.68(5)$} & \change{$2145.0(1)_{+3.6\%}^{-9.5\%}$} & 
\change{$135.904(5)$} & \change{$138.729(8)_{+4.7\%}^{-7.7\%}$}
\\
\hline
\end{tabular}
}
\caption{As in \refta{tab:THDM-H24f}, but for the decay width
$\Gamma^{\PH_1\to4f}$[MeV] of the heavy CP-even Higgs boson $\PH_1$ of the \THDM.}
\label{tab:THDM-H14f}
\end{table}%
Not unexpectedly, the renormalization schemes based on
$\MSbar$-renormalized mixing angles deliver problematic results.
At least for the  scenarios A1, A2, and B1,
the NLO results of the \MSbarPRTS\ scheme still show extrema in
the $\mu$~dependence that may be interpreted as plateaus of stability,
but those regions are rather small and do not cover a range in $\mu$
from $\mu_0/2$ to $2\mu_0$. 
Though the extrema are located near
$\mu_0$, and the NLO results for $\Gamma^{\PH_1\to4f}$ in this region
are compatible with the results obtained in the other stable schemes.
The \MSbarFJTS\ scheme fails to produce stability regions in general,
as a result of the very strong running of $\beta-\alpha$.
For scenario B2, both the \MSbarPRTS\ and the \MSbarFJTS\ scheme fail
badly, as expected already from the bad behaviour observed for the
decay of the light Higgs boson $\PH_2$ above.
The \OS\ and \RSBFM\ schemes, however, mostly deliver NLO results
that are in nice mutual agreement, with the only exception of the
\OSone\ scheme in scenario B1, where the running 
of $\lambda_5$ in the loop corrections introduces a 
scale uncertainty of the order of $20{-}30\%$.
Note that also the NLO correction in this scheme is extremely large.

\subsection{Higgs-boson production processes at the LHC}
\label{se:HZ}

As a second application of our renormalization schemes, we consider the
\change{EW corrections to the production} of BSM Higgs bosons in association
with a vector boson, also known as Higgs-strahlung, and in association with two
jets, usually referred to as vector-boson fusion (VBF). For the numerical
integration we have used a modified version of the Monte Carlo program \HAWKTwo\
\cite{Denner:2014cla}. \HAWKTwo{} is a Monte Carlo integrator for
Higgs-strahlung and VBF in the SM, including the full fixed-order NLO QCD and EW
corrections
\cite{Ciccolini:2007jr,Ciccolini:2007ec,Denner:2011id}.  The
modification of \HAWKTwo{} concerns an interface with the \RecolaTwo{}
library which has been introduced in \citere{Denner:2016etu}.
\RecolaTwo{}, the successor of \Recola \cite{Actis:2016mpe}, is a
highly efficient one-loop amplitude provider for the SM and BSM
theories which is publicly available \cite{Denner:2017wsf}. All
necessary ingredients for the integration of the before-mentioned
processes for BSM theories can be automatically generated by
\RecolaTwo{} which, in turn, upgrades the original \HAWKTwo{} to a
general integrator for BSM Higgs production in Higgs-strahlung and VBF
as long as no external charged Higgs boson is considered. A new
release of \HAWKTwo{} will be made available shortly.

\subsubsection{Cut setup and parameters}
For the analysis of
\change{
Higgs-strahlung we focus on the final state with one charged muon and
a muon--neutrino, $\Pp\Pp \to
\PH \Pm^+ \Pnm + X$ at 13 \TeV.  The muon is not recombined
with collinear photons, and is assumed to be perfectly isolated,
treated as bare muon as described in \citere{Denner:2011id}.  We use
similar cuts to the ones given in \citere{Chatrchyan:2013zna}, \ie we demand the muon
\begin{itemize}
  \item has transverse momentum $p_{\mathrm{T},\Pm^+}> 20\GeV$,
  \item be central with rapidity $\left|y_{\Pm^+}\right| < 2.4$,
\end{itemize}
and require a missing transverse momentum of $p_{\mathrm{T}}^\mathrm{miss}> 25\GeV$.
} 

For the Higgs-boson production in VBF we require two hards jets
emitted from partons $i$ which fulfil
\begin{itemize}
  \item pseudo-rapidity $|\eta_i|  < 5$.
\end{itemize}
The jet definition
is performed using the anti-$k_\mathrm{T}$ algorithm \cite{Cacciari:2008gp} with jet size $D=0.4$.
The jets j$_i, i=1,2$, 
are required to fulfil typical VBF cuts (see e.g.
\citere{deFlorian:2016spz}):
\begin{itemize}
  \item transverse momentum $p_{\mathrm{T},{\mathrm{j}_i}} > 20\GeV $,
  \item rapidity $|y_{{\mathrm{j}_i}}| < 5$,
  \item rapidity difference $|y_{{\mathrm{j}_1}}-y_{{\mathrm{j}_2}}| > 3$,
  \item \change{opposite hemispheres $y_{{\mathrm{j}_1}} y_{{\mathrm{j}_2}} < 0$,}
  \item invariant mass $M_{{\mathrm{j}_1}{\mathrm{j}_2}}> 130\GeV$.
\end{itemize}
For Higgs-strahlung we investigate all the benchmark scenarios given
in \refta{tab:SESMinput} and \refta{tab:THDMinput} in the \HSESM{} and
\THDM, respectively. For VBF we only consider the points A1 and A2 in
\refta{tab:THDMinput} in the \THDM, since in general the results look
pretty similar to those for Higgs-strahlung.

\subsubsection{Higgs-strahlung in the \HSESM}
\label{se:HZ_HSESM}

\begin{table}
\centerline{\renewcommand{\arraystretch}{1.25}
\begin{tabular}{|c||l|l|l|l|l|l|l|l|}
\hline
& \multicolumn{2}{c|}{BHM200$^+$} & \multicolumn{2}{c|}{BHM200$^-$} 
\\
Scheme 
  & \multicolumn{1}{c|}{LO} & \multicolumn{1}{c|}{NLO}  
  & \multicolumn{1}{c|}{LO} & \multicolumn{1}{c|}{NLO}  
\\
\hline
\hline
  \MSbarPRTS 
& $42.61(2)^{+3.2\%}_{-4.5\%}$
& $40.80(2)^{+0.1\%}_{+1.2\%}$
& $42.55(2)^{+3.3\%}_{-4.6\%}$
& $40.80(2)^{+0.1\%}_{+1.3\%}$
\\
\hline
  \MSbarFJTS
& $42.12(2)^{-3.5\%}_{+2.7\%}$
& $40.90(2)^{+1.3\%}_{-0.3\%}$
& $42.23(2)^{-0.6\%}_{+0.7\%}$
& $40.85(2)^{+0.1\%}_{-0.2\%}$
\\
\hline
  \OS 
& $42.95(2)$
& $40.74(2)^{+0.0\%}_{+0.0\%}$
& $42.96(2)$
& $40.74(2)^{-0.0\%}_{-0.0\%}$
\\
\hline
  \RSBFM
& $42.95(2)$
& $40.74(2)$
& $42.96(2)$
& $40.74(2)$
\\
\hline
\multicolumn{5}{c}{}\\
\hline
& \multicolumn{2}{c|}{BHM400} & \multicolumn{2}{c|}{BHM600} 
\\
Scheme & \multicolumn{1}{c|}{LO} & \multicolumn{1}{c|}{NLO}  & \multicolumn{1}{c|}{LO} & \multicolumn{1}{c|}{NLO}  
\\
\hline
\hline
  \MSbarPRTS 
& $43.56(2)^{+0.5\%}_{-0.5\%}$
& $41.49(2)^{-0.0\%}_{+0.0\%}$
& $44.51(2)^{+0.1\%}_{-0.1\%}$
& $42.33(3)^{+0.0\%}_{+0.0\%}$
\\
\hline
\MSbarFJTS
& $43.63(2)^{-2.1\%}_{+1.7\%}$
& $41.48(2)^{+0.4\%}_{+0.1\%}$
& $44.79(2)^{-1.6\%}_{+1.2\%}$
& $42.33(3)^{+0.2\%}_{+0.2\%}$
\\
\hline
  \OS 
& $43.73(2)$
& $41.48(2)^{+0.0\%}_{+0.0\%}$
& $44.63(2)$
& $42.31(3)^{+0.0\%}_{-0.0\%}$
\\
\hline
  \RSBFM 
& $43.78(2)$
& $41.4\change{8}(2)$
& $44.6\change{3}(2)$
& $42.3\change{1}(3)$
\\
\hline
\end{tabular}
}
\caption{LO and NLO integrated cross sections $\sigma$ [pb]
for the production of the light Higgs boson in Higgs-strahlung, $\Pp\Pp\to
\PH_{2}\Pm^+ \Pnm+X$, for various 
\HSESM\ scenarios in different renormalization schemes,
with the \OS\ scheme as input scheme (and full conversion
of the input parameters into the other schemes).
The scale variation (given in percent) corresponds to the 
scales $\mu=\mu_0/2$ and $\mu=2\mu_0$ with central scale $\mu_0=M_{\PH_2}$.}
\label{tab:SESM-HZ-Hl}
\end{table}

\begin{figure}
\centerline{
\includegraphics[width=0.49\textwidth]{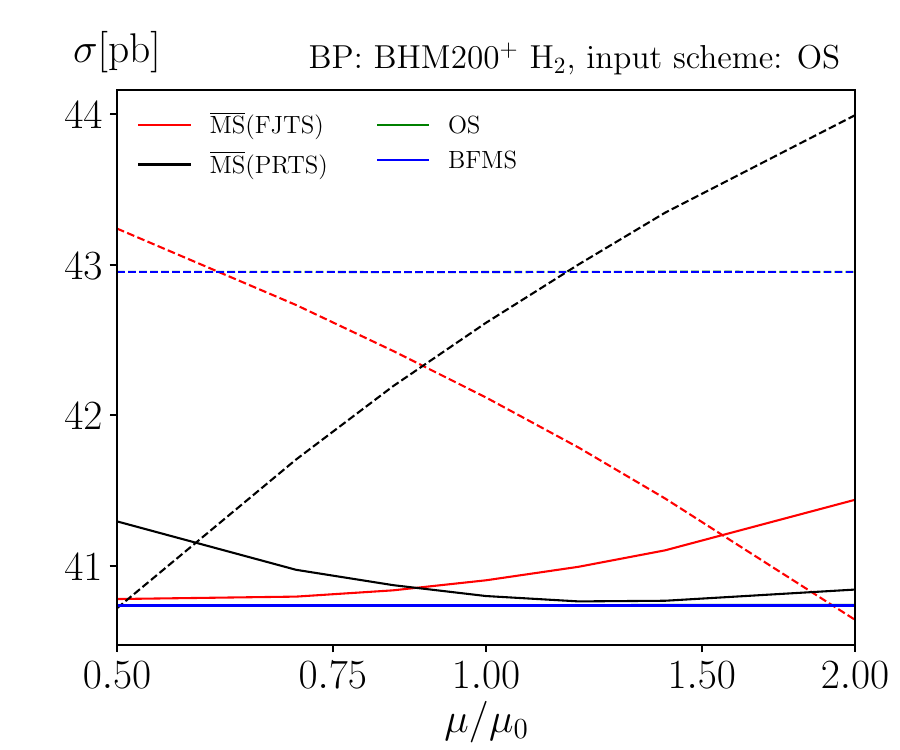}
\includegraphics[width=0.49\textwidth]{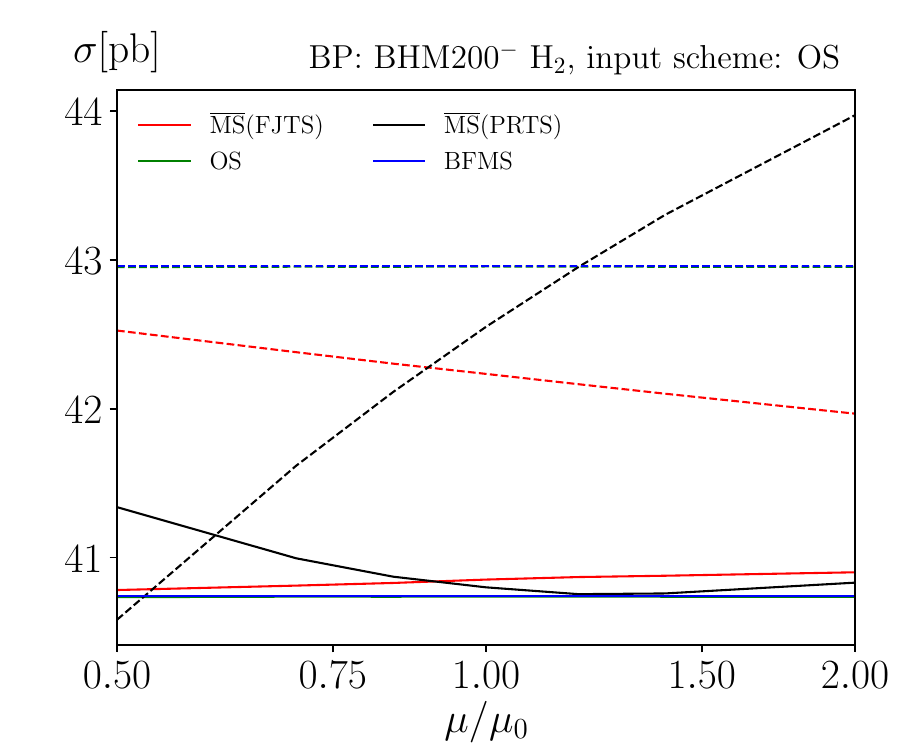}
}
\caption{Scale dependence for light-Higgs-boson production in
  Higgs-strahlung, $\Pp\Pp\to \PH_{2}\Pm^+\Pnm+X$, for the \HSESM\ 
  benchmark scenarios BHM200$^+$ (left) and BHM200$^-$ (right) in
  different renormalization schemes, with the \OS\ scheme as input
  scheme (and full conversion of the input parameters into the other
  schemes).  LO results are shown as dashed, NLO as full lines; the
  central scale is set to $\mu_0=M_{\PH_2}$.}
\label{fig:HS-HZ-H2-BHM200P-BHM200M}
\end{figure}

In \reftas{tab:SESM-HZ-Hl} and \ref{tab:SESM-HZ-Hh} we show the
results for $\Pp\Pp\to \PH\Pm^+ \Pnm+X$ for $\PH = \PHtwo$ and
$\PHone$, respectively.  For the light-Higgs-boson production in
\refta{tab:SESM-HZ-Hl} the LO scale dependence is small, ranging from
$\sim1.5\%$ for BHM400 and BHM600 to $\sim3\%$ for BHM200$^\pm$. The
scale uncertainty gets significantly reduced at NLO by a factor of
$4$--$5$.  The NLO scale uncertainty for all on-shell schemes and the
\MSbarPRTS{} are below $0.05\%$.  Overall, the observed EW corrections
are small, and central results are consistent within all schemes, with
a scheme dependence at the permille level.  A detailed scale
dependence of BHM200$^+$ and BHM200$^-$ is shown in
\reffi{fig:HS-HZ-H2-BHM200P-BHM200M}.

\begin{figure}
\centerline{
\includegraphics[width=0.49\textwidth]{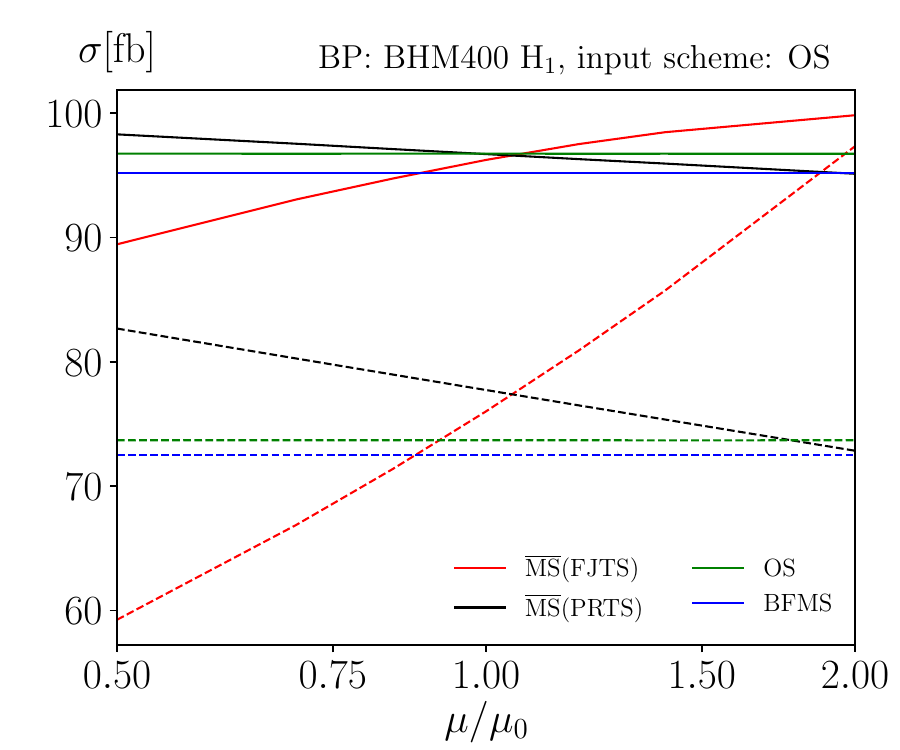}
\includegraphics[width=0.49\textwidth]{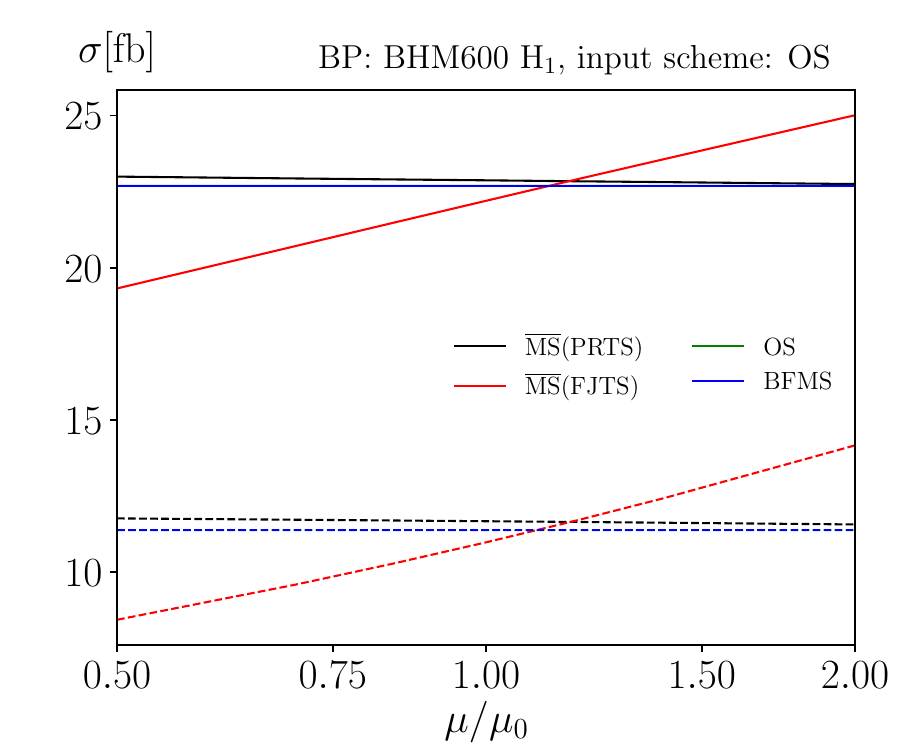}
}
\caption{As in \reffi{fig:HS-HZ-H2-BHM200P-BHM200M}, but for $\PHone$
production in 
the benchmark scenarios BHM400 (left) and BHM600 (right).}
\label{fig:HS-HZ-H1-BHM400-BHM600}
\end{figure}

For heavy-Higgs-boson production $\PHone$ in \refta{tab:SESM-HZ-Hh}
the picture is qualitatively the same with the difference being that
the size of the corrections and the scale uncertainties are amplified
with respect to light-Higgs-boson production.  However,
phenomenologically, the scenario corresponds to an entirely different
situation, since the heavy Higgs boson is only coupling weakly to the
vector bosons and loop contributions significantly contribute to the
production, resulting in large EW corrections. In
\reffi{fig:HS-HZ-H1-BHM400-BHM600} we depict the scale dependence for
BHM400 and BHM600. While for BHM400 the $\MSbar$ schemes show a
significant reduction in the scale uncertainty, for BHM600 the
absolute scale variation does not change from LO to NLO, though a
reduction of the relative scale uncertainty results from the increase
of the integrated cross section.
It is notable that even though the corrections reach up to $100\%$ all
schemes agree at the central scale within less than about 2\%.  We
conclude that also for the heavy-Higgs-boson production all schemes
are mutually consistent with no sign of artificially enhanced
corrections.

\begin{table}
\centerline{\renewcommand{\arraystretch}{1.25}
\begin{tabular}{|c||l|l|l|l|l|l|l|l|}
\hline
& \multicolumn{2}{c|}{BHM200$^+$} & \multicolumn{2}{c|}{BHM200$^-$} 
\\
Scheme 
  & \multicolumn{1}{c|}{LO} & \multicolumn{1}{c|}{NLO}  
  & \multicolumn{1}{c|}{LO} & \multicolumn{1}{c|}{NLO}  
\\
\hline
\hline
  \MSbarPRTS 
& $1039.9(4)^{-32.3\%}_{+44.4\%}$
& $915(1)^{-2.3\%}_{-11.8\%}$
& $1055.4(4)^{-32.8\%}_{+45.3\%}$
& $918(1)^{-2.1\%}_{-12.6\%}$
\\
\hline
  \MSbarFJTS
& $1161.3(5)^{+30.8\%}_{-23.6\%}$
& $893(2)^{-13.3\%}_{+2.6\%}$
& $1132.5(4)^{+5.8\%}_{-6.2\%}$
& $908(2)^{-1.1\%}_{+1.6\%}$
\\
\hline
  \OS 
& $957.8(4)$
& $926.6(7)^{-0.0\%}_{+0.0\%}$
& $957.9(4)$
& $928.8(7)^{+0.0\%}_{+0.0\%}$
\\
\hline
  \RSBFM
& $957.5(4)$
& $926.3(7)$
& $957.7(4)$
& $928.7(7)$
\\
\hline 
\multicolumn{5}{c}{}\\
\hline
& \multicolumn{2}{c|}{BHM400} & \multicolumn{2}{c|}{BHM600} 
\\
Scheme & \multicolumn{1}{c|}{LO} & \multicolumn{1}{c|}{NLO}  & \multicolumn{1}{c|}{LO} & \multicolumn{1}{c|}{NLO}  
\\
\hline
\hline
  \MSbarPRTS 
& $77.73(3)^{-6.3\%}_{+6.4\%}$
& $96.72(9)^{-1.6\%}_{+1.6\%}$
& $11.992(4)^{-2.3\%}_{+2.1\%}$
& $23.20(1)^{-1.3\%}_{+1.3\%}$
\\
\hline
\MSbarFJTS
& $76.01(3)^{+28.1\%}_{-22.1\%}$
& $96.25(8)^{+3.7\%}_{-7.1\%}$
& $10.586(3)^{+33.1\%}_{-25.2\%}$
& $21.769(9)^{+14.8\%}_{-14.8\%}$
\\
\hline
  \OS 
& $73.69(2)$
& $95.54(6)^{-0.0\%}_{+0.0\%}$
& $11.380(3)$
& $22.69(1)^{-0.0\%}_{+0.0\%}$
\\
\hline
  \RSBFM 
& $72.5\change{2}(2)$
& $95.1\change{8}(6)$
& $11.3\change{76}(3)$
& $22.\change{71}(1)$
\\
\hline
\end{tabular}
}
\caption{As in \refta{tab:SESM-HZ-Hl}, but for 
 the cross section  $\sigma$ [fb] of heavy Higgs-boson production in Higgs-strahlung, $\Pp\Pp\to \PH_{1}\Pm^+ \Pnm+X$, in the \HSESM.}
\label{tab:SESM-HZ-Hh}
\end{table}

\subsubsection{Higgs-strahlung in the \THDM}
\label{se:HZ_THDM}

\begin{figure}
\centerline{
\includegraphics[width=0.49\textwidth]{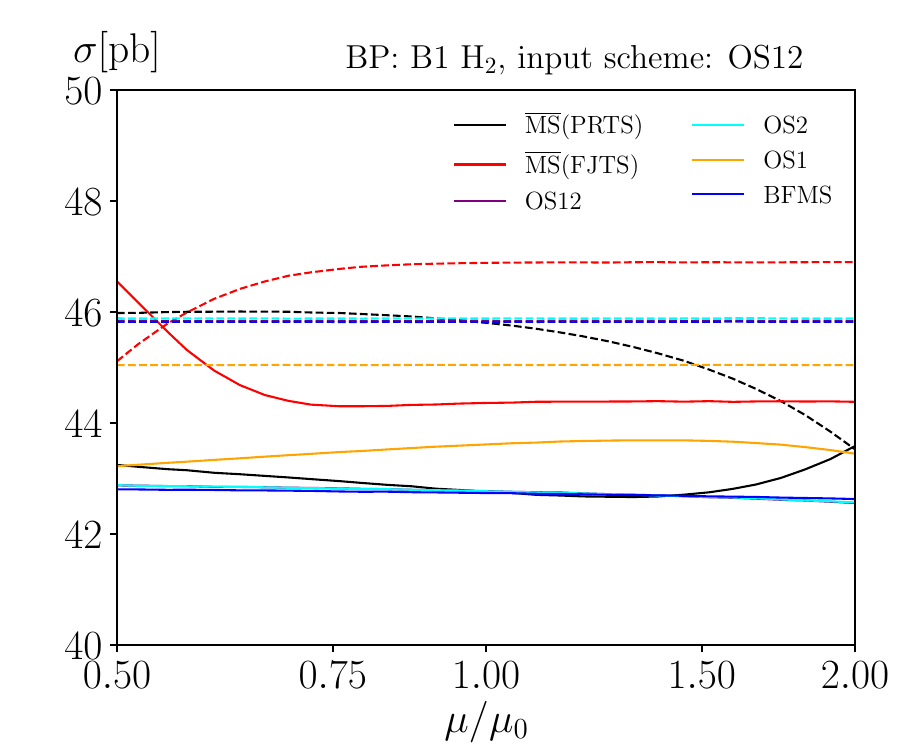}
\includegraphics[width=0.49\textwidth]{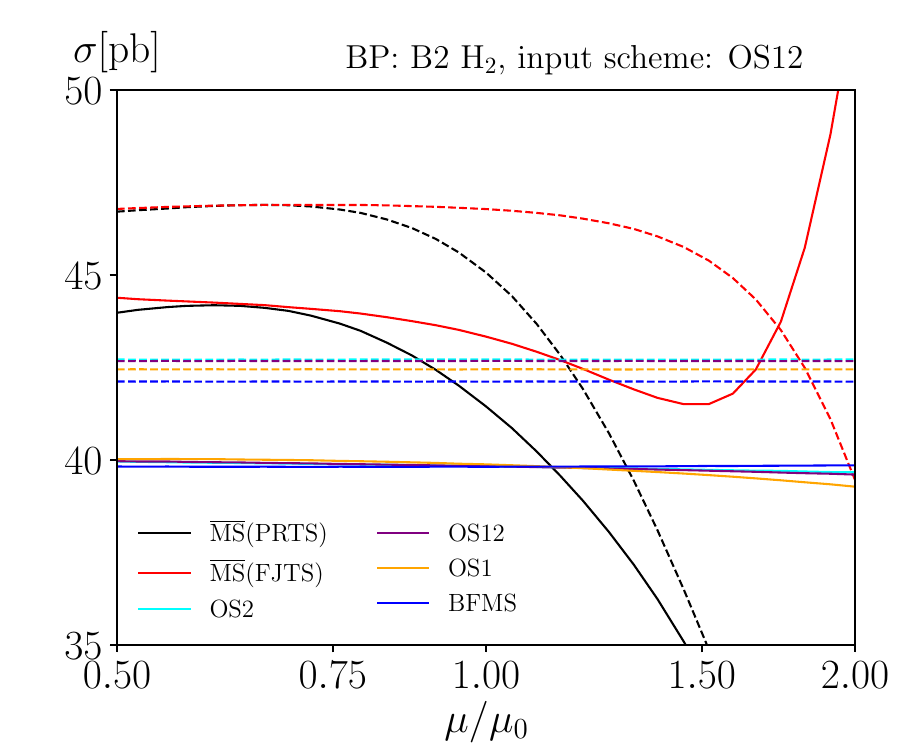}
}
\caption{Scale dependence for light-Higgs-boson production in
  Higgs-strahlung, $\Pp\Pp\to \PH_{2}\Pm^+ \Pnm+X$, for the \THDM\ 
  benchmark scenarios B1 (left) and B2 (right) in different
  renormalization schemes, with the \OSonetwo\ scheme as input scheme
  (and full conversion of the input parameters into the other
  schemes).  LO results are shown as dashed, NLO as full lines; the
  central scale is set to
  $\mu_0=(M_{\PH_2}+M_{\PH_1}+M_{\Ha}+2M_{\PH^+})/5$.}
\label{fig:THDM-HZ-H2-B1-B2}
\end{figure}

\begin{figure}
\centerline{
\includegraphics[width=0.49\textwidth]{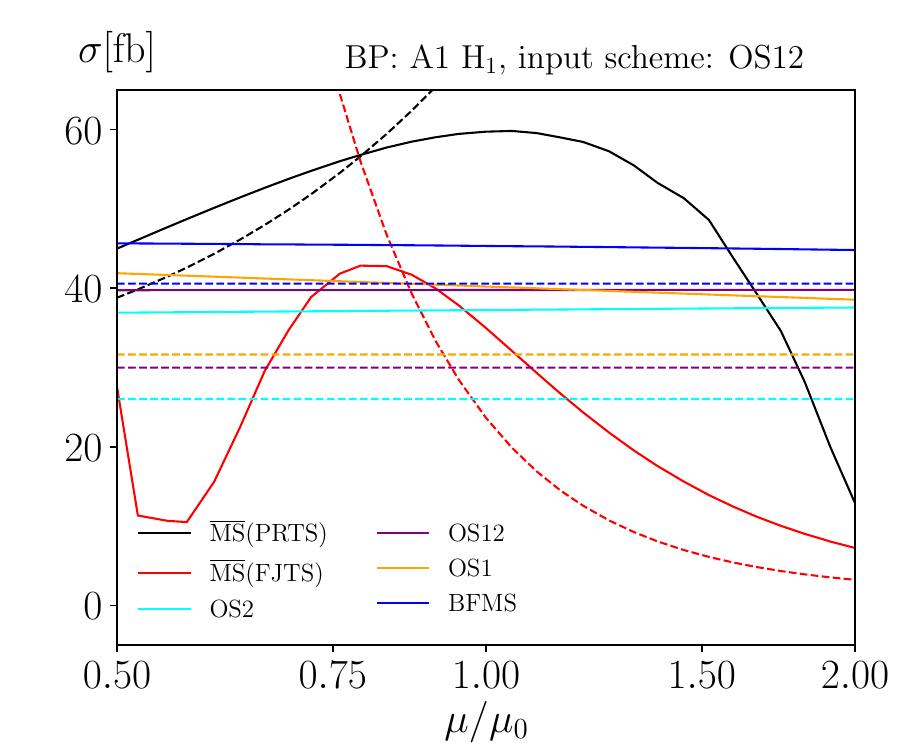}
\includegraphics[width=0.49\textwidth]{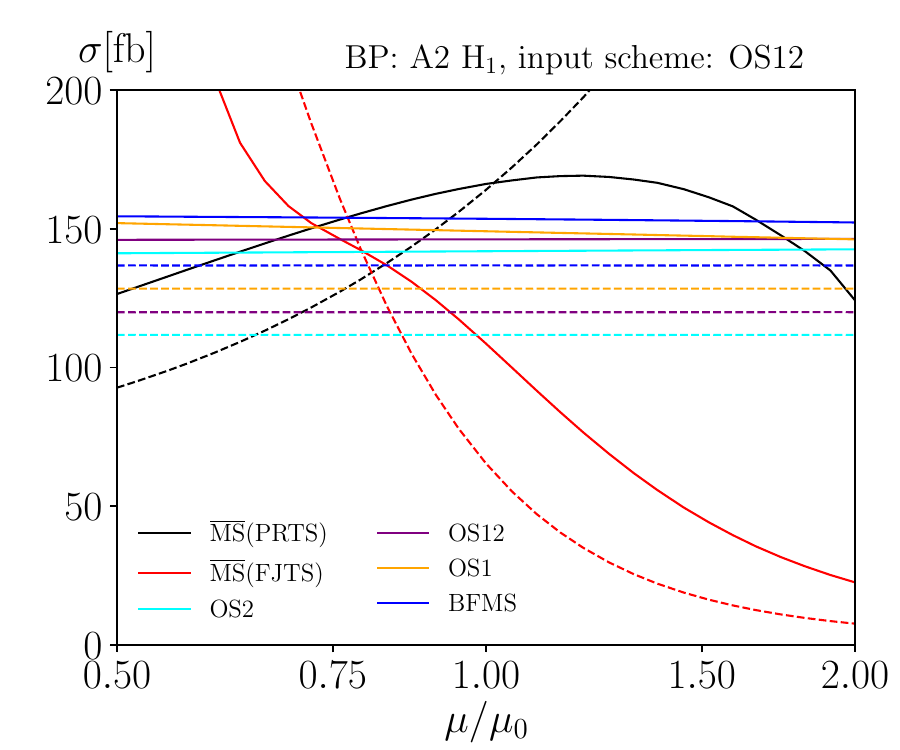}
}
\caption{Scale dependence for heavy-Higgs-boson production in
  Higgs-strahlung, $\Pp\Pp\to \PH_{1}\Pm^+ \Pnm+X$, for the \THDM\ 
  benchmark scenarios A1 (left) and A2 (right) in different
  renormalization schemes, with the \OSonetwo\ scheme as input scheme
  (and full conversion of the input parameters into the other
  schemes).  LO results are shown as dashed, NLO as full lines; the
  central scale is set to
  $\mu_0=(M_{\PH_2}+M_{\PH_1}+M_{\Ha}+2M_{\PH^+})/5$.}
\label{fig:THDM-HZ-H1-A1-A2}
\end{figure}

The results for Higgs production in Higgs-strahlung in the \THDM{}
are shown in \reftas{tab:THDM-HZ-Hl-OS12} and
\ref{tab:THDM-HZ-Hh-OS12} for the light (\PHtwo) and heavy (\PHone)
Higgs-boson production, respectively, with the \OSonetwo as input scheme.
\begin{table}
\centerline{\renewcommand{\arraystretch}{1.25}
\begin{tabular}{|c||l|l|l|l|l|l|l|l|}
\hline
& \multicolumn{2}{c|}{A1} & \multicolumn{2}{c|}{A2} 
\\
Scheme 
  & \multicolumn{1}{c|}{LO} & \multicolumn{1}{c|}{NLO}  
  & \multicolumn{1}{c|}{LO} & \multicolumn{1}{c|}{NLO}  
\\
\hline
\hline
  \MSbarPRTS 
& $45.76(2)^{-3.4\%}_{+1.2\%}$
& $43.67(2)^{+2.1\%}_{+0.4\%}$
& $44.33(2)^{-6.4\%}_{+2.5\%}$
& $42.25(2)^{+2.5\%}_{+1.2\%}$
\\
\hline
  \MSbarFJTS
& $46.53(2)^{+0.7\%}_{-7.2\%}$
& $43.93(3)^{+0.7\%}_{+1.4\%}$
& $45.87(2)^{+2.0\%}_{-10.2\%}$
& $42.90(3)^{+2.6\%}_{-9.2\%}$
\\
\hline
  \OSone
& $46.40(2)$
& $43.88(3)^{-0.1\%}_{+0.0\%}$
& $44.88(2)$
& $42.43(3)^{-0.0\%}_{+0.0\%}$
\\
\hline
  \OStwo
& $46.49(2)$
& $43.90(3)^{-0.1\%}_{+0.1\%}$
& $45.15(2)$
& $42.49(3)^{-0.2\%}_{+0.1\%}$
\\
\hline
  \OSonetwo
& $46.43(2)$
& $43.87(3)^{-0.1\%}_{+0.1\%}$
& $45.02(2)$
& $42.44(3)^{-0.2\%}_{+0.1\%}$
\\
\hline
  \RSBFM
& $\change{46.26}(2)$
& $\change{46.82}(3)^{-0.1\%}_{+0.1\%}$
& $\change{46.76}(2)$
& $\change{46.38}(3)^{-0.1\%}_{+0.1\%}$
\\
\hline
\multicolumn{5}{c}{}\\
\hline
& \multicolumn{2}{c|}{B1} & \multicolumn{2}{c|}{B2} 
\\
Scheme & \multicolumn{1}{c|}{LO} & \multicolumn{1}{c|}{NLO}  & \multicolumn{1}{c|}{LO} & \multicolumn{1}{c|}{NLO}  
\\
\hline
\hline
  \MSbarPRTS 
& $45.81(2)^{-5.0\%}_{+0.4\%}$
& $42.76(3)^{+1.9\%}_{+1.1\%}$
& $45.07(2)^{-44.5\%}_{+3.6\%}$
& $41.45(3)^{-42.1\%}_{+6.1\%}$
\\
\hline
\MSbarFJTS
& $46.89(2)^{+0.0\%}_{-3.8\%}$
& $44.36(3)^{+0.0\%}_{+4.9\%}$
& $46.78(2)^{-15.6\%}_{+0.0\%}$
& $43.33(3)^{+21.4\%}_{+2.4\%}$
\\
\hline
  \OSone
& $45.04(2)$
& $43.61(2)^{-0.4\%}_{-0.9\%}$
& $42.45(2)$
& $39.88(3)^{-1.5\%}_{+0.3\%}$
\\
\hline
  \OStwo
& $45.88(2)$
& $42.77(3)^{-0.5\%}_{+0.3\%}$
& $42.72(2)$
& $39.83(3)^{-0.4\%}_{+0.3\%}$
\\
\hline
  \OSonetwo
& $45.84(2)$
& $42.77(3)^{-0.5\%}_{+0.2\%}$
& $42.68(2)$
& $39.83(3)^{-0.6\%}_{+0.3\%}$
\\
\hline
  \RSBFM 
  & $\change{45.82}(2)$
  & $\change{42.74}(3)^{-\change{0.3}\%}_{+\change{0.2}\%}$
  & $\change{42.12}(2)$
  & $\change{39.82}(2)^{+0.1\%}_{+0.0\%}$
\\
\hline
\end{tabular}
}
\caption{LO and NLO integrated cross sections $\sigma$ [pb]
for the production of the light Higgs boson in Higgs-strahlung,
$\Pp\Pp\to\PH_2\Pm^+ \Pnm+X$, for various 
\THDM\ scenarios in different renormalization schemes,
with the \OSonetwo\ scheme as input scheme (and full conversion
of the input parameters into the other schemes).
The scale variation (given in percent) corresponds to 
the scales $\mu=\mu_0/2$ and $\mu=2\mu_0$ with central
scale $\mu_0=(M_{\PH_2}+M_{\PH_1}+M_{\Ha}+2M_{\PH^+})/5$.}
\label{tab:THDM-HZ-Hl-OS12}
\end{table}
\begin{table}
\centerline{\renewcommand{\arraystretch}{1.25}
\begin{tabular}{|c||l|l|l|l|l|l|l|l|}
\hline
& \multicolumn{2}{c|}{A1} & \multicolumn{2}{c|}{A2} 
\\
Scheme 
  & \multicolumn{1}{c|}{LO} & \multicolumn{1}{c|}{NLO}  
  & \multicolumn{1}{c|}{LO} & \multicolumn{1}{c|}{NLO}  
\\
\hline
\hline
  \MSbarPRTS 
& $72.47(3)^{>+100\%}_{-46.5\%}$
& $59.7(2)^{-79.4\%}_{-24.7\%}$
& $164.02(6)^{>+100\%}_{-43.5\%}$
& $166.2(2)^{-25.4\%}_{-23.9\%}$
\\
\hline
  \MSbarFJTS
& $23.647(9)^{-86.4\%}_{>+100\%}$
& $34.98(3)^{-79.3\%}_{-27.0\%}$
& $65.42(2)^{-88.3\%}_{>+100\%}$
& $108.6(1)^{-79.2\%}_{>+100\%}$
\\
\hline
  \OSone
& $31.64(1)$
& $40.21(2)^{-4.1\%}_{+4.2\%}$
& $128.41(5)$
& $149.10(8)^{-2.0\%}_{+2.0\%}$
\\
\hline
  \OStwo
& $26.03(1)$
& $37.25(3)^{+0.8\%}_{-0.9\%}$
& $111.73(4)$
& $141.89(5)^{+0.5\%}_{-0.5\%}$
\\
\hline
  \OSonetwo
& $29.98(1)$
& $39.78(2)^{+0.0\%}_{-0.1\%}$
& $119.92(4)$
& $146.18(5)^{+0.1\%}_{-0.1\%}$
\\
\hline
  \RSBFM
& $\change{40.58}(1)$
& $\change{45.33}(4)^{-\change{1.1}\%}_{+\change{0.7}\%}$
& $\change{136.78}(5)$
& $\change{153.6}(1)^{-\change{0.9}\%}_{+\change{0.6}\%}$
\\
\hline
\multicolumn{5}{c}{}\\
\hline
& \multicolumn{2}{c|}{B1} & \multicolumn{2}{c|}{B2} 
\\
Scheme & \multicolumn{1}{c|}{LO} & \multicolumn{1}{c|}{NLO}  & \multicolumn{1}{c|}{LO} & \multicolumn{1}{c|}{NLO}  
\\
\hline
\hline
  \MSbarPRTS 
& $5.458(2)^{>+100\%}_{-15.5\%}$
& $11.189(5)^{+20.8\%}_{-22.0\%}$
& $444.2(2)^{>+100\%}_{-90.1\%}$
& $624.4(7)^{>+100\%}_{-92.2\%}$
\\
\hline
\MSbarFJTS
& $0.05296(2)^{-99.8\%}_{>+100\%}$
& $0.1243(1)^{-96.9\%}_{<-100\%}$
& $27.85(1)^{>+100\%}_{-1.3\%}$
& $161.7(7)^{<-100\%}_{<-100\%}$
\\
\hline
  \OSone
& $9.316(5)$   
& $9.20(6)^{-7.7\%}_{+30.2\%}$
& $1079.8(4)$
& $1015.6(9)^{+6.7\%}_{+1.4\%}$
\\
\hline
  \OStwo
& $5.076(2)$
& $10.954(4)^{-0.1\%}_{-0.4\%}$
& $1015.7(4)$
& $1022.6(5)^{-3.9\%}_{+2.1\%}$
\\
\hline
  \OSonetwo
& $5.291(2)$
& $11.051(4)^{+0.0\%}_{-0.0\%}$
& $1025.1(4)$
& $1023.6(5)^{-2.5\%}_{+1.8\%}$
\\
\hline
  \RSBFM 
& $\change{5.386(2)}$
& $\change{11.303}(4)^{-\change{4.4}\%}_{+\change{1.5}\%}$
& $\change{1159.8(5)}$
& $\change{1032(1)}^{-\change{8.4}\%}_{+\change{4.9}\%}$
\\
\hline
\end{tabular}
}
\caption{As in \refta{tab:THDM-HZ-Hl-OS12}, but for 
 the cross section  $\sigma$ [fb] of heavy Higgs-boson production in
 Higgs-strahlung, $\Pp\Pp\to \PH_{1}\Pm^+ \Pnm+X$, in the \THDM.} 
\label{tab:THDM-HZ-Hh-OS12}
\end{table}
Figures \ref{fig:THDM-HZ-H2-B1-B2} and
\ref{fig:THDM-HZ-H1-A1-A2} show the scale dependence for  $\Pp\Pp\to
\PH_{1,2}\Pm^+ \Pnm+X$ for different renormalization schemes.

As illustrated in \reffi{fig:THDM-HZ-H2-B1-B2}, the on-shell schemes
yield stable results for light-Higgs-boson production.
Table~\ref{tab:THDM-HZ-Hl-OS12} shows that neglecting B2 for the
moment and comparing to Higgs-strahlung in the \HSESM, the reduction
of the scale uncertainty is either less strong (e.g.~for A1, A2, B1 in
\MSbarPRTS, and A1 in \MSbarFJTS) or not observed (A2, B1 in
\MSbarFJTS{}).  Nonetheless, the results for the benchmark scenarios
A1, A2, and B1 visibly agree in all schemes within 2\% at NLO, apart
from the \MSbarFJTS scheme for B1.  For B2 the $\MSbar$ schemes show
large scale uncertainties, and no stabilization of the results is
visible when going from LO to NLO.  The central values for the
$\MSbar$ schemes differ by up to $9\%$ from the results for the
on-shell schemes, while the latter mutually agree at the permille
level.

As can be seen in \refta{tab:THDM-HZ-Hh-OS12}, for the
heavy-Higgs-boson production in general the scale uncertainty within
the traditional $[\mu_0/2,2\mu_0]$ window remains large within the
$\MSbar$ schemes.  Nonetheless, extrema or at least regions of smaller
scale dependence show up for the benchmark scenarios A1 and A2 (\cf
\reffi{fig:THDM-HZ-H1-A1-A2}) which can be viewed as narrow plateaus.

Focusing on scenario A1 the scale uncertainty is still large at NLO in the
$\MSbar$ schemes, while the results in the on-shell schemes are well
consistent and their spread decreases at NLO. Yet uncertainties of 4\%
are visible for \OSone{} which are not unexpected due to the genuinely
large corrections for heavy-Higgs-boson production in all on-shell
schemes.  The large corrections are due to the proximity of the
scenario A1 to the alignment limit and also cause large conversion
effects between the on-shell or BFM and $\MSbar$ schemes
(see \refta{tab:conversion-THDM-OS12} in \refapp{se:conversiontables}).
For scenario A2, which is further away from the alignment limit, the picture is
slightly better.

For scenario B1 the conversion effects between the schemes \MSbarPRTS,
\OStwo, \OSonetwo, and \RSBFM{} are small (see
\refta{tab:conversion-THDM-OS12}). This is reflected in a LO
prediction of similar size and a surprisingly good agreement at NLO
between those schemes, even though the corrections exceed $100\%$ in
heavy-Higgs-boson production. The \MSbarFJTS{} scheme fails to give
any reasonable result in heavy-Higgs-boson production due to the large
conversion effects pushing into the alignment limit $\cab\approx0$.
The large conversion effects observed for the \OSone{} scheme can be
traced back to the renormalization of $\beta$, and replacing
\eqref{eq:db_nu1_ons} with \eg \eqref{eq:db_nu2_ons}, but keeping the
renormalization of $\alpha$ fixed, results in small conversion
effects.  While, we could not completely disentangle the reason for
the large effects in \eqref{eq:db_nu1_ons}, this is a least partly
caused by the enhancement proportional to $\tb$ in
\eqref{eq:db_nu1_ons} which is not present in the \OStwo{}
\eqref{eq:db_nu2_ons}, \OSonetwo{} \eqref{eq:db_os12_thdm}, and \RSBFM
schemes.  While the \OStwo{} scheme might still be affected by similar
problems in certain parameter regions, the scheme \OSonetwo{} is free
of such {\it artificial} enhancements and therefore preferable.  The
seemingly fair agreement between the \OSone{} and the other on-shell
and BFM schemes at NLO is just accidental, since for the Higgs decay
(\refta{tab:THDM-H14f}) no good agreement is found.  We conclude that
the schemes \OStwo, \OSonetwo, \RSBFM{}, and \MSbarPRTS{} give
consistent predictions even though the $K$ factor is about $2$.

For scenario B2, conversion effects between $\MSbar$ and on-shell or
BFM schemes (see \refta{tab:conversion-THDM-OS12}) are sizeable, while
within the on-shell and BFM schemes these are small. For the on-shell
and BFM schemes, even for heavy-Higgs-boson production the corrections
are well behaved, 
\change{ranging between $-11\%$ and $+1\%$}.  
The $\MSbar$
schemes suffer from very large scale uncertainties and large
corrections at the central scale.

\subsubsection{Higgs production via vector-boson fusion in the \THDM}
\label{se:VBF_THDM}

\begin{figure}
\centerline{
\includegraphics[width=0.49\textwidth]{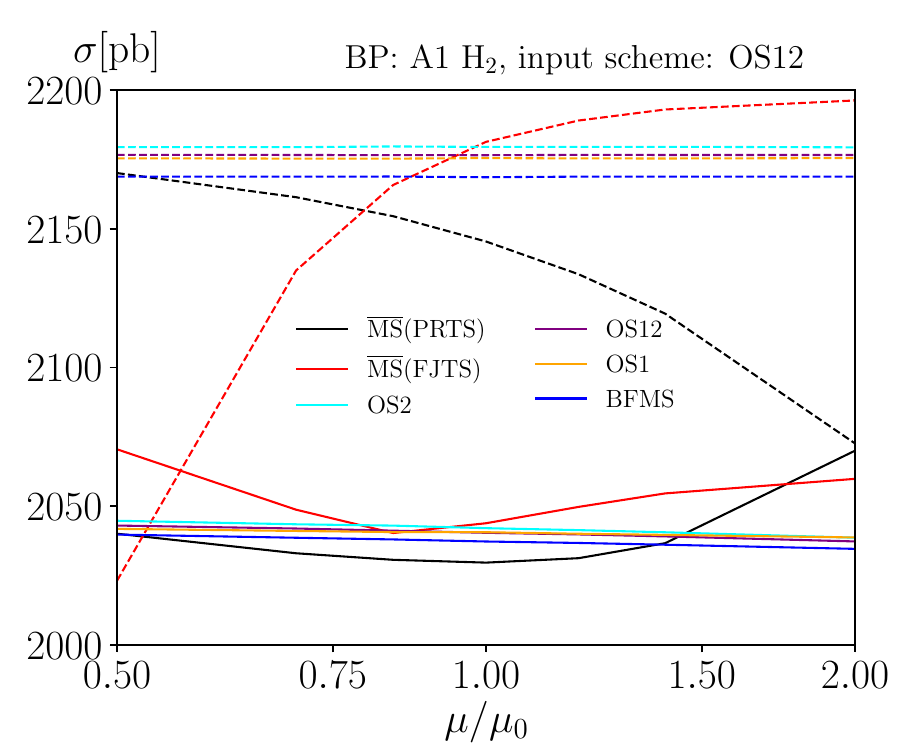}
\includegraphics[width=0.49\textwidth]{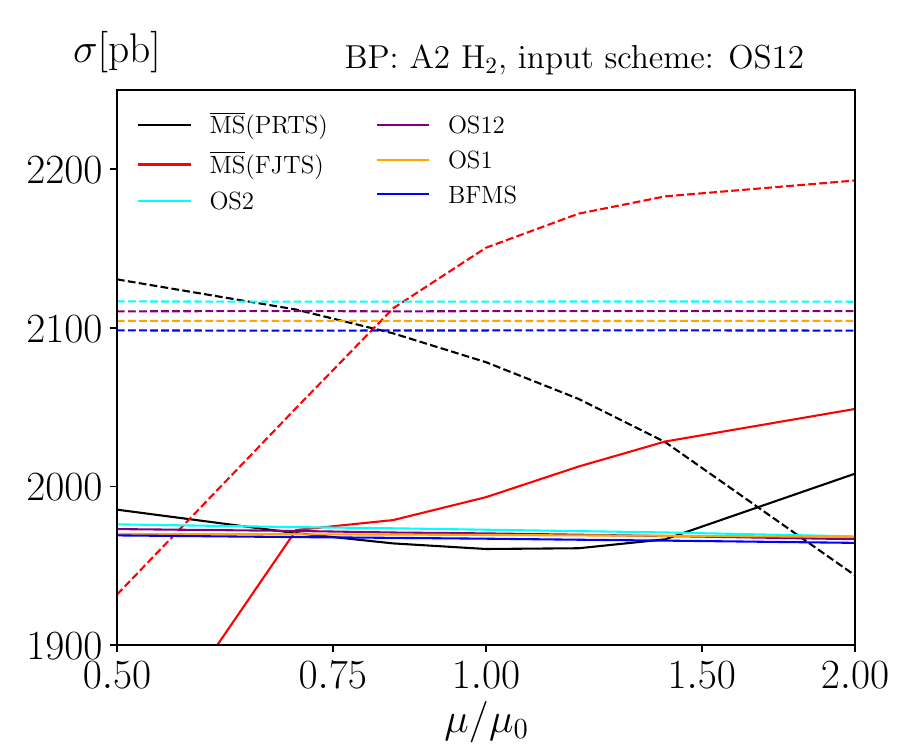}
}
\caption{Scale dependence for light-Higgs-boson production in
  VBF, $\Pp\Pp\to \PH_{2}\Pj\Pj+X$, for the \THDM\ benchmark scenarios
  A1 (left) and A2 (right) in different renormalization schemes, with
  the \OSonetwo\ scheme as input scheme (and full conversion of the
  input parameters into the other schemes).  LO results are shown as
  dashed, NLO as full lines; the central scale is set to
  $\mu_0=(M_{\PH_2}+M_{\PH_1}+M_{\Ha}+2M_{\PH^+})/5$.}
\label{fig:THDM-VBF-H2-A1-A2}
\end{figure}

For Higgs production via VBF we provide only some exemplary results.
We show the scale dependence for light-Higgs-boson production in A1
and A2 in \reffi{fig:THDM-VBF-H2-A1-A2}, and summarize the usual scale
variation in \refta{tab:THDM-HZ-Hl-VBF-OS12}.  Since the behaviour of
these results resembles closely the one for Higgs-strahlung, apart
from the magnitude of the cross sections, we did not investigate any
other benchmark scenarios. We plan to provide more detailed
phenomenological results on VBF elsewhere.
\begin{table}
\centerline{\renewcommand{\arraystretch}{1.25}
\begin{tabular}{|c||l|l|l|l|l|l|l|l|}
\hline
& \multicolumn{2}{c|}{A1} & \multicolumn{2}{c|}{A2} 
\\
Scheme 
  & \multicolumn{1}{c|}{LO} & \multicolumn{1}{c|}{NLO}  
  & \multicolumn{1}{c|}{LO} & \multicolumn{1}{c|}{NLO}  
\\
\hline
\hline
  \MSbarPRTS 
& $2145.4(6)^{-3.4\%}_{+1.1\%}$
& $2029.7(8)^{+2.0\%}_{+0.5\%}$
& $2078.3(6)^{-6.5\%}_{+2.5\%}$
& $1960.5(8)^{+2.4\%}_{+1.3\%}$
\\
\hline
  \MSbarFJTS
& $2181.3(8)^{+0.7\%}_{-7.3\%}$
& $2043.9(9)^{+0.8\%}_{+1.3\%}$
& $2150.5(6)^{+2.0\%}_{-10.2\%}$
& $1993.2(9)^{+2.8\%}_{-9.3\%}$
\\
\hline
  \OSone
& $2175.5(8)$
& $2040.6(9)^{-0.1\%}_{+0.1\%}$
& $2104.4(6)$
& $1969.4(8)^{-0.1\%}_{+0.0\%}$
\\
\hline
  \OStwo
& $2179.5(8)$
& $2042.1(9)^{-0.2\%}_{+0.1\%}$
& $2116.5(6)$
& $1972.7(8)^{-0.2\%}_{+0.2\%}$
\\
\hline
  \OSonetwo
& $2176.5(8)$
& $2040.4(9)^{-0.2\%}_{+0.1\%}$
& $2110.5(6)$
& $1970.4(8)^{-0.2\%}_{+0.1\%}$
\\
\hline
  \RSBFM
& $\change{2168.6}(8)$
& $\change{2037.3(8)}^{-0.1\%}_{+0.1\%}$
& $\change{2098.4}(6)$
& $\change{1967.0}(8)^{-0.1\%}_{+0.1\%}$
\\
\hline
\end{tabular}
}
\caption{LO and NLO integrated cross sections $\sigma$ [fb]
for the production of the light Higgs boson in VBF, $\Pp\Pp\to
\PH_{2}\Pj\Pj+X$, for the 
\THDM\ scenarios A1 and A2 in different renormalization schemes,
with the \OSonetwo\ scheme as input scheme (and full conversion
of the input parameters into the other schemes).
The scale variation (given in percent) corresponds to 
the scales $\mu=\mu_0/2$ and $\mu=2\mu_0$ with central
scale $\mu_0=(M_{\PH_2}+M_{\PH_1}+M_{\Ha}+2M_{\PH^+})/5$.}
\label{tab:THDM-HZ-Hl-VBF-OS12}
\end{table}

\vspace*{4ex}

\section{Conclusions}
\label{se:Conclusions}

Models with extended Higgs sectors are of prime importance for
investigating the mechanism of electroweak symmetry breaking.
Precision investigations of these models require the inclusion of
higher-order corrections at least at NLO and thus renormalization.  In
particular, prescriptions for the renormalization of mixing angles in
the scalar sector are needed.

In this paper we have discussed a variety of renormalization
prescriptions for scalar mixing angles with particular regard to
symmetry, gauge independence, and numerical stability.  
In detail, we have considered and compared three types of
renormalization schemes for mixing angles in the Higgs sector:
\begin{itemize}
\item $\MSbar$ renormalization conditions for mixing angles are easy
  to implement. They depend, however, on the treatment of tadpoles and
  require care in view of gauge dependence,
  but have the benefit that the size of missing higher-order corrections 
  can be investigated by renormalization scale variation.
\item  We have formulated on-shell renormalization conditions for
the mixing angles in the Two-Higgs-Doublet Model and the Higgs-Singlet
Extension of the Standard Model based on combinations of  physical observables.
More precisely, ratios of matrix elements or formfactors depending on the 
desired mixing angles only, deliver appropriate renormalization conditions.
To obtain such ratios, it is often useful to introduce spurious particles
with infinitesimal couplings, which do not change the physical theory.
\item Rigid gauge invariance and/or the background-field method allow
  to introduce renormalization conditions for mixing angles in general
  theories.
\end{itemize}

We numerically studied and compared various renormalization conditions
for mixing angles in the Two-Higgs-Doublet Model and the Higgs-Singlet
Extension of the Standard Model for Higgs decays into four fermions
and for Higgs-production in association with a vector boson or in
vector-boson fusion. While renormalization schemes based on $\MSbar$
subtraction tend to become unstable in delicate scenarios, in
particular for heavy Higgs-boson production in the Two-Higgs-Doublet
Model, the proposed on-shell schemes and the schemes motivated by the
background field method behave decently. They do not allow for a
realistic estimate of missing higher-order corrections via scale
variation, but instead the renormalization-scheme dependence can be
investigated by comparing results obtained with different schemes
after a consistent conversion of input parameters between the schemes.
In general, one should avoid renormalization
schemes that introduce artificial enhancement factors, \eg for
degenerate masses or small mixing angles.

Based on our study, we propose to use on-shell or symmetry-based
schemes for the central predictions, as these turn out to be more
robust. The reliability of the results can be checked by using two or
more different on-shell schemes, where already the consistent
parameter conversion between different schemes provides uncertainty
estimates. Finally, schemes based on $\MSbar$ renormalization can be
used, with due care, to study scale uncertainties.

\change{The schemes introduced in this paper are based on genuine
  on-shell conditions or simple symmetry principles. Thus, their
  generalization to higher orders is well-defined (similar to the \MSbar
  scheme), even if the explicit results for the counterterms might
  become more involved owing to reducible contributions. This is in
  contrast to some of the schemes proposed in the literature that rely
  on the form of the one-loop expressions.}

The proposed schemes are not restricted to the considered models, but
can be generalized to other extensions of the Standard Model,
involving additional scalars, vector bosons, or fermions.

\subsection*{Acknowledgements}

Heidi Rzehak is gratefully acknowledged for useful discussions.
The work of A.D.\ 
is supported by the German Science Foundation (DFG)
under reference number DE~623/5-1,
the work of S.D.\ by the DFG project DI~784/4-1.
S.D.\ furthermore acknowledges 
support by the state of 
Baden-W\"urttemberg through bwHPC and the DFG through 
grant no INST 39/963-1 FUGG.
J.-N.~Lang acknowledges support
from the Swiss National Science Foundation (SNF) under contract
BSCGI0-157722. 
The authors would like to express special thanks to the
Mainz Institute for Theoretical Physics (MITP) for its hospitality and
support.

\clearpage

\appendix
\section*{Appendix}

\section{Translation of conventions for the \HSESM}
\label{app:conventions}

In order to treat the mixing between two Higgs bosons in a generic
way, we deviate somewhat in our notation from the recent literature
on the \HSESM.
Here, we collect translation rules for the fields and parameters from the \HSESM\
formulation of previous works \cite{Denner:2017vms,Altenkamp:2018bcs}:
\beq
\begin{array}{lclcl}
\si &=& -S^{\mbox{\scriptsize\cite{Denner:2017vms}}}\sqrt{2} &=& \sigma^{\mbox{\scriptsize\cite{Altenkamp:2018bcs}}},
\\
\eta_1 &=& -\rho_2^{\mbox{\scriptsize\cite{Denner:2017vms}}} &=& h_1^{\mbox{\scriptsize\cite{Altenkamp:2018bcs}}},
\\
\eta_2 &=& \rho_1^{\mbox{\scriptsize\cite{Denner:2017vms}}} &=& h_2^{\mbox{\scriptsize\cite{Altenkamp:2018bcs}}},
\\
v_1 &=& -v_s^{\mbox{\scriptsize\cite{Denner:2017vms}}} &=& v_1^{\mbox{\scriptsize\cite{Altenkamp:2018bcs}}},
\\
v_2 &=& v^{\mbox{\scriptsize\cite{Denner:2017vms}}} &=& v_2^{\mbox{\scriptsize\cite{Altenkamp:2018bcs}}},
\\
H_1 &=& H_{\mathrm{h}}^{\mbox{\scriptsize\cite{Denner:2017vms}}} &=& H^{\mbox{\scriptsize\cite{Altenkamp:2018bcs}}},
\\
H_2 &=& H_{\mathrm{l}}^{\mbox{\scriptsize\cite{Denner:2017vms}}} &=& h^{\mbox{\scriptsize\cite{Altenkamp:2018bcs}}},
\\
\alpha &=& \alpha^{\mbox{\scriptsize\cite{Denner:2017vms}}}+\frac{\pi}{2} &=& \alpha^{\mbox{\scriptsize\cite{Altenkamp:2018bcs}}},
\\
c_\alpha &=& -s_\alpha^{\mbox{\scriptsize\cite{Denner:2017vms}}} &=& c_\alpha^{\mbox{\scriptsize\cite{Altenkamp:2018bcs}}},
\\
s_\alpha &=& +c_\alpha^{\mbox{\scriptsize\cite{Denner:2017vms}}} &=& s_\alpha^{\mbox{\scriptsize\cite{Altenkamp:2018bcs}}},
\\
\mu_2^2 &=& -m_1^2{}^{\mbox{\scriptsize\cite{Denner:2017vms}}} &=& \mu_2^2{}^{\mbox{\scriptsize\cite{Altenkamp:2018bcs}}},
\\
\mu_1^2 &=& -m_2^2{}^{\mbox{\scriptsize\cite{Denner:2017vms}}} &=& 2\mu_1^2{}^{\mbox{\scriptsize\cite{Altenkamp:2018bcs}}},
\\
\lambda_2 &=& 2\lambda_1^{\mbox{\scriptsize\cite{Denner:2017vms}}} &=& \lambda_2^{\mbox{\scriptsize\cite{Altenkamp:2018bcs}}},
\\
\lambda_1 &=& 2\lambda_2^{\mbox{\scriptsize\cite{Denner:2017vms}}} &=& 16\lambda_1^{\mbox{\scriptsize\cite{Altenkamp:2018bcs}}},
\\
\lambda_{3} &=& \lambda_3^{\mbox{\scriptsize\cite{Denner:2017vms}}} &=& 2\lambda_{12}^{\mbox{\scriptsize\cite{Altenkamp:2018bcs}}}.
\end{array}
\eeq

\section{Translation of self-energies to the background-field method}
\label{app:bfm_se}

In \refse{se:RSsymm} we have 
derived renormalization conditions based on the background-field method
which requires the evaluation of mixing and self-energies in this
framework. Since the expressions differ from those in the conventional
formalism, we provide the differences relevant for the renormalization
of the mixing angles in the \HSESM and the \THDM using the following
notation
\beq\label{eq:deltaSigma}
\Delta\Sigma^{XY}(p^2) = \Sigma^{\hat{X}\hat{Y}}_{\mathrm{BFM}}(p^2) -
\Sigma^{XY}_{\mathrm{conv.}}(p^2).
\eeq
Some corresponding results in the SM can be found in \citere{Denner:1994nh}.

The one-loop tadpoles (one-point functions) do not differ between the
BFM and the conventional formalism, although the individual tadpole
contributions are not the same diagram by diagram.  As a consequence
also all tadpole contributions to the self-energies are the same in
the BFM and the conventional formalism, and the differences reported
in the following only result from 1PI contributions.

In order to use the mixing-angle renormalization based on the BFM,
only the {\it parameter} coun\-ter\-terms have to be calculated in the
BFM while the loop diagrams and the wave-function renormalization
constants can be calculated in the conventional formalism (owing to
gauge independence of the sum of a bare amplitude and its
corresponding wave-function counterterms).  As far as parameter
counterterms are determined from on-shell field renormalization
constants, the latter have to be calculated in the BFM as well.  To
calculate the counterterms in the BFM formalism it is sufficient to
combine the results for the differences \refeq{eq:deltaSigma} given in
this appendix with the conventional self-energies.

In order to distinguish the electromagnetic coupling $\alem$ from the
mixing angle $\al$, we label it with the index ``em''.

\subsection{Higgs sector of the \HSESM}

In the \HSESM the differences for the self-energies relevant for the
renormalization of the mixing angle $\alpha$ read for $H_{i,j}\in\{H_1,H_2\}$

\begin{align}
\Delta\Sigma^{H_iH_j}(p^2) ={}&
 \frac{\alem}{16\pi\cw^2\sw^2} c^{H_iH_j}(s)
      \left[ 2\cw^2 B_0(p^2,\MW,\MW)+B_0(p^2,\MZ,\MZ)\right],
\end{align}
with
\begin{align}
c^{H_1H_1}(s) ={}& 2\sa^2(\MHsone-p^2),\notag\\
c^{H_1H_2}(s) ={}& \ca\sa(\MHsone+\MHstwo-2p^2),\notag\\
c^{H_2H_2}(s) ={}& 2\ca^2(\MHstwo-p^2).
\end{align}

For the pseudo-scalar and charged sector we have
\begin{align}
\Delta\Sigma^{G_0 G_0}(p^2) ={}&
  -\frac{\alem}{8\pi\cw^2\sw^2} p^2
  \biggl[
  2\cw^2 B_0 (p^2,\MW,\MW)
  \notag\\
  &+{\ca^2} B_0 (p^2,\MZ,\MHtwo)
  +
  {\sa^2}B_0 (p^2,\MZ,\MHone)
  \biggr]
  , \\
\Delta\Sigma^{G^\pm G^\mp}(p^2) ={}&
  -\frac{\alem}{8\pi\sw^2}p^2
  \biggl[
    4 \sw^2 B_0 (p^2,0,\MW) +
  \frac{4 \cw^4 - 3 \cw^2 + 1}{\cw^2} B_0(p^2,\MZ,\MW)\notag\\
  &+ {\ca^2} B_0(p^2,\MW,\MHtwo)
  + {\sa^2} B_0(p^2,\MW,\MHone)
  \biggr].
\end{align}

\subsection{Higgs sector of \THDM}

In the \THDM, the differences read for $H_{i,j}\in\{H_1,H_2\}$
\begin{align}
\Delta\Sigma^{H_1H_1}(p^2) ={}&
\frac{\alem}{8\pi\cw^2\sw^2} 
  (\MHsone-p^2)
      \Bigl[ 
  \cab^2 B_0(p^2,\MZ,\MZ) + \sab^2 B_0(p^2,\MZ,M_{\PHa})
\notag\\&\qquad
+2\cw^2 [\cab^2 B_0(p^2,\MW,\MW)+\sab^2 B_0(p^2,\MW,\MHpm)]
\Bigr],\\
\Delta\Sigma^{H_2H_2}(p^2) ={}&
\frac{\alem}{8\pi\cw^2\sw^2} 
(\MHstwo-p^2)
      \Bigl[ 
  \sab^2 B_0(p^2,\MZ,\MZ) +\cab^2 B_0(p^2,\MZ,M_{\PHa})
\notag\\&\qquad
+2\cw^2 [\sab^2 B_0(p^2,\MW,\MW)+\cab^2 B_0(p^2,\MW,\MHpm)]
\Bigr],\\
\Delta\Sigma^{H_1 H_2}(p^2) ={}&
-\frac{\alem}{16\pi\cw^2\sw^2}
\cab\sab(\MHsone+\MHstwo-2p^2)
      \Bigl[ 
  B_0(p^2,\MZ,\MZ) - B_0(p^2,\MZ,M_{\PHa})
\notag\\&\qquad
+2\cw^2 [B_0(p^2,\MW,\MW)-B_0(p^2,\MW,\MHpm)]
\Bigr],
\end{align}
and
\beq
\cab=\cos(\al-\be),\qquad\sab=\sin(\al-\be).
\eeq
For the pseudo-scalar fields we find
\begin{align}
\Delta\Sigma^{G_0 \Ha}(p^2) ={}&
  \frac{\alem}{16\pi\cw^2\sw^2}
  \,\cab\sab(\MHsa-2p^2)
  \left[B_0(p^2,\MZ,\MHone)-B_0(p^2,\MZ,\MHtwo)\right],\\
\Delta\Sigma^{\Ha \Ha}(p^2) ={}&
  \frac{\alem}{8\pi\cw^2\sw^2} 
  (\MHsa-p^2)
  \Bigl[
       2 \cw^2 B_0(p^2,\MW,\MHpm) 
 \notag\\ & 
    + \sab^2 B_0(p^2,\MZ,\MHone)
    + \cab^2 B_0(p^2,\MZ,\MHtwo)
      \Bigr],\\
\Delta\Sigma^{G_0 G_0}(p^2) ={}&
  -\frac{\alem}{8\pi\sw^2\cw^2} p^2
  \Bigl[
  2\cw^2 B_0 (p^2,\MW,\MW)
  \notag\\
  &+{\sab^2} B_0 (p^2,\MZ,\MHtwo)
  +
  {\cab^2}B_0 (p^2,\MZ,\MHone)
  \Bigr],
\end{align}
and for the charged ones we have
\begin{align}
  \Delta\Sigma^{G^\pm H^\mp}(p^2) ={}&
  \frac{\alem}{8\pi\sw^2}
\,\cab\sab(\MHspm-2p^2)
      \left[B_0(p^2,\MW,\MHone)-B_0(p^2,\MW,\MHtwo)\right],\\
\Delta\Sigma^{H^+ H^-}(p^2) ={}&
  \frac{\alem}{8\pi\sw^2} 
  (\MHspm-p^2)
  \biggl[
         B_0(p^2,\MW,\MHa) 
 \notag\\&
  +      4 \sw^2 B_0(p^2,0,\MHpm) 
  +      \frac{(\cw^2-\sw^2)^2}{\cw^2} B_0(p^2,\MZ,\MHpm) 
 \notag\\&
  +  \sab^2 B_0(p^2,\MW,\MHone)
  + \cab^2 B_0(p^2,\MW,\MHtwo)
      \biggr],\\
  \Delta\Sigma^{G^+ G^-}(p^2)  ={}&
  -\frac{\alem}{8\pi\sw^2}p^2
  \biggl[
    4 \sw^2 B_0 (p^2,0,\MW) +
  \frac{4 \cw^4 - 3 \cw^2 + 1}{\cw^2} B_0(p^2,\MZ,\MW)\notag\\
  &+ {\sab^2} B_0(p^2,\MW,\MHtwo)
  + {\cab^2} B_0(p^2,\MW,\MHone)
  \biggr].
\end{align}

\section{Tadpole contributions to scalar mixing and self-energies}
\label{app:tadpoles}

In general, the vertex functions contain explicit and implicit tadpole
contributions. The explicit tadpole contributions, \ie tadpole loop
diagrams, result from the expansion of the effective action about a
point that does not correspond to the stationary point. Implicit
tadpole contributions or tadpole counterterms originate from tadpole
terms in the Lagrangian or shifts in the fields. Since the treatment
of tadpoles involves some freedom, and as we employ two $\MSbar$
schemes differing in the treatment of tadpole counterterms, we briefly
summarize their properties and provide, for completeness, the
tadpole-counterterm expressions for mixing and self-energies that are
needed in those renormalization schemes.

In the \PRTS, the expansion is performed about the stationary point at
the one-loop level, and the bare (squared) masses of the physical
fields are defined as the coefficients of the terms quadratic in the
corresponding fields.  Moreover, we require vanishing mixing between
the physical (mass-eigenstate) scalar fields, \ie there are no
implicit tadpoles in the mixing energies of these ``physical fields''.
In the \THDM we define the mixing angle $\beta$ from the ratio of the
true vevs, $\tan\beta=\vtwo/\vone$.  In this scheme, the tadpole terms
result exclusively from tadpole counterterms in the Lagrangian and are
absent, by definition, in two-point functions that involve only
physical fields.

In the \FJTS, all bare parameters are defined in terms of the bare
parameters of the symmetric Lagrangian. The bare scalar fields are shifted,
$H_{\rB,i}\to H_{\rB,i}+\Delta v_i$, so that the shifted fields
describe excitations about the stationary point. The tadpole terms
result exclusively from the shifts $\Delta v_i$ of the scalar fields.
Tadpole terms appear in almost all vertex functions, apart from the
one-point functions.

In the following we provide the implicit tadpole counterterms for the scalar
self-energies and mixing energies in the \THDM and \HSESM both in the \PRTS and
the \FJTS, using the following convention for the tadpole counterterms,
\beq
\Sigma_{\mathrm{tadpole}}^{HH'} = t_{HH'},
\eeq
corresponding to the second terms on the r.h.s.\ in \refeq{eq:2ptadoles}.
While in the \PRTS the sum of the third and fourth terms in
\refeq{eq:2ptadoles} is zero owing $\dthhat=-\Thhat$,
in the \FJTS this condition must not necessary be fulfilled but is
convenient.

\subsection{Tadpole counterterms in the \FJTS}

In the \FJTS{} the Feynman rules for two-point tadpole counterterms
are easily obtained according to the formula
\begin{align}
  t_{X Y} = - \frac{\dtHone}{\MHone^{2}} C_{X Y \PH_1} -
\frac{\dtHtwo}{\MHtwo^{2}} C_{X Y \PH_2},
\end{align}
where $X,Y$ are fields, and 
$\ri C_{X Y \PH_1}, \ri C_{X Y \PH_2}$ represent
the couplings appearing in the Feynman
rules involving those fields and either $\PH_1$ or $\PH_2$, respectively. 

For example, in the \HSESM{} we derive the following expressions:
\begin{align}
\tHoneHone ={}& 3\dtHone\left(
           \frac{\sa^3}{\vtwo}+\frac{\ca^3}{\vone}\right)
           +\dtHtwo
          \sa\ca
          \left(\frac{\sa}{\vtwo}-\frac{\ca}{\vone}\right) 
           \left(1+\frac{2\MHone^2}{\MHtwo^2}\right)     
               ,\notag\\
\tHtwoHtwo ={}& 3\dtHtwo\left(
          \frac{\ca^3}{\vtwo}-\frac{\sa^3}{\vone}\right)
          +\dtHone
          \ca\sa
          \left(\frac{\ca}{\vtwo}+\frac{\sa}{\vone}\right) 
          \left(1+\frac{2\MHtwo^2}{\MHone^2}\right)     
               ,\notag\\
  \tHoneHtwo ={}&
  \ca\sa\left[
          \dtHone
           \left(\frac{\sa}{\vtwo}-\frac{\ca}{\vone}\right)
           \left(2+\frac{\MHtwo^2}{\MHone^2}\right)
         +\dtHtwo
           \left(\frac{\ca}{\vtwo}+\frac{\sa}{\vone}\right)
           \left(2+\frac{\MHone^2}{\MHtwo^2}\right)
           \right]
           ,\notag\\
  \tGzGz ={}& \tGpmGpm = \frac{\dtHone\sa + \dtHtwo\ca}{\vtwo}.
\end{align}

For the \THDM{} the results for the tadpole counterterms in the \FJTS
can be found in App.~B of \citere{Denner:2016etu}.

\subsection{Tadpole counterterms in the \PRTS}

In the \HSESM{} the tadpole contributions
in the \PRTS{} for the scalar
self-energies and mixing energies read:
\begin{align}
\tHoneHone = {}& \tHtwoHtwo = \tHoneHtwo =   0
,\notag\\
  \tGzGz ={}& \tGpmGpm = \frac{\dtHone\sa + \dtHtwo\ca}{\vtwo}
  \label{eq:tg0_hs_mdts},
\end{align}
where 
$\vtwo={2\sw\MW}/{e}.$

In the \THDM{} they can be extracted from
\citeres{Denner:2016etu,Altenkamp:2017ldc}
and read:
\begin{align}
\tHoneHone ={}& \tHoneHtwo =  \tHtwoHtwo = \tHaHa = \tHpmHpm =0,\notag\\
\tGzGz ={}& \tGpmGpm = 
\frac{1}{v}\left(\dtHone \cab-\dtHtwo \sab
           \right) ,
\notag\\
\tGzHa ={}& \tGpmHpm = 
\frac{1}{v}\left( \dtHone\sab+\dtHtwo\cab 
         \right)
          \label{eq:tgha_mdts},
\end{align}
where 
$v=\sqrt{\vone^2+\vtwo^2}=2\sw\MW/e.$

\section{Vertex corrections for on-shell schemes}
\label{app:vcos}

In this appendix we provide explicit results for relative vertex
corrections entering the on-shell renormalization conditions
introduced in \refse{se:os_thdm} for the \THDM.

\begin{align}
\delta_{\Hone\nu_1\nu_1}=
  -\frac{\alem \cbe}{4 \pi \sw^2\ca}\biggl[
  &  \frac{\cab}{2} \left(1 - \frac{4 \MW^{2}}{\MHone^{2}}\right) B_0(\MHone^2,\MW,\MW)\notag\\
  &+ \frac{\cab}{4 \cw^{2}} \left(1 - \frac{4 \MZ^{2}}{\MHone^{2}}\right) B_0(\MHone^2,\MZ,\MZ) \notag\\
  &- \frac{\sab}{2} \tb B_0(\MHone^2,\MW,\MHpm)
  - \frac{\sab \tb}{4 \cw^{2}} B_0(\MHone^2,\MZ,\MHa)\notag\\
  &+  \left( -\frac{\ca}{\cbe}+\frac{2 \MW^2}{\MHone^2}\cab \right) B_0(0,\MW,0)\notag\\
  &+ \frac{1}{2 \cw^2} \left( -\frac{\ca}{\cbe}+\frac{2 \MZ^2}{\MHone^2}\cab \right) B_0(0,\MZ,0)\notag\\
  &- \cab \MW^{2} \left(1 - \frac{2 \MW^{2}}{\MHone^{2}}\right) C_0(\MHone^2,0,0,\MW,\MW,0) \notag\\
  &- \frac{\MZ^{2} \cab}{2 \cw^{2}} \left(1 - \frac{2 \MZ^{2}}{\MHone^{2}}\right)C_0(\MHone^2,0,0,\MZ,\MZ,0) \notag\\
  &+ \sab \tb \MHpm^{2} C_0(\MHone^2,0,0,\MW,\MHpm,0)\notag\\
  &+ \frac{\MHa^{2} \sab \tb}{2 \cw^{2}} C_0(\MHone^2,0,0,\MZ,\MHa,0) 
\biggr], \\
  \delta_{\Htwo\nu_1\nu_1}=
  \frac{\alem \cbe}{4 \pi \sw^2\sa}  \biggl[
  &- \frac{\sab}{2} \left(1 - \frac{4 \MW^{2}}{ \MHtwo^{2}}\right) B_0(\MHtwo^2,\MW,\MW)\notag\\
  &-\frac{\sab}{4 \cw^{2}} \left(1 - \frac{4 \MZ^{2}}{\MHtwo^{2}}\right) B_0(\MHtwo^2,\MZ,\MZ)\notag\\
  &- \frac{\cab \tb}{2} B_0(\MHtwo^2,\MW,\MHpm) 
  - \frac{\cab \tb}{4 \cw^{2}} B_0(\MHtwo^2,\MZ,\MHa)\notag\\
  &+  \left( \frac{\sa}{\cbe}-\frac{2 \MW^{2}}{\MHtwo^{2}}\sab \right) B_0(0,\MW,0)\notag\\
  & + \frac{1}{2 \cw^{2}}  \left( \frac{\sa}{\cbe}-\frac{2 \MZ^{2}}{\MHtwo^2}\sab \right)  B_0(0,\MZ,0)\notag\\
   &+ \sab \MW^{2} \left(1 - \frac{2 \MW^{2}}{\MHtwo^{2}}\right)C_0(\MHtwo^2,0,0,\MW,\MW,0)\notag\\
  &+ \frac{\sab \MZ^{2}}{2 \cw^{2}} \left(1 - \frac{2 \MZ^{2}}{\MHtwo^{2}
  }\right)C_0(\MHtwo^2,0,0,\MZ,\MZ,0)\notag\\
   &+ \cab \tb \MHpm^{2} C_0(\MHtwo^2,0,0,\MW,\MHpm,0) \notag\\
  &+ \frac{\cab \tb \MHa^{2} }{2 \cw^{2}}C_0(\MHtwo^2,0,0,\MZ,\MHa,0)
  \biggr],\\
\delta_{\Ha\nu_1\nu_1}=
 -  \frac{\alem}{4 \pi \sw^2 \tb}  \biggl[
  &  \frac{\tb}{2} B_0(\MHa^2,\MW,\MHpm) \notag\\
  &- \frac{\sab\ca}{4 \cw^{2}\cbe}  B_0(\MHa^2,\MZ,\MHone)
   + \frac{\cab\sa}{4 \cw^{2}\cbe}  B_0(\MHa^2,\MZ,\MHtwo) \notag\\ 
  &- \tb B_0(0,\MW,0)
  - \frac{\tb}{2 \cw^{2}} B_0(0,\MZ,0)\notag\\
  &-  \MHpm^{2} \tb C_0(\MHa^2,0,0,\MW,\MHpm,0)\notag\\
  &+ \frac{\MHone^{2} \sab\ca}{2 \cw^{2}\cbe} C_0(\MHa^2,0,0,\MZ,\MHone,0)\notag\\
  &- \frac{\MHtwo^{2} \cab\sa}{2 \cw^{2}\cbe} C_0(\MHa^2,0,0,\MZ,\MHtwo,0)
\biggr],
\\
\delta_{\Hone\nu_2\nu_2}=
  -\frac{\alem \sbe}{4 \pi \sw^2\sa}\biggl[
  &  \frac{\cab}{2} \left(1 - \frac{4 \MW^{2}}{\MHone^{2}}\right)
  B_0(\MHone^2,\MW,\MW) \notag\\
&{}+ \frac{\cab}{4 \cw^{2}} \left(1 - \frac{4 \MZ^{2}}{\MHone^{2}}\right)B_0(\MHone^2,\MZ,\MZ)\notag\\
  &+ \frac{\sab}{2 \tb} B_0(\MHone^2,\MW,\MHpm) 
   + \frac{\sab}{4 \cw^{2} \tb} B_0(\MHone^2,\MZ,\MHa)\notag\\
  &  +\left(-\frac{\sa}{\sbe}+\frac{2 \MW^{2}}{\MHone^2}\cab \right) B_0(0,\MW,0)\notag \\
  &+ \frac{1}{2 \cw^{2}}  \left(-\frac{\sa}{\sbe}+\frac{2 \MZ^{2}}{\MHone^2}\cab\right)  B_0(0,\MZ,0)\notag\\
  &- \MW^{2} \cab \left(1 - \frac{2 \MW^{2}}{\MHone^{2}}\right)C_0(\MHone^2,0,0,\MW,\MW,0)\notag\\
  &- \frac{\MZ^{2} \cab}{2 \cw^{2}} \left(1 - \frac{2 \MZ^{2}}{\MHone^{2} }\right)C_0(\MHone^2,0,0,\MZ,\MZ,0) \notag\\
  &- \frac{\sab \MHpm^{2}}{\tb} C_0(\MHone^2,0,0,\MW,\MHpm,0)\notag\\
  &- \frac{\MHa^{2} \sab}{2 \cw^{2} \tb}C_0(\MHone^2,0,0,\MZ,\MHa,0)
  \biggr],\\
\delta_{\Htwo\nu_2\nu_2}=
  -\frac{\alem \sbe}{4 \pi \sw^2\ca}\biggl[
  &- \frac{\sab}{2 } \left(1 - \frac{4 \MW^{2}}{\MHtwo^{2}}\right)B_0(\MHtwo^2,\MW,\MW) \notag\\
  &- \frac{\sab}{4 \cw^{2}} \left(1 - \frac{4 \MZ^{2}}{\MHtwo^{2} }\right)B_0(\MHtwo^2,\MZ,\MZ) \notag\\
  &+ \frac{\cab}{2 \tb} B_0(\MHtwo^2,\MW,\MHpm)
  + \frac{\cab}{4 \cw^{2} \tb} B_0(\MHtwo^2,\MZ,\MHa)\notag\\
  &-  \left( \frac{\ca}{\sbe}+\frac{2 \MW^{2}}{\MHtwo^{2}}\sab\right)  B_0(0,\MW,0)\notag\\
  & - \frac{1}{2 \cw^{2}} \left(\frac{\ca}{\sbe}+\frac{2 \MZ^{2}}{\MHtwo^2} \sab\right)  B_0(0,\MZ,0)\notag\\
  &+ \MW^{2} \sab \left(1 - \frac{2 \MW^{2}}{\MHtwo^{2}}\right)C_0(\MHtwo^2,0,0,\MW,\MW,0)\notag\\
  &+ \frac{\MZ^{2} \sab}{2 \cw^{2}} \left(1 - \frac{2 \MZ^{2}}{\MHtwo^{2}
  }\right)C_0(\MHtwo^2,0,0,\MZ,\MZ,0)\notag\\
  &- \frac{\MHpm^{2}\cab}{\tb} C_0(\MHtwo^2,0,0,\MW,\MHpm,0)\notag\\
  &- \frac{\MHa^{2} \cab}{2 \cw^{2} \tb}C_0(\MHtwo^2,0,0,\MZ,\MHa,0)
\biggr],\\
\delta_{\Ha\nu_2\nu_2}=
  -\frac{\alem \tb}{4 \pi\sw^2}\biggl[
    & \frac{1}{2 \tb} B_0(\MHa^2,\MW,\MHpm)\notag\\
    &+  \frac{\sab\sa}{4 \cw^{2} \sbe}  B_0(\MHa^2,\MZ,\MHone) 
     +  \frac{\cab\ca}{4 \cw^{2} \sbe}  B_0(\MHa^2,\MZ,\MHtwo)\notag\\
    &- \frac{1}{\tb} B_0(0,\MW,0)
    - \frac{1}{2 \cw^{2} \tb} B_0(0,\MZ,0)\notag\\
    &- \frac{\MHpm^{2}}{\tb} C_0(\MHa^2,0,0,\MW,\MHpm,0)\notag\\
    &- \frac{\MHone^{2} \sab\sa}{2 \cw^{2} \sbe}  C_0(\MHa^2,0,0,\MHone,\MZ,0) \notag\\
    &-  \frac{\MHtwo^{2} \cab\ca}{2 \cw^{2} \sbe}  C_0(\MHa^2,0,0,\MHtwo,\MZ,0) 
    \biggr].
\end{align}
For the $B_0$ and $C_0$ functions we use the conventions of \citere{Denner:1991kt}. 

\section{Background-field Ward identities}
\label{app:wi}

In the BFM, the invariance of the effective action under background
gauge transformations gives rise to simple Ward identities (WIs) for the
vertex functions of the background fields \cite{Denner:1994xt}.
These WIs depend on the treatment of tadpoles.%
\footnote{In this appendix the parameters $\MZ$, $\cw$, $\sw$ should
  be understood as bare parameters and the WIs are those for bare
  vertex functions.}

In general, the generating functional of vertex functions $\Gamma$ is
related to the generating functional $ T_{\mathrm{c}}$ of connected Green
functions via the Legendre transformation
\beq\label{eq:Legendre1}
T_{\mathrm{c}}[\{J_{\Xhat}\}] = \ri\sum_{\hat X} \int \rd^4 x J_{\Xhat}(x) \Xhat(x) 
+\ri \Gamma\bigl[\bigl\{\Xhat\bigr\}\bigr],
\eeq
where
\beq\label{eq:Legendre2}
\frac{\delta T_{\mathrm{c}}}{\ri\delta J_{\Xhat}(x)}=\Xhat(x)
\quad\text{or}\quad
\frac{\delta \Gamma}{\delta \Xhat(x)}=-J_{\Xhat}(x).
\eeq
The arguments of $\Gamma$ related to the various (possibly shifted)
fields of the theory are denoted by $\Xhat$ and the corresponding
arguments of $ T_{\mathrm{c}}$ by $J_{\Xhat}$.  As usual, the
connected 2-point Green functions are the inverse of the 2-point
vertex functions
\beq\label{eq:vfproprel}
\frac{\delta^2 T_{\mathrm{c}}}{\ri\delta J_{\Xhat}(x)\ri\delta J_{\Yhat}(y)}=
-\left(\frac{\delta^2 \ri\Gamma}{\delta \Xhat(x)\delta \Yhat(y)}\right)^{-1}.
\eeq
This relation implies, in particular, that the presence of tadpoles in
the propagators is directly connected to the presence of tadpoles in
the 2-point vertex functions. Accordingly, we define the self-energies
$\Sigma$ as the higher-order contributions to the 2-point vertex
functions (full inverse propagators) including all relevant tadpole
contributions (\cf Eq.~\refeq{eq:2ptadoles}). The 1PI contribution to
the self-energies $\Sigma_{\mathrm{1PI}}$ includes only the first term
on the r.h.s.\ of \refeq{eq:2ptadoles}.

\subsection{A Standard Model example}

We illustrate the influence of the tadpole treatment on the BFM WIs
using an example for the SM.  Invariance of the effective action
$\Gamma$ under background gauge transformations related to the
parameter $\hat\theta^Z$ yields in general
\begin{align}
  0={}\delta_{\hat\theta^Z} \Gamma
    ={}&- \partial^x_\mu \frac{\delta \Gamma}{\delta \Zhat_\mu(x)}\bigl[\bigl\{\Xhat\bigr\}\bigr]
  - \frac{e}{2 \cw \sw} \left(\vshift + \Hhat(x)\right) \frac{\delta \Gamma}{\delta \chihat(x)}\bigl[\bigl\{\Xhat\bigr\}\bigr]
\notag\\&  +
   \frac{e}{2\cw\sw}  \chihat(x) 
  \frac{\delta \Gamma}{\delta \Hhat(x)}\bigl[\bigl\{\Xhat\bigr\}\bigr]
  + \ldots\;,
  \label{eq:gfwi}
\end{align}
where $\{\Xhat\}$ represents the set of all background fields of the
SM and we suppressed some terms that are irrelevant in the following.
Here, $\vshift + \Hhat(x)$ denotes the decomposition of the Higgs-boson
field into a constant vev $\vshift$ and the field excitation $\Hhat(x)$,
which will be made more precise for the different renormalization schemes.

The WIs are obtained by taking derivatives of \refeq{eq:gfwi} with
respect to some background fields and evaluating at specific field
values $\Xhat(x)=\overline{\Xhat}$. Differentiating for instance with
respect to the would-be Goldstone-boson field $\hat\chi$ yields
\begin{align}
  0={}& {-\partial^x_\mu} \frac{\delta^2 \Gamma}{\delta \Zhat_\mu(x)
    \delta \chihat(y)} \bigl[\bigl\{\overline{\Xhat}\bigr\}\bigr]
  - \frac{e}{2 \cw \sw} \left(\vshift+ \overline{\Hhat} \right) \frac{\delta^2 \Gamma}{\delta \chihat(x) \delta
  \chihat(y)}\bigl[\bigl\{\overline{\Xhat}\bigr\}\bigr]
\notag\\&  + 
   {\frac{e}{2 \cw \sw}} \delta(x-y) \frac{\delta \Gamma}{\delta\Hhat(x)}\bigl[\bigl\{\overline{\Xhat}\bigr\}\bigr]
+
   {\frac{e}{2\cw\sw}} \, \overline{\chihat}\, 
  \frac{\delta^2 \Gamma}{\delta \Hhat(x) \delta \chihat(y)}\bigl[\bigl\{\overline{\Xhat}\bigr\}\bigr]
  + \ldots\; .
  \label{eq:wi1}
\end{align}

Expanding the vertex functional about the point
$\Xhat=\overline{\Xhat}=0$ for  $\vshift=v_{\rB}$, \ie for vanishing fields
and without any extra shift in the fields, the identity
\refeq{eq:wi1} turns into
\begin{align}
  0 ={}& {-\partial^x_\mu} \frac{\delta^2 \Gamma}{\delta \Zhat_\mu(x)
    \delta\chihat(y)} \bigl[\bigl\{0\bigr\}\bigr]
  - \frac{e}{2 \cw \sw} v_{\rB}\frac{\delta^2 \Gamma}{\delta \chihat(x) \delta
  \chihat(y)}\bigl[\bigl\{0\bigr\}\bigr] + 
   {\frac{e}{2 \cw \sw}} \delta(x-y) \frac{\delta \Gamma}{\delta\Hhat(x)}\bigl[\bigl\{0\bigr\}\bigr].
  \label{eq:wi1naive}
\end{align}
Transforming to momentum space, expanding to one-loop order and using
$\MZ=ev/(2\cw\sw)$ implies for the one-loop contributions%
\footnote{For the definition of the Lorentz decomposition of the
  self-energies we use the conventions of \citere{Denner:1994xt}.
  Note, however, that at variance with \citere{Denner:1994xt}, where
  all self-energies $\Sigma$ are 1PI, we use self-energies $\Sigma$
  based on complete 2-point functions (\cf Eq.~\refeq{eq:2ptadoles})
  and denote their 1PI parts by $\Sigma_{\mathrm{1PI}}$.}
\begin{align}
  0 ={}& p^2 \Sigma_{\mathrm{1PI}}^{\Zhat\chihat}(p^2) - \ri\MZ\Sigma_{\mathrm{1PI}}^{\chihat\chihat}(p^2)
 + 
   \ri\frac{e}{2 \cw \sw} \Thhat,
  \label{eq:wi1senaive}
\end{align}
which reproduces the WI (30) of \citere{Denner:1994xt}.
The Legendre transform \refeq{eq:Legendre2} yields in this case
\beq
\left.\frac{\delta T_{\mathrm{c}}}{\ri\delta J_{\Xhat}(x)}\right|_{\hat Y=0} =\Xhat(x) = 0,
\eeq
\ie the connected one-point functions, the explicit tadpoles, vanish
and consequently the vertex functions are one-particle irreducible
(1PI) as indicated in \refeq{eq:wi1senaive}. Moreover, no implicit
tadpoles occur since we expand about the bare vacuum.  Note, however,
that the so-defined vertex functions do not correspond to a generating
functional of connected Green functions for vanishing sources and
cannot be used to calculate the $S$-matrix in a straight-forward way
unless additional tadpole contributions are included properly.

In the case of spontaneous symmetry breaking, the correct generating
functional of connected Green functions is related to the vertex
functional at the stationary point $\overline{\Xhat}_\mathrm{s}$, defined by
\begin{align}
  \frac{\delta \Gamma}{\delta \Xhat} [\{\overline{\Xhat}_\mathrm{s}\}] = 0,
  \label{eq:statcon}
\end{align}
which according to \refeq{eq:Legendre2} corresponds to $J_{\Xhat}(x)=0$.
On the other hand, \refeq{eq:statcon} implies that all explicit tadpoles
vanish. 
The expansion about the  stationary point can be implemented in
different ways as detailed in the following.

In the \PRTS, the fields are expanded about the one-loop vev $\vshift=v$. The
vertex functions are defined at the stationary point 
$\overline{\Xhat}_\mathrm{s}\equiv 0$ for all fields,
and consequently all explicit tadpoles vanish. The WI \refeq{eq:wi1}
becomes
\begin{align}
  0={}& {-\partial^x_\mu} \frac{\delta^2 \Gamma}{\delta \Zhat_\mu(x)
    \delta\chihat(y)} \bigl[\bigl\{0\bigr\}\bigr]
  - \frac{ev}{2 \cw \sw} \frac{\delta^2 \Gamma}{\delta \chihat(x) \delta
  \chihat(y)}\bigl[\bigl\{0\bigr\}\bigr].
  \label{eq:wi1mdts}
\end{align}
Transforming to momentum space implies for the one-loop self-energies
\begin{align}
  0 ={}& p^2 \Sigma^{\Zhat\chihat}(p^2) - \ri\MZ\Sigma^{\chihat\chihat}(p^2).
\end{align}
Splitting the self-energies into irreducible parts and
implicit tadpole terms results in
\begin{align}
  0 ={}& p^2 \Sigma_{\mathrm{1PI}}^{\Zhat\chihat}(p^2) 
-
\ri\MZ\left[\Sigma_{\mathrm{1PI}}^{\chihat\chihat}(p^2)+\frac{\dthhat}{v}\right]
\notag\\={}&
 p^2 \Sigma_{\mathrm{1PI}}^{\Zhat\chihat}(p^2) 
-
\ri\MZ\Sigma_{\mathrm{1PI}}^{\chihat\chihat}(p^2)
+\ri\frac{e}{2\cw\sw} {\Thhat}.
\label{eq:wi1mdtsse1pi}
\end{align}
In the last step we used the fact that the Higgs tadpole is cancelled
by the corresponding counterterm (implicit tadpole), \ie
$\dthhat=-\Thhat$. The last line of Eq.~\refeq{eq:wi1mdtsse1pi} coincides with \refeq{eq:wi1senaive} and  the WI (30) in
\citere{Denner:1994xt}.

In the \FJTS, two different approaches can be used. %
In the first approach, which is our default, the fields are expanded
about the stationary point, but the full vev consists of its
tree-level part $v_{\rB}$ and the field 
shift $\Delta v$,
\beq
v=v_{\rB}+\Delta v.
\label{eq:vshift}
\eeq 
Attributing this shift to the vev, the stationary point corresponds to
$\overline{\Xhat}_\mathrm{s}=0$ for all fields. All explicit tadpoles
vanish, but owing to the shift 
in \refeq{eq:vshift} extra terms
proportional to $\Delta v$ appear in the WI, and \refeq{eq:wi1}
becomes
\begin{align}
  0 ={}& {-\partial^x_\mu} \frac{\delta^2 \Gamma}{\delta \Zhat_\mu(x)
    \delta\chihat(y)} \bigl[\bigl\{0\bigr\}\bigr]
  - \frac{e}{2 \cw \sw} (v_{\rB}+\Delta v)\frac{\delta^2 \Gamma}{\delta \chihat(x) \delta
  \chihat(y)}\bigl[\bigl\{0\bigr\}\bigr].
  \label{eq:wi1fjtsa}
\end{align}
Transforming to momentum space and expanding to one-loop order yields
\begin{align}
  0 ={}& p^2 \Sigma^{\Zhat\chihat}(p^2) - \ri\MZ\Sigma^{\chihat\chihat}(p^2)
 - \ri\MZ p^2 \frac{\Delta v}{v_{\rB}}.
  \label{eq:wi1sefjtsa}
\end{align}
Upon a perturbative expansion and transformation to momentum space,
the condition \refeq{eq:statcon} for $\overline{\Xhat}_\mathrm{s}=0$
fixes $\Delta v$:
\beq
\Delta v = -\frac{\dthhat}{\MH^2} = \frac{\Thhat}{\MH^2}.
\eeq
Splitting the self-energies into irreducible parts and
implicit tadpole terms results in
\begin{align}
  0 ={}& p^2 \left[\Sigma_{\mathrm{1PI}}^{\Zhat\chihat}(p^2) 
        +\ri\MZ \frac{\Delta v}{v_{\rB}}\right]
-
\ri\MZ\left[\Sigma_{\mathrm{1PI}}^{\chihat\chihat}(p^2)-\MH^2
  \frac{\Delta v}{v_{\rB}}\right] 
 - \ri\MZ p^2 \frac{\Delta v}{v_{\rB}}
\notag\\={}&
 p^2 \Sigma_{\mathrm{1PI}}^{\Zhat\chihat}(p^2) 
-
\ri\MZ\Sigma_{\mathrm{1PI}}^{\chihat\chihat}(p^2)
+\ri\frac{e}{2\cw\sw} {\Thhat}.
  \label{eq:wi1se1pifjts}
\end{align}
Thus, in terms of irreducible self-energies we again recover
\refeq{eq:wi1senaive}, \ie the WI (30) in \citere{Denner:1994xt}.
 
Alternatively, the \FJTS can be used without the shift $\Delta v$, \ie
for $\vshift=v_\rB$. Then, the stationary point of the vertex functional is
determined from \refeq{eq:statcon}, 
so that
\begin{align}
0=\frac{\delta \Gamma}{\delta
  \Hhat(x)}\bigl[\bigl\{\overline{\Xhat}_\mathrm{s}\bigr\}\bigr]
={}&
\frac{\delta \Gamma}{\delta
  \Hhat(x)}\bigl[\bigl\{0\bigr\}\bigr]  
+ 
\int\rd^4 z\frac{\delta^2 \Gamma}{\delta
  \Hhat(x)\delta\Hhat(z)}\bigl[\bigl\{0\bigr\}\bigr]\,
  \overline{\Hhat}_\mathrm{s}(z)
  +\mathcal{O}\Bigl(\overline{\Hhat}_\mathrm{s}^2\Bigr).
\label{eq:tadexp}
\end{align}
Together with \refeq{eq:vfproprel} and after a perturbative expansion
of the propagator this yields
\begin{align}
\overline{\Hhat}_\mathrm{s}(x) &{} =
\ri\int\rd^4 z
\frac{\delta T_{\mathrm{c}}}{\ri\delta J_{\Hhat}(x)\ri\delta J_{\Hhat}(z)}\bigl[\bigl\{0\bigr\}\bigr]
\frac{\delta\Gamma}{\delta
  \Hhat(z)}\bigl[\bigl\{0\bigr\}\bigr]
  +\mathcal{O}\Bigl(\overline{\Hhat}_\mathrm{s}^2\Bigr), \notag\\
&{} =
\ri\,\frac{\ri}{-\MH^2}\,\Thhat +\mathcal{O}\Bigl(\overline{\Hhat}_\mathrm{s}^2\Bigr),
\end{align} 
where the second line results after transformation to momentum space.
Thus, $\overline{\Hhat}_\mathrm{s}$ corresponds to the tadpole with
external propagator attached, so that we get at NLO
\beq
\overline{\Hhat}_\mathrm{s} = \frac{\Thhat}{\MH^2},
\label{eq:statpointfjts}
\eeq
while 
\beq
\overline{\Xhat}_\mathrm{s} = 0, \quad \Xhat\ne \Hhat.
\eeq
In this formulation, \refeq{eq:wi1} turns into
\begin{align}
  0 ={}& {-\partial^x_\mu} \frac{\delta^2 \Gamma}{\delta \Zhat_\mu(x)
    \delta\chihat(y)} \bigl[\bigl\{\overline{\Xhat}_\mathrm{s}\bigr\}\bigr]
  - \frac{e}{2 \cw \sw} (v_{\rB}+\overline{\Hhat}_\mathrm{s})\frac{\delta^2 \Gamma}{\delta \chihat(x) \delta
  \chihat(y)}\bigl[\bigl\{\overline{\Xhat}_\mathrm{s}\bigr\}\bigr].
  \label{eq:wi1fjtsb}
\end{align}
While the WI \refeq{eq:wi1fjtsb} and the vertex functions
$\displaystyle\frac{\delta^2 \Gamma}{\delta
  \Xhat(x)\delta\Yhat(y)}\bigl[\bigl\{\overline{\Xhat}_\mathrm{s}\bigr\}\bigr]$
appearing therein evaluated at the stationary point
$\overline{\Xhat}_\mathrm{s}$ do not contain explicit tadpoles, a perturbative
expansion yields
\begin{align}
\frac{\delta^2 \Gamma}{\delta
  \Xhat(x)\delta\Yhat(y)}\bigl[\bigl\{\overline{\Xhat}_\mathrm{s}\bigr\}\bigr]
={}&
\frac{\delta^2 \Gamma}{\delta
  \Xhat(x)\delta\Yhat(y)}\bigl[\bigl\{0\bigr\}\bigr]  \notag\\
&{}+ \int\rd^4 z\frac{\delta^3 \Gamma}{\delta
  \Xhat(x)\delta\Yhat(y)\delta\Hhat(z)}\bigl[\bigl\{0\bigr\}\bigr]
  \overline{\Hhat}_\mathrm{s}(z)
  +\mathcal{O}\Bigl(\overline{\Hhat}_\mathrm{s}^2\Bigr),
\label{eq:VFexp}
\end{align}
where the terms $\mathcal{O}\Bigl(\overline{\Hhat}_\mathrm{s}^2\Bigr)$
are beyond one-loop order.  The first term on the r.h.s.\ of
\refeq{eq:VFexp} delivers the two 1PI contributions appearing in
\refeq{eq:wi1naive}, the second term corresponds to an explicit
tadpole contribution to the $\hat X\hat Y$ self-energy.  Thus, in
terms of usual building blocks the vertex functions are composed of
1PI terms including contributions from explicit tadpoles. The
corresponding WIs in momentum space are obtained upon replacing
$\Delta v \to \Thhat/\MH^2$ in \refeq{eq:wi1sefjtsa} and
\refeq{eq:wi1se1pifjts}, \ie they are equivalent to those equations.

\subsection{A \THDM example}

We give a second example%
\footnote{For other examples of BFM Ward identities in the \THDM see
  footnote 5 of \citere{Altenkamp:2017ldc}.}
 that is relevant for the renormalization of
the mixing angle $\beta$ in the \THDM. Starting from the analogue of
the WI \refeq{eq:gfwi} in the \THDM and differentiating with respect
to the physical pseudoscalar field $\Ha$ yields when expanding about
the stationary point:
\begin{align}
  0 ={}& {-\partial^x_\mu} \frac{\delta^2 \Gamma}{\delta \Zhat_\mu(x)
    \delta\Hahat(y)} \bigl[\bigl\{\overline{\Xhat}_\mathrm{s}\bigr\}\bigr]
  - \frac{e}{2 \cw \sw} \left(\cbe\vone+\sbe\vtwo \right) \frac{\delta^2 \Gamma}{\delta \Gzhat(x) \delta
  \Hahat(y)}\bigl[\bigl\{\overline{\Xhat}_\mathrm{s}\bigr\}\bigr]
\notag\\&  +
   {\frac{e}{2 \cw \sw}}  \left(\sbe\vone-\cbe\vtwo \right)
   \frac{\delta^2 \Gamma}{\delta\Hahat(x)\delta\Hahat(y)}\bigl[\bigl\{\overline{\Xhat}_\mathrm{s}\bigr\}\bigr]
+\ldots \;.
  \label{eq:wi2}
\end{align}

In the \PRTS, where $\vone$ and $\vtwo$ are the one-loop vevs, 
$\tb=\vtwo/\vone$, and $\overline{\Xhat}_\mathrm{s}=0$, this becomes
\begin{align}
  0 ={}& {-\partial^x_\mu} \frac{\delta^2 \Gamma}{\delta \Zhat_\mu(x)
    \Hahat(y)} \bigl[\bigl\{0\bigr\}\bigr]
  - \MZ\frac{\delta^2 \Gamma}{\delta \Gzhat(x) \delta
  \Hahat(y)}\bigl[\bigl\{0\bigr\}\bigr]
  \label{eq:wi2prts},
\end{align}
where we used
\beq
\MZ= \frac{e}{2 \cw \sw}v =  \frac{e}{2 \cw \sw}(\cbe\vone+\sbe\vtwo).
\eeq
Transforming to momentum space and setting $p^2=0$, this leads to 
\begin{align}
  0 ={}& \Sigma^{\Gzhat\Hahat}(0),
\label{eq:wi_chiHa_mdts}
\end{align}
or in terms of 1PI mixing energies
\begin{align}
  0 ={}& \Sigma_{\mathrm{1PI}}^{\Gzhat\Hahat}(0) + 
\frac{e}{2 \MW \sw}
\left(\sab\dtHone+\cab\dtHtwo\right).
  \label{eq:wi21pi}
\end{align}

In the \FJTS, on the other hand, where $\vone=\vone{}_\rB+\Delta\vone$,
$\vtwo=\vtwo{}_\rB+\Delta\vtwo$, and $\overline{\Xhat}_\mathrm{s}=0$,
we find
\begin{align}
  0 ={}& {-\partial^x_\mu} \frac{\delta^2 \Gamma}{\delta \Zhat_\mu(x)
    \Hahat(y)} \bigl[\bigl\{0\bigr\}\bigr]
  -
  \MZ\left(1+\cbe\frac{\Delta\vone}{v_{\rB}}+\sbe\frac{\Delta\vtwo}{v_{\rB}}
\right)
\frac{\delta^2 \Gamma}{\delta \Gzhat(x) \delta\Hahat(y)}\bigl[\bigl\{0\bigr\}\bigr]
\notag\\&
  + \MZ\left(
\sbe\frac{\Delta\vone}{v_{\rB}}
-\cbe\frac{\Delta\vtwo}{v_{\rB}}
\right)
\frac{\delta^2 \Gamma}{\delta \Hahat(x) \delta\Hahat(y)}
\bigl[\bigl\{0\bigr\}\bigr],
  \label{eq:wi2fjts}
\end{align}
where we used
\beq
\MZ= \frac{e}{2 \cw \sw}v_{\rB} =  \frac{e}{2 \cw \sw}(\cbe\vone{}_\rB+\sbe\vtwo{}_{\rB})
\eeq
and $\cbe$, $\sbe$, $\tb=\vtwo{}_\rB/\vone{}_\rB$ are bare parameters.

Transformation to momentum space, setting $p^2=0$ and using the
explicit LO contributions to the vertex functions, results in 
\begin{align}
  0 ={}& \Sigma^{\Gzhat\Hahat}(0) 
-
\frac{1}{v_{\rB}}\left(\cbe{\Delta\vtwo}-\sbe{\Delta\vone}\right)
\MHa^2.
\label{eq:wi_chiHa_fjts}
\end{align}
Splitting the mixing energy into 1PI parts and implicit counterterms
and expressing $\Delta v_i$ by tadpole counterterms, we obtain again 
\refeq{eq:wi21pi}.

\section{Parameter conversion tables}
\label{se:conversiontables}

In this appendix we give results for the parameter conversion from the
on-shell input schemes \OS and \OSonetwo in the \HSESM{} and \THDM,
respectively, to the other renormalization schemes in two different
variants, as used for the results presented in \refse{se:results}.
The two variants are full conversions, as explained in
\refse{se:conversion}, are equivalent at NLO, but differ in higher
orders due to the tadpole counterterm scheme and the precise form of
the input parameters.  For the details see the caption of the tables
in the following.

\subsection{\HSESM\ scenarios of \refta{tab:SESMinput}}

In \refta{tab:conversion-HSESM-OS} the results for the full conversion
of parameters from the \OS as input scheme to other schemes for the
considered benchmark scenarios of \refta{tab:SESMinput} are shown. No
large conversion effects are observed, and the two different
conversion variants mutually agree.

\begin{table}
\centerline{\renewcommand{\arraystretch}{1.25}
\begin{tabular}{|c||c|c|c|c|c|c|c|c|c|c|c|c|c|c|c|c|}
\hline
Conversion 1 & \multicolumn{2}{c|}{BHM200$^+$} & \multicolumn{2}{c|}{BHM200$^-$} 
& \multicolumn{2}{c|}{BHM400} & \multicolumn{2}{c|}{BHM600} 
\\
Scheme 
  & \multicolumn{1}{c|}{\sa} & \multicolumn{1}{c|}{$\frac{\lambda_3}{2}$}  
  & \multicolumn{1}{c|}{\sa} & \multicolumn{1}{c|}{$\frac{\lambda_3}{2}$}  
  & \multicolumn{1}{c|}{\sa} & \multicolumn{1}{c|}{$\frac{\lambda_3}{2}$}  
  & \multicolumn{1}{c|}{\sa} & \multicolumn{1}{c|}{$\frac{\lambda_3}{2}$}  
\\
\hline
\hline
  \OS         & $0.29$&  $0.07$ & $-0.29$ &$-0.07$ & $0.26$ & $0.17$ & $0.22$ & $0.23$\\
\hline
\hline
  \MSbarPRTS  & $0.302$& $0.073$& $-0.304$& $-0.073$ & $0.267$ & $0.175$ & $0.226$ & $0.236$\\
\hline
  \MSbarFJTS  & $0.321$& $0.077$& $-0.316$& $-0.076$ & $0.264$ & $0.173$ & $0.212$ & $0.222$\\
\hline
  \RSBFM  & $0.290$&  $0.070$ & $-0.290$ & $-0.070$ & \change{$0.258$} & \change{$0.172$} & $0.218$ & \change{$0.237$} \\
\hline
\multicolumn{9}{c}{}\\
\hline
Conversion 2 & \multicolumn{2}{c|}{BHM200$^+$} & \multicolumn{2}{c|}{BHM200$^-$} 
& \multicolumn{2}{c|}{BHM400}     & \multicolumn{2}{c|}{BHM600} 
\\
Scheme 
  & \multicolumn{1}{c|}{\sa} & \multicolumn{1}{c|}{$\frac{\lambda_3}{2}$}  
  & \multicolumn{1}{c|}{\sa} & \multicolumn{1}{c|}{$\frac{\lambda_3}{2}$}  
  & \multicolumn{1}{c|}{\sa} & \multicolumn{1}{c|}{$\frac{\lambda_3}{2}$}  
  & \multicolumn{1}{c|}{\sa} & \multicolumn{1}{c|}{$\frac{\lambda_3}{2}$}  
\\
\hline
\hline
  \OS & $0.29$&  $0.07$ & $-0.29$ &$-0.07$ & $0.26$ & $0.17$ & $0.22$ & $0.23$\\
\hline
\hline
  \MSbarPRTS  & $0.302$ & $0.073$ & $-0.304$ & $-0.073$ & $0.267$ & $0.174$ & $0.226$ & $0.236$\\
\hline
  \MSbarFJTS  & $0.319$ & $0.077$ & $-0.315$ & $-0.076$ & $0.264$ & $0.172$ & $0.212$ & $0.222$ \\
\hline
  \RSBFM  & $0.290$ & $0.070$ & $-0.290$ & $-0.070$ & $0.258$ & $\change{0.172}$
  & $0.220$ & $\change{0.238}$\\
\hline
\end{tabular}
}
  \caption{Conversion of parameters from the \OS scheme as
    input scheme to other renormalization schemes at the central scale
    $\mu=\MHtwo$ performed in the \PRTS{} (upper table, used for the Higgs
    decays in \refse{se:h4f_HSESM}) and \FJTS{} (lower table, used for
    Higgs-production processes in \refse{se:HZ_HSESM}) tadpole scheme.
  Besides the choice of tadpole counterterm scheme, also the choice of bare
  parameters for the matching differs. In the upper table $\tan
  \alpha_\mathrm{B}$ and $\lambda_3$ are used while in the lower one
  $\alpha_\mathrm{B}$ and $\tan \beta_\mathrm{B}$ are used. }
\label{tab:conversion-HSESM-OS}
\end{table}

\subsection{\THDM\ scenarios of \refta{tab:THDMinput}}

In \refta{tab:conversion-THDM-OS12} the corresponding conversion in
the \THDM{} from the \OSonetwo as input scheme to other
renormalization schemes is shown.
\begin{table}
\centerline{\renewcommand{\arraystretch}{1.25}
\begin{tabular}{|c||c|c|c|c|c|c|c|c|c|c|c|c|c|c|c|c|}
\hline
Conversion 1 & \multicolumn{2}{c|}{A1} & \multicolumn{2}{c|}{A2} 
& \multicolumn{2}{c|}{B1} & \multicolumn{2}{c|}{B2} 
\\
Scheme 
  & \multicolumn{1}{c|}{\cab} & \multicolumn{1}{c|}{\tb}  
  & \multicolumn{1}{c|}{\cab} & \multicolumn{1}{c|}{\tb}  
  & \multicolumn{1}{c|}{\cab} & \multicolumn{1}{c|}{\tb}  
  & \multicolumn{1}{c|}{\cab} & \multicolumn{1}{c|}{\tb}  
\\
\hline
\hline
  \OSonetwo  & $0.1$ & $2.0$ & $0.2$ & $2.0$ & $0.15$ & $4.5$  & $0.3$ & $3.0$ \\
\hline
\hline
  \MSbarPRTS  & $0.137$ & $1.90$ & $0.225$ & $1.91$ & $0.153$ & $4.40$  & $0.176$ & $2.83$\\
\hline
  \MSbarFJTS  & $0.090$ & $1.93$ & $0.157$ & $1.91$ & $0.061$ & $3.82$ & $-0.031$ & $3.70$ \\
\hline
  \OSone  & $0.102$ & $1.92$ & $0.205$ & $1.91$ & $0.208$ & $3.46$    & $0.309$ & $2.87$\\
\hline
  \OStwo  & $0.096$ & $2.02$ & $0.195$ & $2.03$ & $0.145$ & $4.61$    & $0.298$ & $3.02$\\
\hline
\RSBFM    & \change{$0.110$} & \change{$1.91$} & \change{$0.209$} & \change{$1.92$} & 
\change{$0.149$} & \change{$4.42$}     & \change{$0.323$} & \change{$2.82$}\\
\hline
\multicolumn{9}{c}{}\\
\hline
Conversion 2 & \multicolumn{2}{c|}{A1} & \multicolumn{2}{c|}{A2} 
& \multicolumn{2}{c|}{B1} & \multicolumn{2}{c|}{B2} 
\\
Scheme 
  & \multicolumn{1}{c|}{\cab} & \multicolumn{1}{c|}{\tb}  
  & \multicolumn{1}{c|}{\cab} & \multicolumn{1}{c|}{\tb}  
  & \multicolumn{1}{c|}{\cab} & \multicolumn{1}{c|}{\tb}  
  & \multicolumn{1}{c|}{\cab} & \multicolumn{1}{c|}{\tb}  
\\
\hline
\hline
  \OSonetwo  & $0.1$ & $2.0$ & $0.2$ & $2.0$ & $0.15$ & $4.5$  & $0.3$ & $3.0$ \\
\hline
\hline
  \MSbarPRTS  & $0.155$ & $1.92$ & $0.234$ & $1.92$ & $0.152$ & $4.51$ & $0.197$ & $2.80$ \\
\hline
  \MSbarFJTS  & $0.089$ & $1.93$ & $0.148$ & $1.89$ & $-0.015$ & $2.34$  & $0.049$ & $3.22$\\
\hline
  \OSone      & $0.102$ & $1.92$ & $0.207$ & $1.91$ & $0.199$ & $4.61$  & $0.308$ & $2.89$ \\
\hline
  \OStwo      & $0.093$ & $2.02$ & $0.193$ & $2.02$ & $0.147$ & $4.44$  & $0.299$ & $3.01$\\
\hline
  \RSBFM      & $\change{0.116}$ & $\change{1.92}$ & $\change{0.214}$ &
  $\change{1.92}$ & $\change{0.151}$ & $\change{4.44}$ & $\change{0.319}$ &
  $\change{2.83}$\\
\hline
\end{tabular}
}
  \caption{Conversion of parameters from the \OSonetwo scheme as
    input scheme to other renormalization schemes at the central scale $\mu_0$
    performed in the \PRTS{} (upper table, used for the Higgs decays in
    \refse{se:h4f_THDM}) and \FJTS{} (lower table, used for Higgs-production
    processes in \refses{se:HZ_THDM} and \ref{se:VBF_THDM}) tadpole scheme.
  Besides the choice of tadpole counterterm scheme, also the bare parameters for
  the matching differs. In the upper table $\tan \alpha_\mathrm{B}$ and $\tan
  \beta_\mathrm{B}$ are used while in the lower one $\alpha_\mathrm{B}$ and
  $\beta_\mathrm{B}$ are used. }
\label{tab:conversion-THDM-OS12}
\end{table}
As compared to the \HSESM, the conversion effects for the scenarios in
\refta{tab:THDMinput} are more pronounced.
Generically the conversion effects are $\lsim5\%$ for the on-shell
schemes and for the \RSBFM{} scheme, with the only exception being the
\OSone scheme in scenario~B1 with conversion effects of the order of
$\sim40\%$. The conversion effects to the \MSbar schemes are typically
larger than those to the on-shell schemes, and the conversion to
\MSbarFJTS{} becomes perturbatively unstable (and thus fails
completely) for scenarios B1 and B2 while for \MSbarPRTS only B2 is
unstable.

Comparing the two conversions, the differences in the scenarios A1 and
A2 to \MSbarPRTS (\MSbarFJTS) amount to $13\%(6\%)$ and $1\%(1\%)$ for
$c_{\alpha\beta}$ and $\tb$, respectively.  The different conversions
to the on-shell schemes and the \RSBFM scheme agree on the level of
$4\%$, with scenario B1 being again the only exception where we find a
difference of $33\%$ in the conversion of $\tb$.

\bibliographystyle{JHEPmod}
\bibliography{mixing} 

\end{document}